\documentclass[draft]{agujournal2019}
\usepackage{url} 
\usepackage{lineno}
\usepackage[inline]{trackchanges} 

\usepackage{tabularray}
\usepackage{threeparttable}
\usepackage{amssymb}

\usepackage{soul}

\newcommand{\ang}[1]{#1^{\circ}}
\newcommand{\sgn}{\text{sgn}}
\newcommand{\Prob}{\mathbb{P}\,}
\linenumbers

\usepackage{enumitem}

\usepackage{xcolor} 
\usepackage{amsmath} 

\newcounter{subitem}[enumi]

\newcommand{\minewa}[1]{{\color{black}{#1}}}
\newcommand{\minewb}[1]{{\color{black}{#1}}}
\newcommand{\minewc}[1]{{\color{black}{#1}}}

\newcommand{\eqmiold}[1]{%
  {\textcolor{gray}{%
    \hbox to 0pt{\rule[0.5ex]{\textwidth}{0.4pt}\hss}%
    #1%
  }}%
}

\renewcommand{\textsection}{Section~}

\usepackage{amsmath}

\draftfalse

\AtBeginDocument{%
	\pagestyle{empty} 
}
\nolinenumbers

\begin{document}

%
%



\title{Automated Detection of Short-term Slow Slip Events in Southwest Japan}

%
%





\authors{Yiming Ma\affil{1\minewc{,3}}\thanks{\minewc{Auckland}, New Zealand}, \minewc{Andreas Anastasiou\affil{2},} Fabien Montiel \affil{1}}


\affiliation{1}{Department of Mathematics and Statistics, University of Otago, Dunedin, New Zealand}
\minewc{\affiliation{2}{Department of Mathematics and Statistics, University of Cyprus, Nicosia, Cyprus}}
\minewc{\affiliation{3}{Department of Mathematical Sciences, Auckland University of Technology, Auckland, New Zealand}}





\correspondingauthor{Yiming Ma}{yiming.ma@aut.ac.nz}




\begin{keypoints}
\item We develop a change-point detection method for identifying automatically the start and end times of short-term SSEs in GPS data.
\item Synthetic tests verified its validity and demonstrated that the new method outperforms two existing methods. 
\item We illustrate the effectiveness of the method in detecting short-term SSEs in Southwest Japan.
\end{keypoints}

%
%

%
%


\begin{abstract}
Inferring from the occurrence pattern of slow slip events (SSEs) the probability of triggering a damaging earthquake within the nearby velocity weakening portion of the plate interface is critical for hazard mitigation. Although robust methods exist to detect long-term SSEs consistently and efficiently, detecting short-term SSEs remains a challenge. In this study, we propose a novel statistical approach, called singular spectrum analysis isolate-detect (SSAID), for automatically estimating the start and end times of short-term SSEs in GPS data. The method recasts the problem of detecting SSEs as that of identifying change-points in a piecewise non-linear signal. This is achieved by obscuring the deviation from piecewise-linearity in the underlying SSE signals using added noise. We verify its effectiveness on a range of model-generated synthetic SSE data with different noise levels, and demonstrate its superior performance compared to two existing methods. We illustrate its capability in detecting short-term SSEs in observed GPS data from $36$ stations in southwest Japan via the co-occurrence of non-volcanic tremors, hypothesis tests and fault estimation.
\end{abstract}

\section*{Plain Language Summary}
[SSEs, a type of slow earthquakes, are thought to play an important role in releasing strain in subduction zones, and affect the occurrence of large earthquakes, although their exact connection remains unclear. Detecting accurately the start and end times of SSEs is one prerequisite to illuminate their interactions with large earthquakes. However, no robust detection method has been well developed so far. SSEs are widely recorded by GPS network, part of the Global Navigation Satellite System (GNSS). Most undetected SSEs in GPS data are short-term SSEs, i.e. SSEs with short durations ranging from days to weeks, since the amplitude changes in the GPS data trend from short-term SSEs are somewhat small, close to (or even lower than) the background noise. Therefore, more urgent efforts should be devoted to developing a rapid automated method for detecting short-term SSEs in GPS data. In this study, we develop a change-point detection method for piecewise signals to detect automatically the start and end times of short-term SSEs in GPS data. We demonstrate its effectiveness on both simulated and observed GPS data. The results show that the detection performance of our method regarding the number of estimated change-points and their locations outperform two existing methods.]

\section{Introduction}
Slow slip events (SSEs) are fault slips occurring at the subduction interface between tectonic plates. They are roughly categorized into short-term SSEs (in the order of days to weeks) and long-term SSEs (in the order of months to years) \cite{Obara2020}. They constitute a type of slow earthquakes \cite{Hirose1999,Mitsui2006,Obara2016,Obara2020}. SSEs play a vital role in releasing stress along subduction interfaces. The associated episodic stress perturbations on the seismogenic zone have been linked to the occurrence of larger natural earthquakes \cite{Segall2006,Ito2013,Bartlow2014,Radiguet2016,Voss2018,Bletery2020}. SSEs might also prevent the rupture of large earthquakes from propagating further along the subduction interface, while large earthquakes can also initiate SSEs in the nearby transition zone \cite{Hirose2012,Yarai2013,Nishikawa2019,Wallace2020,Nishimura2021}. Here the transition zone refers to the area where SSEs occur along the subduction interface. Understanding the process governing SSEs could potentially help us forecast impending earthquakes, although the underlying geophysical mechanism for forming SSEs remains elusive \cite{Mazzotti2004,Jordan2010,Lohman2013,Beeler2014,Obara2016,Barbot2019,Obara2020}. 

Detecting SSEs accurately could be the key to determine the mechanism generating SSEs and understand their connection to large earthquakes \cite{Ikari2013,Saffer2015,Ozawa2019,Nishimura2021}. SSEs are generally recorded through geodetic measurements such as Global Navigation Satellite System (GNSS), tiltmeters and strainmeters. Among these, the Global Positioning System (GPS; one type of GNSS) network is the most popular way of recording ground movements with the intention of uncovering SSEs, because it is relatively inexpensive, easily accessible and sufficiently precise \cite{Melbourne2005,Smith2009,Vergnolle2010,Jiang2012,Cavalie2013,He2017}. Developing a robust method for detecting SSEs in GPS data is crucial, despite the many challenges it presents \cite{Nishimura2013,Nishimura2014,Rousset2017,Takagi2019,Nishikawa2019,Haines2019,Nishimura2021,Okada2022}. For ease of presentation, we refer to GPS data recording SSEs as SSE data thereafter. 

Numerous methods have been proposed to detect the occurrence times of SSEs in GPS data (hereafter referred to as SSE detections). The first group of approaches is based on Kalman filter of state vector, which model the recorded GPS time series as the sum of coherent signals from various sources and estimation errors \cite{Granat2013,Ji2013,Lohman2013,Walwer2016}. These existing approaches include Network Inversion Filter \cite{Segall1997b,Segall2000,Miyazaki2003,McGuire2003}, Monte Carlo Mixture Kalman Filter \cite{Fukuda2004,Fukuda2008}, Network Strain Filter \cite{Ohtani2010}, and further improvements on the above Kalman-filter-based methods \cite{Ji2013,Riel2014,Bedford2018}. \minewa{These methods aim to extract the SSE signal from noisy GPS data, but they rely on different assumptions about the state vectors they estimate. However, these assumptions are debated because the underlying mechanisms that govern SSEs are not yet fully understood} \cite{Obara2016,Obara2020}. 

Another group of approaches consists of estimating the time evolution of the slip distribution on the fault by inverting the recorded GPS data at different sites, so that the occurrence times of SSEs can be simultaneously estimated \cite{Mccaffrey2009,Bartlow2014,Williams2015,Wallace2017,Wallace2018}. One commonly used tool for such detection is TDEFNODE, which is a nonlinear time-dependent inversion code \cite{Mccaffrey2009}. This tool utilizes simulated annealing to downhill simplex minimization, which has been applied to invert various recorded GPS data for detecting SSEs. Two free parameters in this method are the occurrence times and the associated amplitude of SSEs \cite{Mccaffrey2009}. TDEFNODE needs \textit{a priori} information on the functional form (e.g. exponential or Gaussian) of the temporal evolution of SSEs on the fault. However, the selection of a suitable form remains enigmatic, and is generally determined by trial tests \cite{Wallace2017}. In addition, the geometry of the subduction zone must be known to use TDEFNODE, thus its application is affected by the availability of geometrical knowledge in the observed data.



Singular Spectrum Analysis (SSA), a univariate time series analysis method \cite{Ghil2002}, can remedy this latter shortcoming. SSA is designed to extract information from noisy time series and thus, provides insight into the underlying dynamics \cite{Ghil2002}. The key feature of this method is that it does not need any \textit{a priori} knowledge of the underlying pure signal, and the trends obtained in this way are not necessarily linear \cite{Ghil2002,Chen2013}. SSA typically decomposes the noisy data into reconstructed components (RCs). These RCs are sorted in a descending order according to their corresponding eigenvalues, which denote their proportions of the total variance of the original data. Low-order RCs in the queue are regarded as effective signals related to the underlying dynamics, while high-order RCs are taken as noise, and are typically discarded. This is the common way to extract pure SSEs from noisy data by SSA. To determine a threshold between pure signal RCs and noise RCs is relatively subjective. When the signal-to-noise ratio (SNR) is low, SSA normally fails to distinguish signal from noise. \citeA{Chen2013} demonstrated that SSA is a viable and complementary tool for extracting modulated oscillations from GPS time series. 

\citeA{Walwer2016} introduced a more powerful form of SSA, Multichannel Singular Spectrum Analysis (M-SSA), to extract SSEs. M-SSA can simultaneously make use of the spatial and temporal correlations to explore the spatiotemporal variability of the data set. Although M-SSA was shown to outperform many existing detection methods, it still has drawbacks. This method only aims at extracting SSEs without detecting the occurrence times of SSEs, so a follow-up detection to determine the start and end times of SSEs is needed. The size of the lag covariance matrix in M-SSA also grows rapidly with the size of the GPS network considered, leading to computational issues for large-scale networks. M-SSA cannot operate on a single data basis, which limits its applicability to cases where the signals lack spatial coherence, for example, when there are not enough GPS stations, or the stations are too close to each other. Relative Strength Index (RSI), a single-station technique from the stock market \cite{Crowell2016}, is able to solve all the aforementioned issues, but it only applies to long-term SSEs. 

Compared to long-term SSEs, the duration and recurrence interval of short-term SSEs are much smaller, in the order of several days or weeks. The amplitude change in the GPS data caused by a short-term SSE is also relatively small. It can be close to, or even lower than, the background noise, so most short-term SSEs remain undetected \cite{Nishimura2021,Yano2022}. Therefore, more urgent efforts should be devoted to rapid automated methods for detecting short-term SSEs \cite{Hirose2020,Obara2020,Okada2022}, which is the focus of our current study. Linear regression, combined with Akaike's Information Criterion (AIC), is widely used to detect short-term SSEs for large-scale GPS networks \cite{Nishimura2013,Nishimura2014,Nishimura2021,Okada2022}. This method fits linear functions with and without an offset, and then uses AIC to judge which function is a better fit considering a number of free parameters. In this method, the length of the designed sliding window and the user-defined detection threshold determine the detection accuracy. In practice, it is hard to select reasonable values for these subjective parameters \cite{Nishimura2013,Yano2022}. A new method developed by \citeA{Yano2022} can overcome this deficiency, approximating SSE data as piecewise-linear signals by using $l_1$ trend filtering combined with Mallows' $C_p$. The knots in the fitted piecewise-linear signal are then taken as the occurrence times of SSEs. The applications to both synthetic and observed SSE data demonstrated that this method obtained better performance than the linear regression method. However, it is not clear that the assumption that SSE data can be regarded as piecewise-linear signals with the knots being the occurrence times of SSEs is reasonable, since the specific form of the underlying SSE signal remains unknown \cite{Obara2016,Obara2020}.

In this study, we develop a new method, called Singular Spectrum Analysis Isolate-Detect (SSAID), to automatically detect the start and end times of short-term SSEs in GPS data. Our method regards the detection of short-term SSEs in GPS data as a problem of detecting change-points in piecewise non-linear signals, in which the start and end times of SSEs are change-points to be detected. The prominent advantage of SSAID is that it does not require prior knowledge of the exact form of the underlying SSE signal. SSAID aims to obscure the differences between the nonlinear SSE signal and a piecewise-linear model, so that existing change-point detection methods for piecewise-linear signals can be directly applied to detect the start and end times of short-term SSEs. This is done by (i) decomposing the noisy SSE data into spectral components through SSA \cite{Ghil2002} and reconstructing these components into new noisy data signals; (ii) adding noise to these reconstructed signals, and (iii) conducting the detection by Isolate-Detect \cite<ID;>[]{Anastasiou2022}. We conduct a range of simulations to evaluate the detection performance of SSAID using both simulated and observed SSE data.

In \textsection{\ref{sec:data_and_processing}}, we introduce the observed SSE data in southwest Japan and the associated data processing procedures. In \textsection{\ref{sec:method}}, we introduce the method SSAID along with some assumptions. In \textsection{\ref{sec:syn_tests}}, we show results of applying SSAID to a range of simulated SSE data and compare the results with two existing detection methods (i.e. linear regression with AIC; and $l_1$ trend filtering). In \textsection{\ref{sec:application}}, we demonstrate our method's capability in detecting short-term SSEs in observed GPS data. Discussions and conclusions are in \textsection{\ref{sec:conclusions}}. 

\section{Data and processing}
\label{sec:data_and_processing}
\begin{figure}[!ht]	
	\centering
	\includegraphics[width=\textwidth]{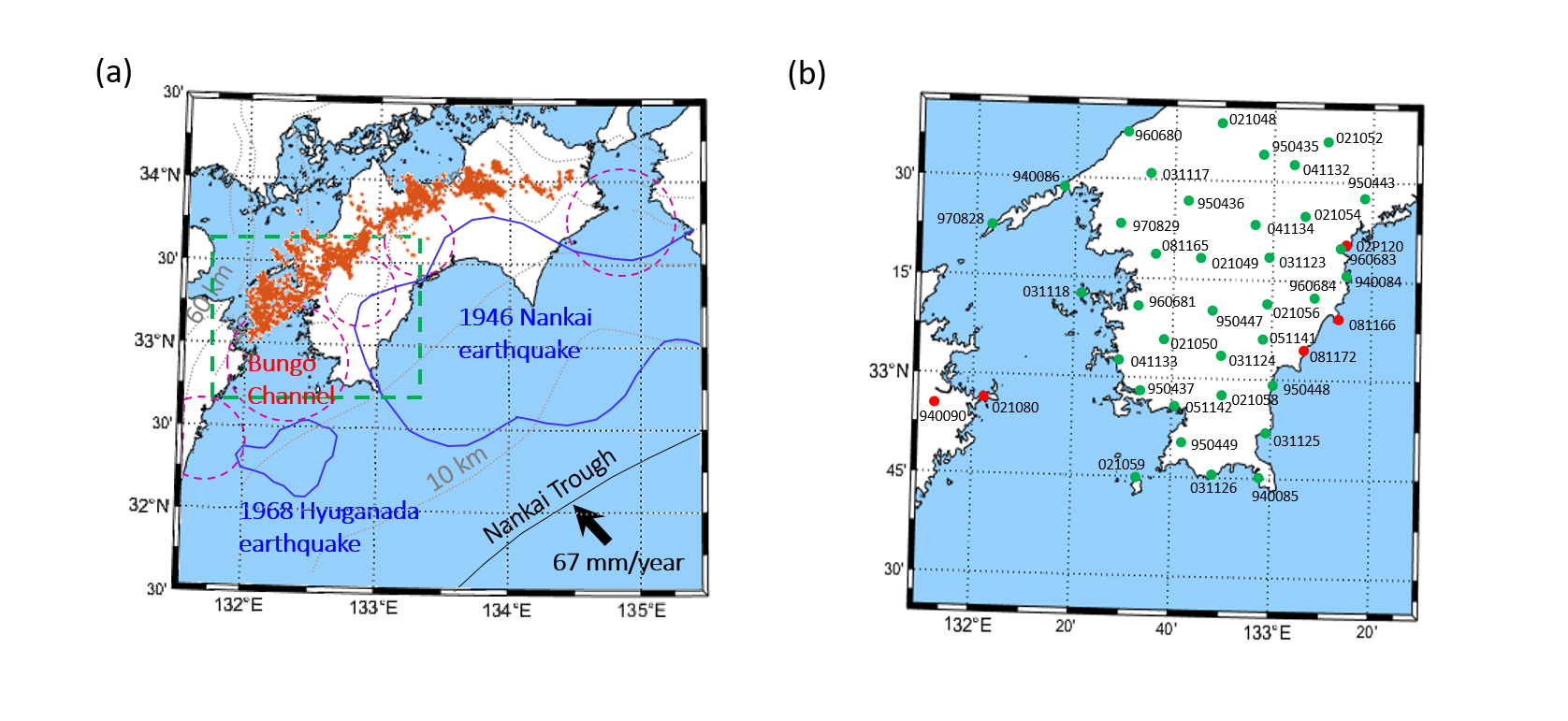}
	\caption{(a) The distribution map of earthquakes in the study area of southwest Japan. The magenta dashed circles and the blue contours denote the source areas of long-term SSEs and megathrust earthquakes, respectively. The orange dots show the epicenters of tremors. Gray dashed lines indicate the depth of the subducting Philippine Sea plate. (b) The distribution map of $36$ GPS stations utilized in the current case study (see \textsection{\ref{sec:application}}). This area is outlined by the dashed green box in panel (a). Both red and green circles indicate the location of GPS stations, and the numbers near to circles refer to the GPS station names. Note that we apply SSAID to detect change-points in SSE data recorded by GPS stations identified as green filled circles in the case study reported in \textsection{\ref{sec:application}}.}
	\label{fig01}	
\end{figure}

We use SSE data from the Nankai subduction zone which has a dense geodetic observation network. In southwestern Japan, the Amurian plate overriding the Philippine Sea plate converges to N$\ang{50}$W at a rate of about $67$ mm/year \cite{Miyazaki2001,Nishimura2014,Kano2020,Obara2020}. Both long-term and short-term SSEs occur across the Nankai Trough \cite{Obara2020} (see Fig. \ref{fig01} (a)). Short-term SSEs in southwest Japan generally exist in the deeper extension of long-term SSE regions.

Our SSE data are obtained from $36$ GPS stations of the GNSS Earth Observation Network System (GEONET) operated by the Geospatial Information Authority of Japan (GSI). These GPS stations are distributed in the Shikoku region along the Bungo Channel (see Fig. \ref{fig01} (b)). The analysis period for this study is from 1 January 2008 to 30 June 2009. The vector of coordinates at each GPS station, containing east, north and upward displacement, has been transformed to the 2005 International Terrestrial Reference Frame (ITRF2005), and can be generally modelled as a sum of different processes \cite{Nikolaidis2002,Davis2012,He2017,Bedford2018}, that is
\begin{linenomath*}
\begin{equation}
\mathbf{u}(t) = \mathbf{d_0} + \mathbf{m_0}t + \sum_{j=1}^{n_o}\mathbf{b_j}H(t-t_j)+\sum_{i=1}^{n_s}\mathbf{h_i}(t-t_i)+\boldsymbol{\xi_\minewa{seas}}(t)+\boldsymbol{\xi_\minewa{u}}(t)+\boldsymbol{\xi_\minewa{SSE}}(t)+\boldsymbol{\epsilon}(t),
\label{eq01}
\end{equation}
\end{linenomath*}



\noindent where $t$ is the time, $\mathbf{d_0}$ and $\mathbf{m_0}$ refer to vectors describing the position of the reference site and the secular velocity, respectively. Here, we refer to the displacement rate of the linear process without the occurrence of other fault slips as the secular velocity, which represents the secular tectonic motions of two contacting plates of the subduction zone. The third term $\sum_{j=1}^{n_o}\mathbf{b_j}H(t-t_j)$ describes the vector of offsets due to non-tectonic changes such as antenna or other instrument changes, where $n_0$ is the number of non-tectonic changes, $t_j$ is the time when the $j$-th non-tectonic change occurs, and $H(t)$ is the Heaviside step function. The fourth term $\sum_{i=1}^{n_s}\mathbf{h_i}(t-t_i)$ represents the vector of coseismic and postseismic movements from ambient regular earthquakes, where $n_s$ is the number of ambient regular earthquakes, $t_i$ is the time at which the $i$-th regular earthquake occurs, and $\mathbf{h_i}$ refers to the coseismic and postseismic movements from the $i$-th regular earthquake \cite{Wdowinski1997,Elgharbawi2015}. The other vectors \minewa{$\boldsymbol{\xi_{seas}}(t)$,  $\boldsymbol{\xi_u}(t)$, $\boldsymbol{\xi_{SSE}}(t)$} and $\boldsymbol{\epsilon}(t)$ describe the movements from seasonal motions, unknown sources, SSEs and noise, respectively.

These SSE data have been pre-processed by \citeA{Nishimura2013} to remove known effects from non-SSE processes. We now briefly illustrate the data processing procedures conducted on the raw GPS data \cite{Nishimura2013,Nishimura2014,Fujita2019,Nishimura2021}. Firstly, \minewc{Nishimura et al.}\ eliminated the coseismic offsets from six ambient large earthquakes (see the detailed catalogue therein), which are estimated by the difference in the $10$-day averages of the daily coordinates before and after the earthquakes. Secondly, the spatial filtering technique of \citeA{Wdowinski1997} was applied to suppress the common mode errors for these stations, which are a major type of spatially correlated noise sources in GPS data \cite{Dong2006}. Finally, the offsets from non-tectonic changes (i.e. the third term in Eq. \eqref{eq01}) such as antenna maintenance were removed by the same method as that used to remove coseismic offsets. Note that the post-seismic deformations from nearby large earthquakes were not removed (i.e. the fourth term in Eq. \eqref{eq01}), however their impacts are negligible in our current application as no obvious large earthquakes were identified in the period analyzed (i.e. from January 1 2008 to June 30 2009) in the research area \cite{Nishimura2013}. 

We denote the processed daily cumulative displacement vector at each station as
\begin{linenomath*}
\begin{equation}
    \mathbf{\bar{u}}(t) = \mathbf{\bar{b}_0}t + \boldsymbol{\bar{\xi}_\minewa{seas}}(t)+ \boldsymbol{\bar{\xi}_\minewa{u}}(t)  + \boldsymbol{\bar{\xi}_\minewa{SSE}}(t) + \boldsymbol{\bar{\epsilon}}(t),
\label{eq02}
\end{equation}
\end{linenomath*}
\noindent where $\mathbf{\bar{b}_0}$ is the vector of coefficients quantifying the secular movement, and \minewa{$\boldsymbol{\bar{\xi}_{seas}}(t)$, $\boldsymbol{\bar{\xi}_{u}}(t)$, $\boldsymbol{\bar{\xi}_{SSE}}(t)$ and $\boldsymbol{\bar{\epsilon}}(t)$} are vectors of daily cumulative displacements of seasonal motions, unknown sources, SSEs and noises, respectively. The daily cumulative displacement $\mathbf{\bar{u}}(t)$ contains three components along different directions (i.e. east, north and upward), which are denoted as $\bar{u}_e$, $\bar{u}_n$, $\bar{u}_z$, respectively. In the following application, we concentrate on the N$\ang{50}$W component of the daily cumulative displacement at each station, denoted by $X_t$, which is parallel to the plate convergence direction of the Nankai Trough (see Fig. \ref{fig01} (a)). This is done by rotating two horizontal components (i.e. east and north) using the following equation,
\begin{linenomath*}
\begin{equation}
    X_t = \bar{u}_e\sin{\bar{\delta}_0} - \bar{u}_n\cos{\bar{\delta}_0},
\label{eq03}
\end{equation}
\end{linenomath*}
\noindent where $\bar{\delta}_0$ is the azimuth angle of the plate convergence direction (see the black arrow in Fig. \ref{fig01} (a); $\bar{\delta}_0\approx{\ang{50}}$ in Nankai Trough). In the following applications, we further remove the daily secular motions and outliers from $X_t$ at each station, through linear least squares and the four-sigma limit, respectively \cite{Nishimura2021}. Note that when conducting hypothesis tests in \textsection{\ref{subsubsec:null_hypothesis}}, we do not remove the daily secular motions, as they can be used to investigate the sign change of the displacement rate from the secular velocity when SSEs arise \cite{Yano2022}. 

\section{Method}
\label{sec:method}
\begin{figure}[!ht]	
	\centering
	\includegraphics[width=\textwidth]{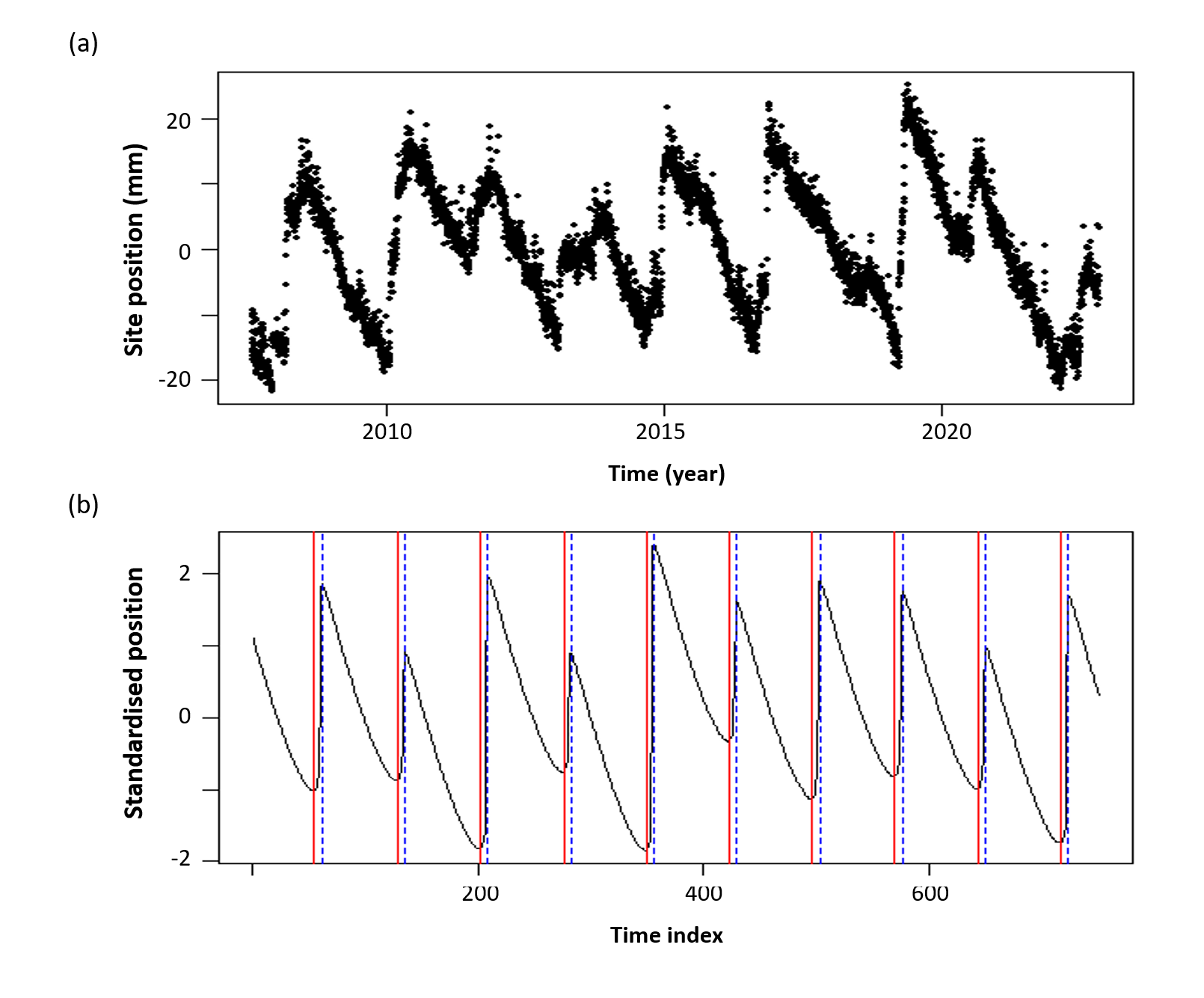}
	\caption{(a) Observed SSE data recorded by the east component of a GPS station (MAHI), in the Hikurangi subduction zone, New Zealand; (b) Synthetic SSE data with $10$ SSEs in a two-year period, which are simulated by a deterministic subduction slip model (see the supplement). Red vertical lines: the start times of SSEs; blue dotted vertical lines: the end times of SSEs.}
	\label{fig02}	
\end{figure}

We propose a new method to detect change-points in univariate time series with continuous, piecewise non-linear structure. Here, change-points refer to the times at which the pattern of the underlying dynamics (i.e. pure signal) changes from one state to a different one. Fig. \ref{fig02} (a) shows an example of observed SSE data from the Hikurangi subduction zone, New Zealand. In periods with no SSEs, the overall trend of the signal is linear and decreasing. The trend is then redirected to a different state  (increasing here) when an SSE starts. Once the SSE ends, the trend reverses back to its original linear decreasing state. The start and end times of SSEs can therefore be regarded as change-points in GPS data. Our method, called Singular Spectrum Analysis Isolate Detect (SSAID), seeks to detect the start and end times of SSEs in noisy GPS data without prior knowledge of the underlying structure of the signal. \minewa{Note that the linear trend of the presented SSE data has been removed, which is not necessarily equivalent to the true secular plate motion at a given GPS site.} Here, we only summarize its underlying assumptions and main features.\minewc{ A full exposition of the methodology can be found in the appendix and the supplement.}

Let us assume that the deviation in the pure SSE signal from a piecewise-linear function can be obscured by noise as long as the noise level is within a suitable range\minewa{, so that SSE data with this range of noise levels can simply be taken as piecewise-linear signals}. If \minewa{the condition is met}, an existing change-point detection method specifically designed for piecewise-linear signals can be directly applied to detect change-points in SSE data\minewa{. If the condition is not met, the existing change-point detection method for piecewise-linear signals will overestimate the number of change-points for low noise levels and underestimate them for high noise levels}. This assumption was validated using numerical tests \minewc{(see Text S2 in the supplement)}, in which various change-point detection methods for piecewise-linear signals were shown to successfully detect change points after different levels of Gaussian noise were added to the signal. Of all the methods considered, Isolate-Detect \cite<ID;>[]{Anastasiou2022} showed the best performance and was therefore selected for application to SSE data. The noise level within a suitable range, i.e. allowing successful change-point detection, is referred to as a suitable noise level (SNL). 

\minewc{For the remainder of this paper, } we define a successful \minewc{\emph{cumulative detection}} when two conditions are met: (1) the number of estimated change-points is exactly the number of true change-points and (2) the root mean squared error (RMSE) of the detected change-point locations is less than a predefined threshold value \minewc{$v$, here $v=3$ days (see Text S2 and Fig. $S8$ (b) in the supplement for a justification)}. 

As the SNL varies with signal types \minewc{(see Fig. $S9$ in the supplement)}, 
it is not possible to predetermine if the raw data has an SNL. By decomposing the raw data and systematically adding Gaussian noise, SSAID generates new time series with SNL (referred to as in-SNL data), greatly improving the probability of successful change-point detection.

\begin{figure}[!ht]	
	\centering
	\includegraphics[width=\textwidth]{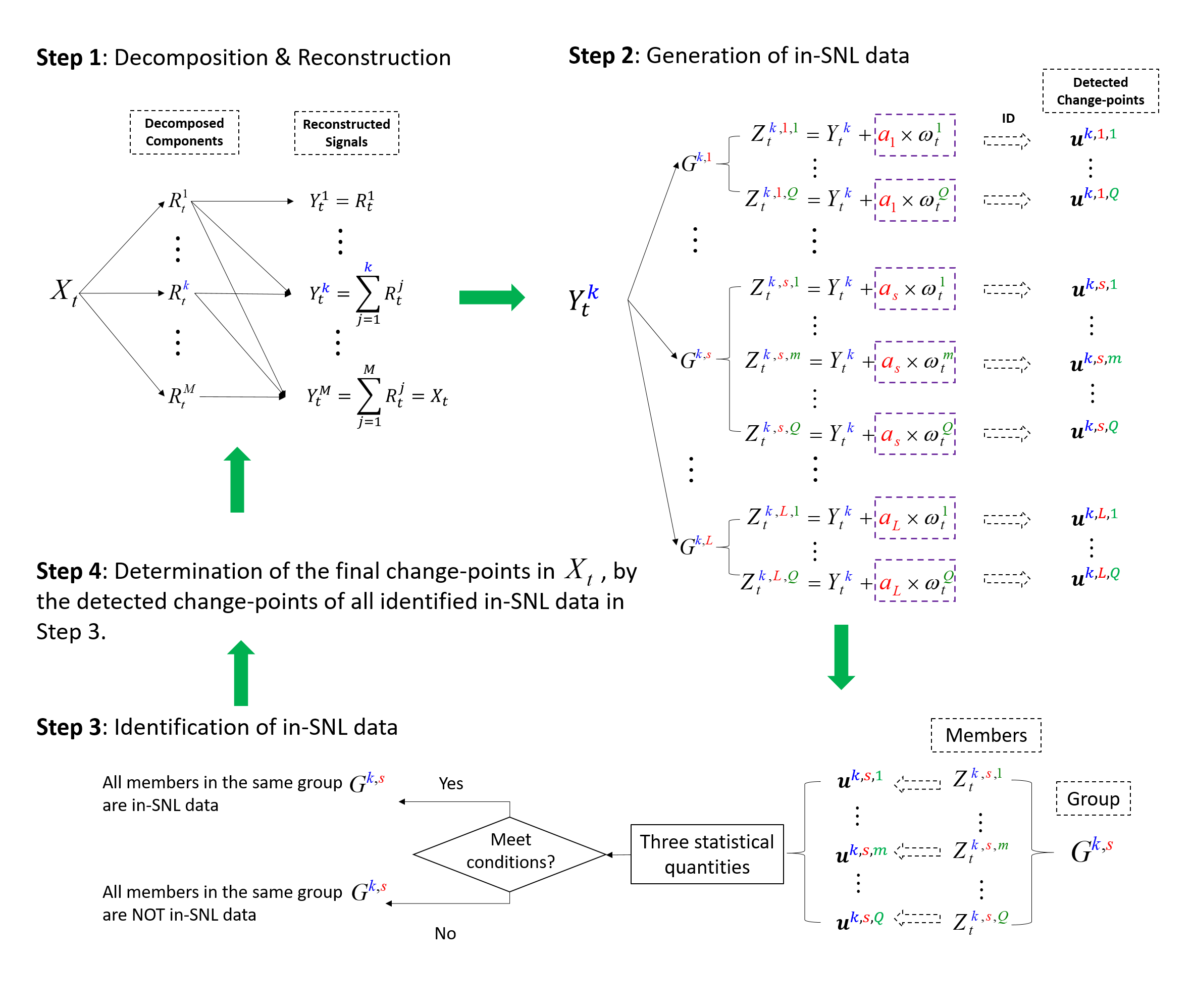}
	\caption{\minewb{The full workflow of SSAID showing how to find the change-points in the noisy time series $X_t$ step by step.}}
	\label{fig02_workflow}	
\end{figure}

SSAID contains four main steps: (1) decomposing and reconstructing the signal using SSA; (2) adding Gaussian noise with different noise levels to reconstructed signals; (3) detecting change-point candidates in SSE data via \minewc{the Isolate-Detect algorithm \cite{Anastasiou2022}} and identifying in-SNL data and (4) determining the final change-points to best characterize the start and end times of SSEs. 
\minewb{Fig. \ref{fig02_workflow} summarizes the workflow of the method.}
Brief descriptions for each step are provided below. The reader is referred to \minewc{the appendix} for a full exposition of the method.

\begin{enumerate}
    \item \textbf{Signal decomposition and reconstruction:} \minewb{We decompose the input data $X_t$ (see Eq. \eqref{eq07}) into $M$ components $R_t^j$ ($j=1,\cdots,M$) using SSA, sorted by their correlation with the underlying dynamics. Components with smaller $j$ values are important for the signal, while larger $j$ values mostly contain noise. We then reconstruct $M$ new data sequences in the form of cumulative sums: $Y_t^k = \sum_{j=1}^{k} R_t^j$ $(k=1,\cdots,M)$.} \minewa{As $k$ increases, $Y_t^k$ gets closer to $X_t$, with $Y_t^M = X_t$.}

    \item \textbf{Generation of in-SNL data:} We add Gaussian noise with different noise levels \minewb{to each} reconstructed data $Y_t^k$\minewb{, defined as} $Z_t^{k,s,m} = Y_t^k + a_s\omega_t^m$ \minewb{for $ s = 1, \cdots, L $ and $ m = 1, \cdots, Q $}, where $\omega_t^m$ are independent, random variables sampled from the standard normal distribution.\minewb{ Here, $ a_s $ represents the noise level, and $ L $ and $ Q $ denote the number of realizations and noise levels considered, respectively.} \minewb{This step ensures the presence} of in-SNL data among these newly created $Z_t^{k,s,m}$ time series. \minewb{For simplicity, we refer to the set of all realizations $ \textbf{\textit{G}}^{k,s} = \{ Z_t^{k,s,1}, \cdots, Z_t^{k,s,Q} \} $ as a group for presentation in the next step.}
    
   \item \textbf{Identification of in-SNL data:} \minewb{ We apply the ID methodology to detect change-points $\textbf{u}^{k,s,m}$ for each newly created time series $Z_t^{k,s,m}$. Subsequently, we compute three statistical quantities for $\textbf{u}^{k,s,m}$ in each group and apply specific conditions (see appendix for details), so that if these conditions are satisfied, all realizations $Z_t^{k,s,m}$ within the same group are classified as in-SNL data; otherwise, they are classified as not in-SNL data.} 

   \item \textbf{\text{Determination of change-points in $X_t$}:}\minewb{We identify change-points in the input data $X_t$ through a majority voting rule based on the identified in-SNL data. This process involves two sub-steps: (1) determining the number of change-points in $X_t$, denoted as $\hat{N}_X$, using the counts of the estimated change-points from the identified in-SNL data; and (2) locating the change-points by finding the mode or the average of each column in a matrix. This matrix comprises the selected $\textbf{u}^{k,s,m}$, with each $\textbf{u}^{k,s,m}$ containing the locations of the $\hat{N}_X$ change-points.}
\end{enumerate}

\section{Tests on synthetic data}
\label{sec:syn_tests}
We now evaluate the detection performance of our method for a range of simulated noisy SSE data $X_t$, which are generated in the following form,
\begin{linenomath*}
\begin{equation}
X_t = f_t + C_{wn}\times{\epsilon_t}, \quad (t=1,\cdots,T),
\label{eq07}
\end{equation}
\end{linenomath*}
where $T$ is the length of the noisy data, and $f_t$ is the simulated pure SSE data (see Fig. \ref{fig02} (b)) from a deterministic subduction slip model (see details in the supplement, Text S1), which is standardised through the Z-score normalisation. The number of true change-points in the simulated pure SSE signal is $N_0=20$. The second term $C_{wn}\times{\epsilon_t}$ in Eq. \eqref{eq07} denotes the noise model contained in $X_t$. We assume that $\epsilon_t$ are independent, Gaussian random variables with mean zero and variance one. The noise level $C_{wn}$ is the standard deviation of the noise model, varing from \minewc{$1\%$} to $100\%$, with increments of $1\%$. Fig. \ref{fig03} (c) and (d) show two examples of simulated noisy SSE data with different noise levels. We create $100$ data sequences of independent standard Gaussian random variables $\epsilon_t$ $(t=1,2,\ldots, T)$. In total, we have $100\times{100}$ noisy time series $X_t$ $(t=1,2,\ldots,T)$. The detection performance of SSAID is controlled by three parameters: the number of SSA components $M$, the number of realisations $Q$, and the highest level of added noise levels in percentage $L$. \minewa{The selection of the parameter $M$ should consider a balance between the quantity of information extracted and the degree of statistical confidence in that information, avoiding values that are excessively small or large \cite{Ghil2002,Chen2013}. Parameters $L$ and $Q$ should be set to sufficiently high values. A larger $L$ ensures the presence of in-SNL data, while a larger $Q$ enhances the detection success rate. However, it is crucial to impose upper limits on both $L$ and $Q$ to manage computing costs, as the computational demands increase significantly with higher values of these parameters.} Based on numerical studies \minewc{(see Text S4 in the supplement)}, we choose the default values $M=100$, $L=80$ and $Q=40$ to ensure optimal performance.


\subsection{Detection results}
\label{subsec:syn_detection_results}
\begin{figure}	
	\centering
	\includegraphics[width=\textwidth]{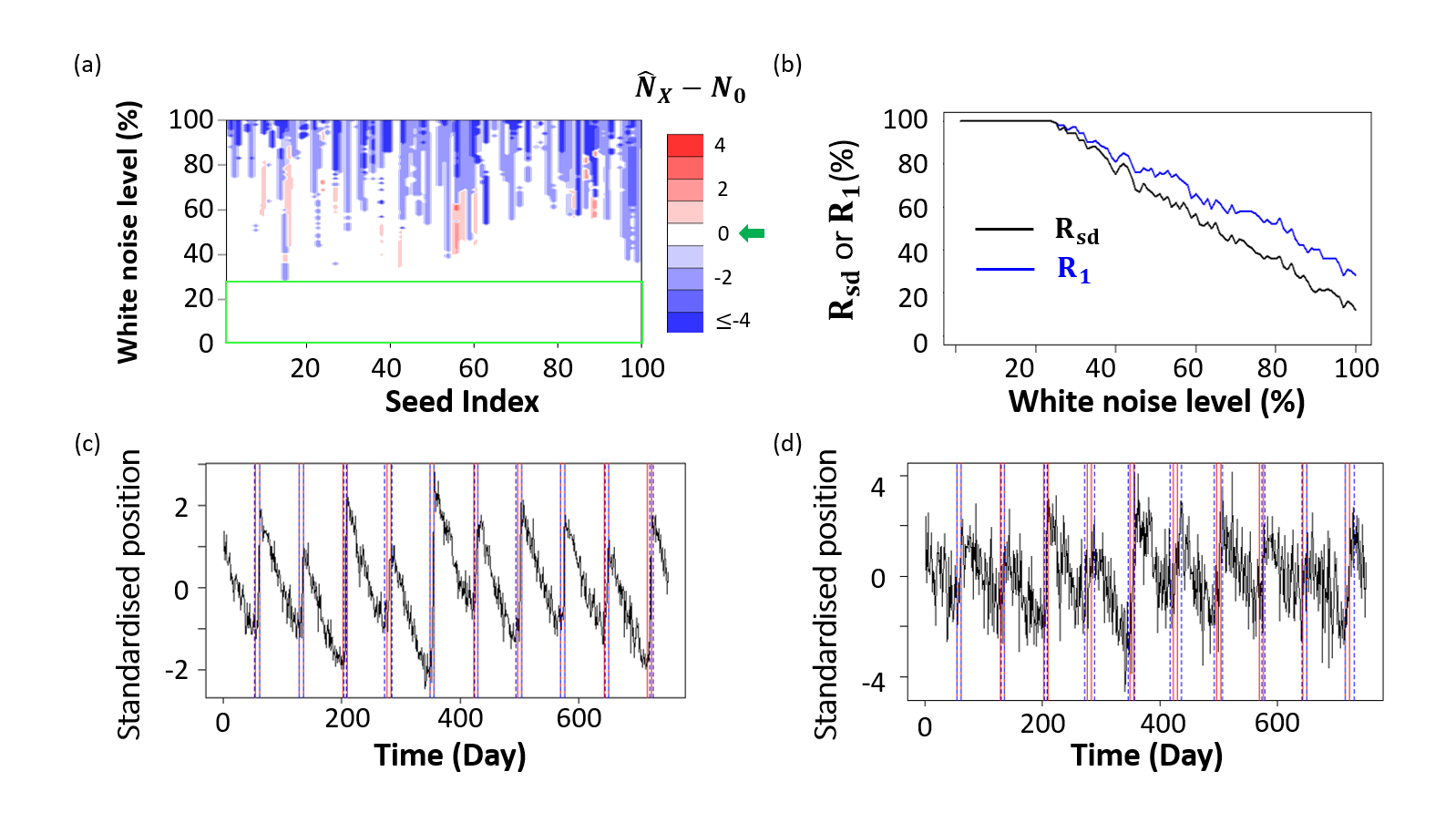}
	\caption{(a) The error between the number of estimated change-points $\hat{N}_X$ by SSAID and the number of true change-points $N_0$ in each simulated noisy data. The error of zero is highlighted by a green arrow in the color bar. (b) The percentage $R_1$ and $R_{sd}$ (see definitions in Eq. \eqref{eq08}) as a function of white noise level $C_{wn}$, calculated from 100 seeds. The locations of the change-points in two simulation examples with different noise levels are shown in (c) $C_{wn}=25\%$; (d) $C_{wn}=100\%$. Blue vertical dotted lines: estimated change-points by SSAID; red vertical lines: true change-points.}
	\label{fig03}	
\end{figure}

Fig. \ref{fig03} (a) shows the error between the number of estimated change-points $\hat{N}_X$ by SSAID and the number of true change-points $N_0$ for each noisy time series. We can observe that SSAID correctly estimates the number of true change-points in over $70\%$ of all cases analyzed. In particular, the number of estimated change-points is correct for all the cases with noise levels lower than $25\%$ (see green box in Fig. \ref{fig03} (a)). To quantify the detection performance of SSAID, we define
\begin{linenomath*}
\begin{equation}
R_{sd} = \frac{\alpha}{\xi} \quad \text{and} \quad R_1=\frac{\beta}{\xi},
\label{eq08}
\end{equation}
\end{linenomath*}
\noindent where $\xi$ is the number of simulations for each noise level (i.e. $\xi=100$ here), $\alpha$ is the number of successful cumulative detections, as defined in \textsection{\ref{sec:method}}), and $\beta$ is the number of cumulative detections for which the number of estimated change-points, $\hat{N}_X$, is equal to the number of true change-points $N_0$ (i.e. $\hat{N}_X=N_0=20$ here), but not with the RMSE requirements imposed on $\alpha$. 

Fig. \ref{fig03} (b) shows that $R_{sd}$ and $R_1$ are different. They are both $100\%$ when $C_{wn}<{25\%}$, and then decrease with increasing $C_{wn}$ values. This implies that the successful \minewc{cumulative} detection rate is higher when the GPS data have a smaller noise level, with $100\%$ success rate if the noise level is less than $25\%$. $R_{sd}$ decreases faster than $R_1$ when $C_{wn}$ increases, indicating that the accuracy of the detected change-point locations fades with increasing $C_{wn}$ values. Fig. \ref{fig03} (c) demonstrates the high accuracy of the change-points detected using our method for data with a low noise level. Fig. \ref{fig03} (d) shows \minewc{a simulated time series with high noise level} ($C_{wn}=100\%$) \minewc{for which cumulative detection was unsuccessful (correct number of change-points, but too large error). Even though the locations of some detected change-points are not as accurate as for lower noise levels, SSAID remains relatively performant in terms of both number of change-points and locations.} 

\subsection{Comparison with two existing methods}
We now compare the detection performance of SSAID with two existing detection methods for short-term SSEs. The first one is linear regression combined with AIC proposed by \citeA{Nishimura2013}, which has been widely applied in different areas \cite{Nishimura2013,Nishimura2014,Nishimura2021,Okada2022}. This method (1) uses a sliding window with a fixed width; (2) fits a linear model to the data in the window; (3) divides the data in the window into equal halves and fits a linear model to each half, and (4) calculates the AIC difference (i.e. $\Delta$AIC) between the single linear model and the two-line model at the middle point of the window. If that midpoint is a change-point, e.g. the start- or end-point of an SSE, the two-line model fits the observational data better than a single linear model, thus resulting in a negative $\Delta$AIC. As a negative $\Delta$AIC does not always correspond to change-points in SSE signals, we must specify an appropriate threshold, denoted by $\zeta$, in order to detect change-points of SSEs. If $\Delta$AIC is lower than $\zeta$, its corresponding time is regarded as a change-point. The detection performance of the linear regression approach is mainly controlled by the length of the sliding window and the specified threshold $\zeta$, however, and selecting appropriate values for the two parameters is subjective \cite{Nishimura2013,Nishimura2021}. 

\begin{figure}	
	\centering
 	\includegraphics[width=\textwidth]{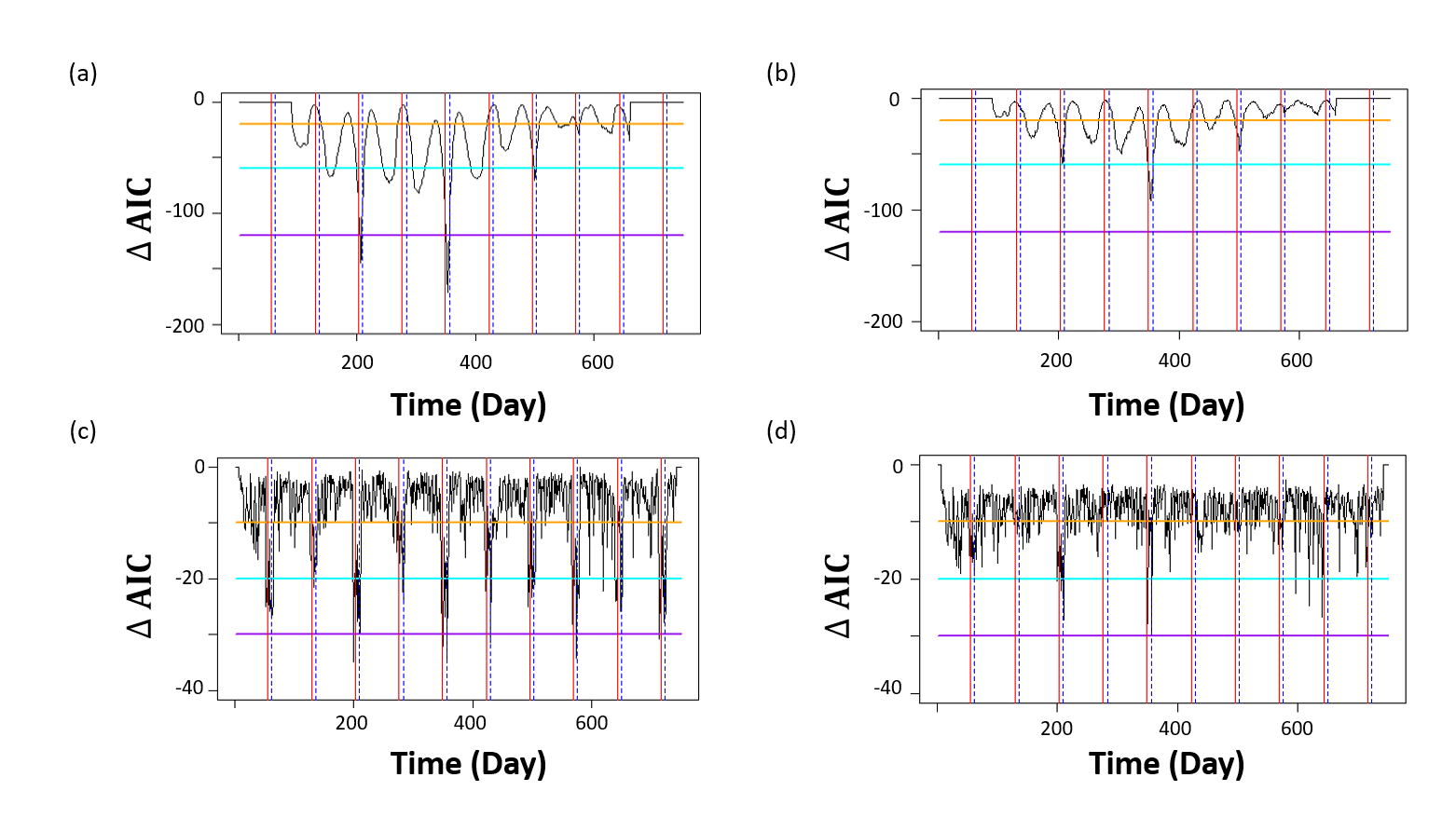}
	\caption{The calculated $\Delta{AIC}$ for different noisy data with different sliding windows. Panel (a) and (b) are plotted for the noisy data shown in Fig. \ref{fig03} (c) and (d) with a sliding window of $180$ days, respectively. While panel (c) and (d) are the same as (a) and (b) but with a sliding window of $15$ days. Horizontal solid and dotted lines are associated with different thresholds to identify change-points of SSEs: high threshold (orange); medium threshold (cyan); low threshold (purple). The intersections between horizontal lines and $\Delta{AIC}$ curve are considered as change-points. Vertical red lines: start times of SSEs; vertical blue dashed lines: end times of SSEs.}
	\label{fig04}	
\end{figure}

In our comparison tests, we first take a sliding time window of $180$ days, which is consistent with that of \citeA{Nishimura2013}, to calculate $\Delta$AIC for each data point of the simulated SSE data in Fig. \ref{fig03} (c) and (d). Fig. \ref{fig04} (a) and (b) show $\Delta$AIC values across the time series with three threshold values\minewc{: $\zeta=-10, -20$ and $-30$ (later referred to as low, medium and high, respectively, in absolute value)}. We observe that the change-points at both ends of the simulated data \minewc{cannot be detected} regardless of the selected threshold due to the excessive length of the sliding window. This demonstrates that a smaller sliding window is needed \cite{Yano2022}. We then decrease the sliding window to $15$ days to calculate $\Delta$AIC for each data point again, and we have a much shorter blinded interval of $7$ days at both ends of the simulated period. In Fig. \ref{fig04} (c) and (d), we also observe that none of the detection thresholds considered succeeds in finding all the true change-points accurately. When $\zeta$ is too low, only the most significant SSEs can be detected, while for larger $\zeta$, the detection generally overestimates the number of change-points. The selection of the threshold value depends on the signal itself, making it impossible to detect all the change-points in multiple time series or even within a single time series by using a single threshold.

We then apply the method proposed by \citeA{Yano2022} to the synthetic data considered in Fig. \ref{fig03}. The method (1) applies $l_1$ trend filtering to the raw data \minewa{$X_i (i=1,\cdots,T)$} with a range of hyperparameters $\lambda$; (2) obtains a fitted piecewise-linear signal \minewa{$\hat{X}_t (t=1,\cdots,T)$} for each $\lambda$; (3) calculates the associated Mallows' $C_p$ for each $\lambda$\minewa{, which is defined by  $\sum_{t=1}^T(X_t-\hat{X}_t)^2/\hat{\sigma}_s^2 + N_{knots} + 2$, with $\sigma_s^2$ and $N_{knots}$ representing the noise variance of $X_t$ and the number of knots in $\hat{X}_t$, respectively}; (4) chooses the one with the minimum Mallows' $C_p$ as the best piecewise-linear approximation to characterize the raw data; and (5) takes the knots of the chosen piecewise-linear model as the occurrence times of SSEs. \minewa{Fig. $S13$ illustrates an example of determining a suitable $\lambda$ value for a noisy time series.} This method is similar to other change-point detection methods for piecewise-linear signals, for which \minewc{we} have demonstrated that they cannot be directly applied to detect SSEs in GPS data \minewc{(see Text S2 in supplement)}. Fig. \ref{fig05} (a) and (b) show that in most cases $l_1$ trend filtering overestimates the number of change-points in simulated SSE data and its associated successful \minewc{cumulative} detection ratio $R_{sd}$ for each noise level is much lower than that of SSAID, regardless of the noise level.

\begin{figure}	
	\centering
	\includegraphics[width=\textwidth]{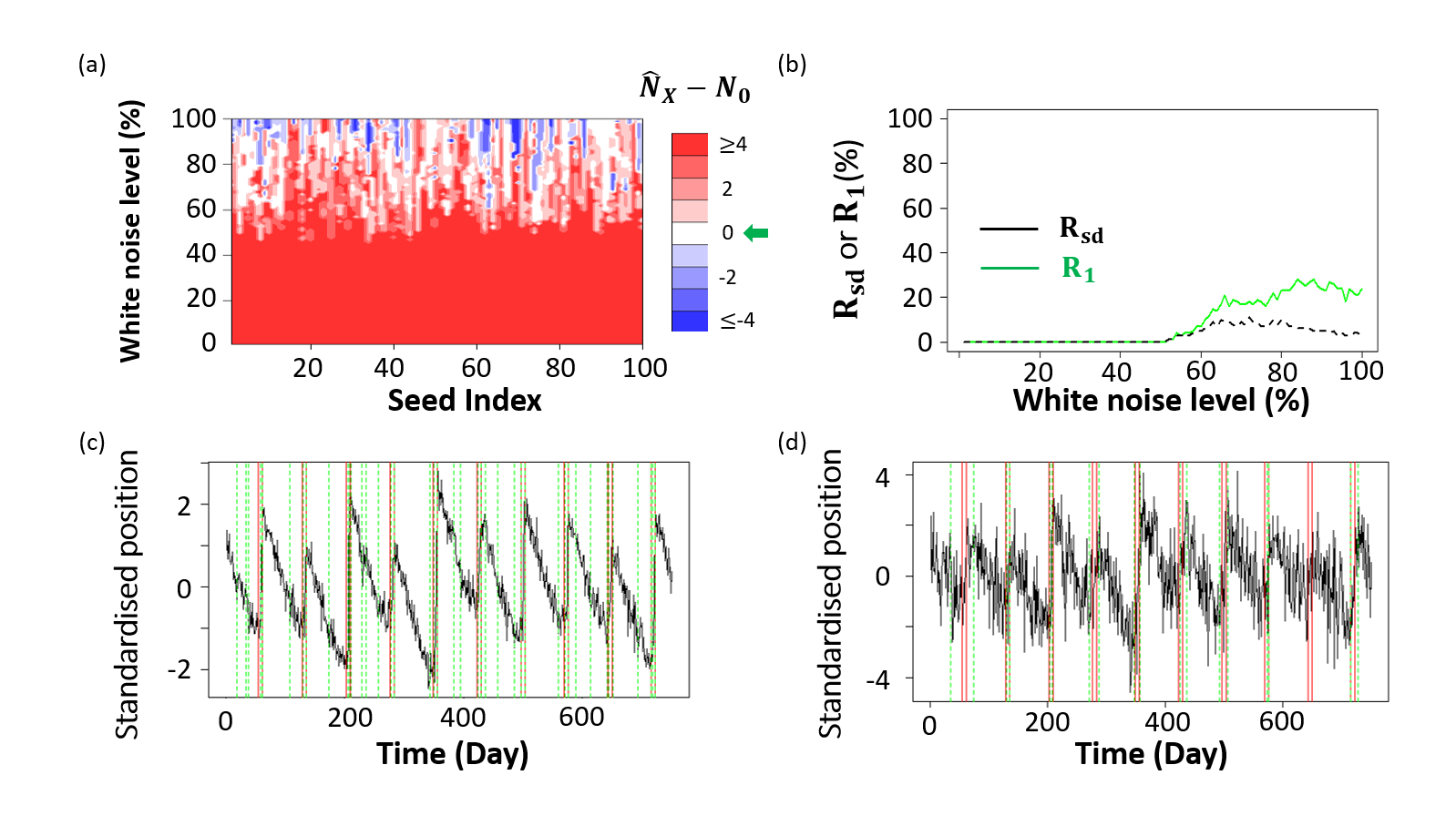}
	\caption{Same as Fig. \ref{fig03} but using $l_1$ trend filtering to detect change-points in simulated SSE data.}
	\label{fig05}	
\end{figure}

\begin{figure}	
	\centering
 	\includegraphics[width=\textwidth]{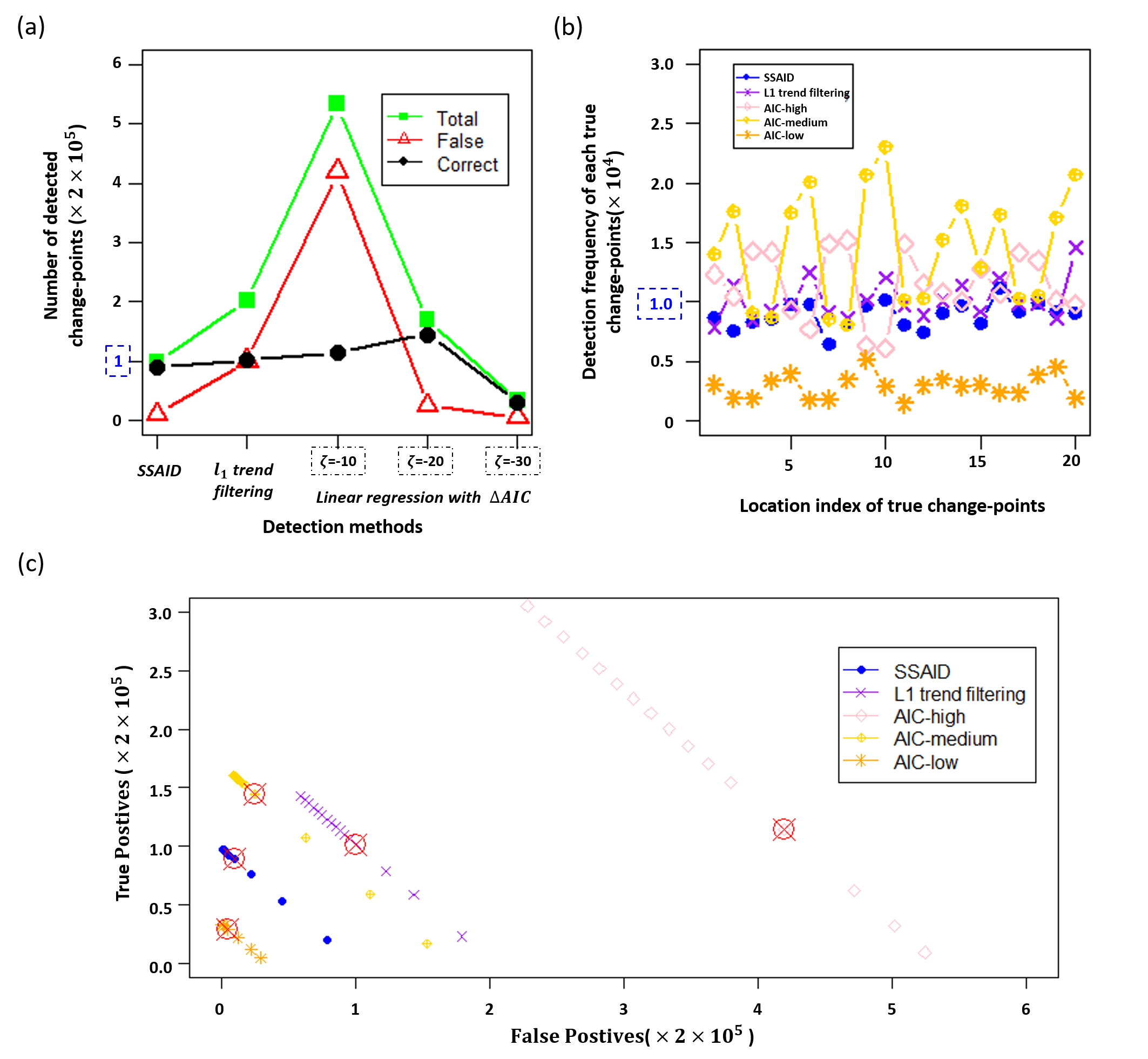}
	\caption{(a) Number of different detected change-points by various methods; (b) detection frequency of each true change-points by different methods. 
    The total number of detected change-points and the number of correctly detected change-points are expected to be $20\times{10,000}$, while the expected detection frequency of each true change-point is $10,000$. These expected values are highlighted by the blue dotted boxes.\minewa{(c) Plots of false-true positives for each detection method with different thresholds of predefined acceptable error, ranging from 1 to 20 days. The dots indicated by the red circles with a cross correspond to an acceptable error of 3 days.}}
	\label{fig04_comparison}	
\end{figure}

We now compare the performance of the aforementioned methods quantitatively by calculating the total number of detected change-points across all considered scenarios (i.e. all noise levels and all seeds), as well as the counts of correct and false detections. An estimated change-point is considered correct if its error is no more than $3$ days from any true change-point location \minewc{(as previously justified)}; otherwise, it is regarded as false. Both the total number of detected change-points and the number of correctly detected change-points are expected to be $20\times{10,000}$. In Fig. \ref{fig04_comparison} (a), we can see that the method SSAID aligns well with the expected values, exhibiting a satisfactory total number of detected change-points and a considerable number of correct detections, with minimal false detections. However, when using the $l_1$ trend filtering method, we observe that the total number of detected change-points is about twice the expected value, indicating \minewc{a severe overdetection issue.} The results obtained with the method of linear regression with $\Delta{AIC}$ underscore the significant influence of the chosen threshold $\zeta$ on the success of detection. Setting the threshold to a low value results in a large number of false detections. Conversely, choosing the threshold $\zeta$ to a medium value (see $-20$ in Fig. \ref{fig04_comparison} (a)) can significantly reduce false detections, but leads to a notable overestimation of true change-points. Further \minewc{changing $\zeta$} to a higher \minewc{threshold level} causes the majority of detections to miss the true change-points. 

We also analyze the \minewc{detection frequency} for each true change-point in the simulated data, which we should expect to be $10,000$. Fig. \ref{fig04_comparison} (b) shows that the detection results obtained by SSAID exhibit slight oscillations around the expected values, indicating greater stability compared to the other methods. We conducted further analysis on the histograms of the detected change-points for all the simulated noisy SSE data from all the different seeds and noise levels by these detection methods (see Fig. \minewc{$S14$-$S15$} in the supplement). The results indicate that most SSAID detections tend to converge to accurate locations with minimal errors, while the other methods, despite exhibiting similar behaviors, either suffer from a higher number of false detections and larger errors, or miss the majority of true change-points. This further demonstrates the superior detection performance of SSAID. 

\minewa{To provide a clear visual comparison of the performance of different methods, we now create a plot similar to an ROC curve as shown in Fig. \ref{fig04_comparison}(c). In this plot, a correct detection (an error of no more than 3 days from any true change-point) is defined as a true positive, while a false detection (an error of more than 3 days from any true change-point) is defined as a false positive.
The (0,1) point (with $2\times10^5$ factored out) corresponds to the successful detection of all change-points.
It is evident that the SSAID method is the closest to our expectation. Furthermore, since the definition of true and false positives depends on the predefined threshold of accepted error (3 days in our tests, indicated by the red circle with a cross), we varied this threshold from $1$ to $20$. The results consistently show that the detection of SSAID remains the closest to the (0,1) point, further verifying its good performance.}

\minewa{\subsection{The effect of color noise on SSAID detection performance}
We now investigate how color noise influences the detection performance of SSAID. In GPS time series, noise typically comprises both white noise and color noise, the latter being temporally correlated \cite{Dmitrieva2015}. This temporal correlation is often described using power-law models, where spectral amplitude changes according to $F(f)\propto{f}^{-n}$, with $f$ representing frequency and $n$ being the power-law index \cite{Agnew1992}. In the realm of GPS time series, color noise is often conceptualized as a combination of flicker noise ($n=1$) and random walk ($n=2$), or with a non-integer power-law index \cite{Zhang1997, Mao1999}. Most studies indicate that the optimal representation of time-dependent GPS noise is flicker noise, with little or no random walk component \cite{Williams2004, Hackl2011, Zhang1997, Amiri2007,Dmitrieva2015}. Consequently, in our subsequent analyses, we consider color noise to consist solely of flicker noise. 

To simulate synthetic noisy SSE data incorporating both white and color noise, we augment Eq. \eqref{eq07} with an additional term, i.e. 
$${X_t = f_t + C_{wn}\times{\epsilon_t} + C_{cn}\times{\epsilon_t^*}},$$ 
where $\epsilon_t^*$ and $C_{cn}$ represent the flicker noise model and its noise level, respectively. The synthetic test results in Section \ref{subsec:syn_detection_results} revealed that the SSAID detection performance diminishes as the white noise level increases. Consequently, we confine the variation of noise levels to a lower range, spanning $1\%$ to $40\%$  with an increment of $2\%$. We generate $100$ data sequences for the same white noise and color noise levels, utilizing different seeds for each. In total, we obtain $20\times20\times{100}$ noisy time series. Fig. \ref{fig04_color_noise} (a) shows the percentage of successful cumulative detection $R_{sd}$ for simulated noisy time series with different white and color noise levels. Notably, when $C_{wn}\leq{25\%}$ and $C_{cn}\leq{15\%}$, $R_{sd}$ can reach a maximum of $100\%$. This underscores SSAID's capability to maintain high performance even in the presence of color noise. However, as $C_{cn}$ approaches $30\%$, $R_{sd}$ decreases to $20\%$, independently from the white noise level.
In addition, $R_{sd}$ decreases to approximately $80\%$ when $C_{wn}\geq{30\%}$ and $C_{cn}\leq{15\%}$. This highlights the sensitivity of SSAID performance to noise levels, particularly to color noise.

\begin{figure}	
	\centering
	\includegraphics[width=\textwidth]{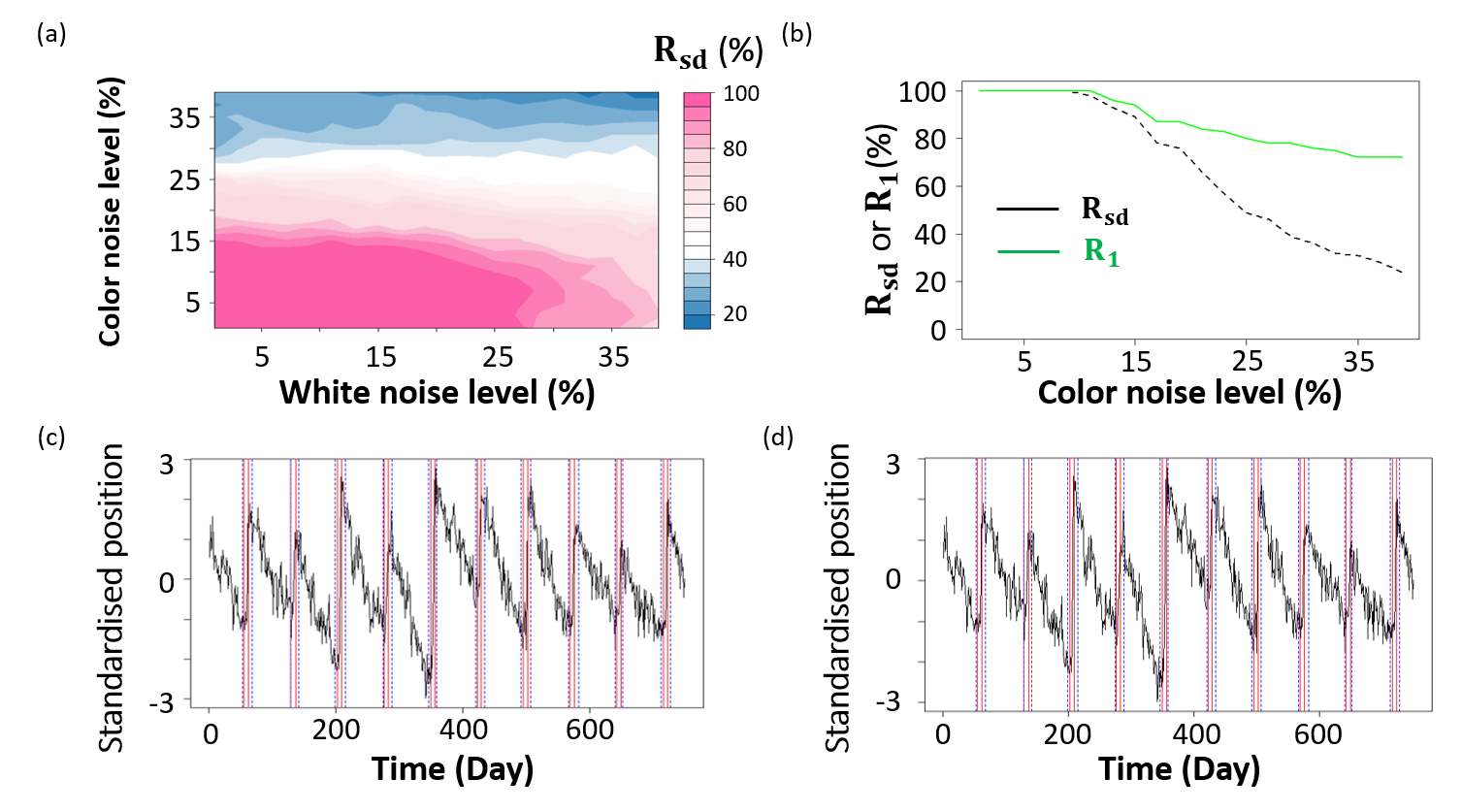}
	\caption{\minewa{(a) The percentage of successful cumulative detection $R_{sd}$ (see definitions in Eq. \eqref{eq08}) for each simulated noisy data with different white and noise levels. For the same $C_{wn}$ and $C_{cn}$, we generate $100$ noisy data sequences using $100$ seeds. (b) The percentage $R_1$ and $R_{sd}$ as a function of color noise level $C_{cn}$ when the white noise level $C_{wn}$ is fixed at $21\%$. The locations of the change-points in two simulation examples with different noise levels are shown in (c) $C_{wn}=21\%$ and $C_{cn}=11\%$; (d) $C_{wn}=25\%$ and  $C_{cn}=31\%$. Blue vertical dotted lines: estimated change-points by SSAID; red vertical lines: true change-points.}}
	\label{fig04_color_noise}	
\end{figure}
}

\section{Application to Observed Data}
\label{sec:application}
\subsection{SSE detection via hypothesis testing}
\label{subsec:SSE_detection_via_hypothsis}
\begin{figure}[!ht]	
	\centering
	\includegraphics[width=\textwidth]{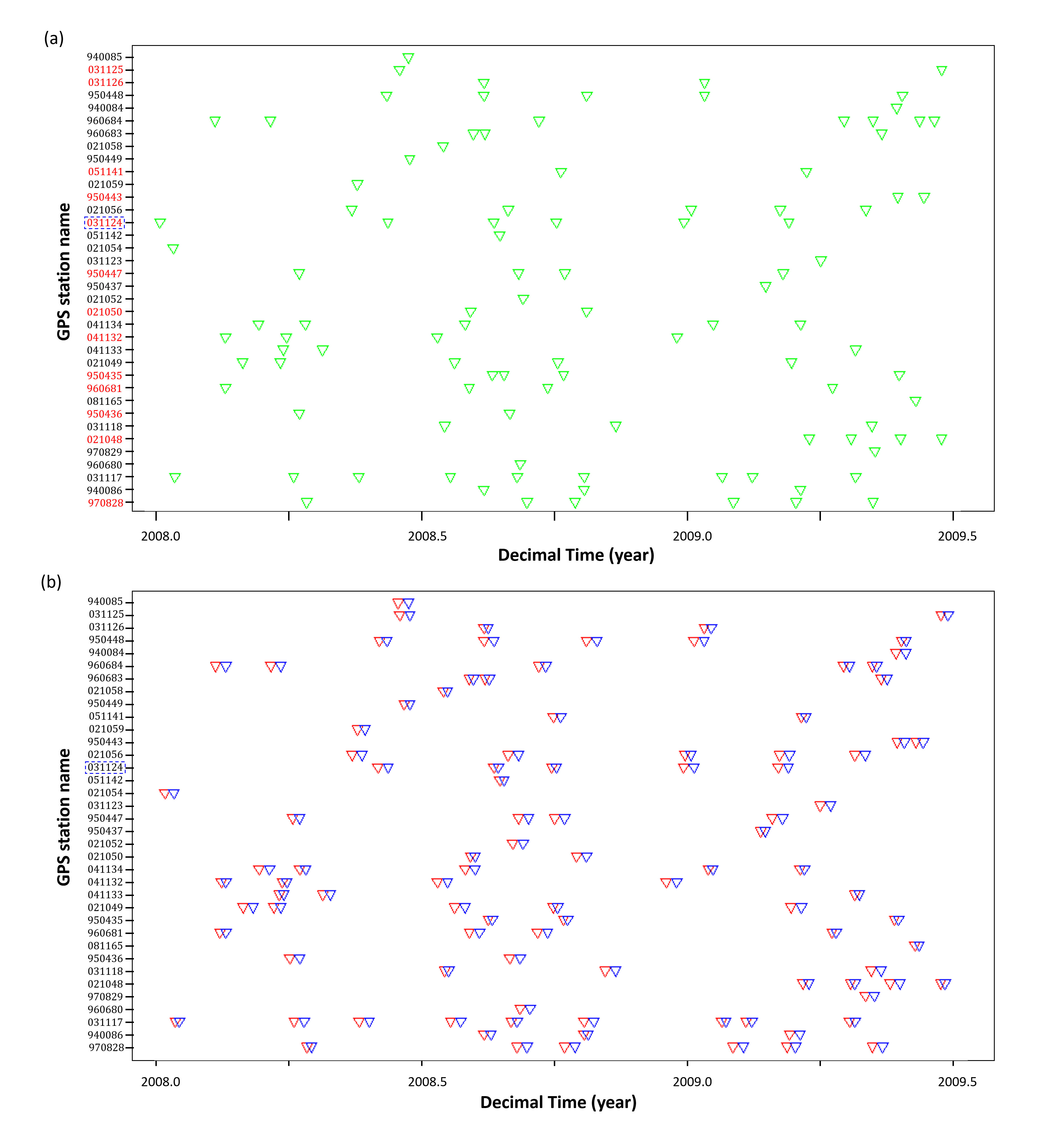}
	\caption{(a) Detected change-points by SSAID in GPS data recorded by the $36$ GPS stations, shown in Fig. \ref{fig01} (b). Station names for which the number of detected change-points is even are highlighted in red. (b) Pre-processed results of detected change-points shown in panel (a). Red triangles: starting change-points; blue triangles: ending change-points. \minewa{\textit{The figure, and subsequent similar figures, is re-plotted by sorting the stations along the direction of N$\ang{50}$W, which is perpendicular to the Nankai Trough.}}}
	\label{fig06}	
\end{figure}

We first present the raw results of detected change-points in the SSE data introduced in \textsection{\ref{sec:data_and_processing}}. The change-points at each station, shown in Fig. \ref{fig06} (a) (see green triangles), do not seem to exhibit a consistent pattern at first sight. In contrast to simulated SSE data (see \textsection{\ref{sec:syn_tests}}), we do not know \textit{a priori} when an SSE starts and ends to validate the detection. However, we can quantify the confidence that a detected change-point corresponds to an SSE by using a hypothesis test, based on the sign change of the displacement rate at the start times of SSEs from the secular displacement rate \cite{Yano2022}. To apply the hypothesis test, we need to know the start and end times of a potential SSE, indicating a pair of change-points are needed to define an SSE. Thereafter, we refer to change-points associated with the start and end times of potential SSEs as starting and ending change-points, respectively.

\subsubsection{Pre-processing\minewc{ for hypothesis testing}}
\label{subsubsec:preprocessing_cp}
\begin{figure}[!ht]	
	\centering
	\includegraphics[width=\textwidth]{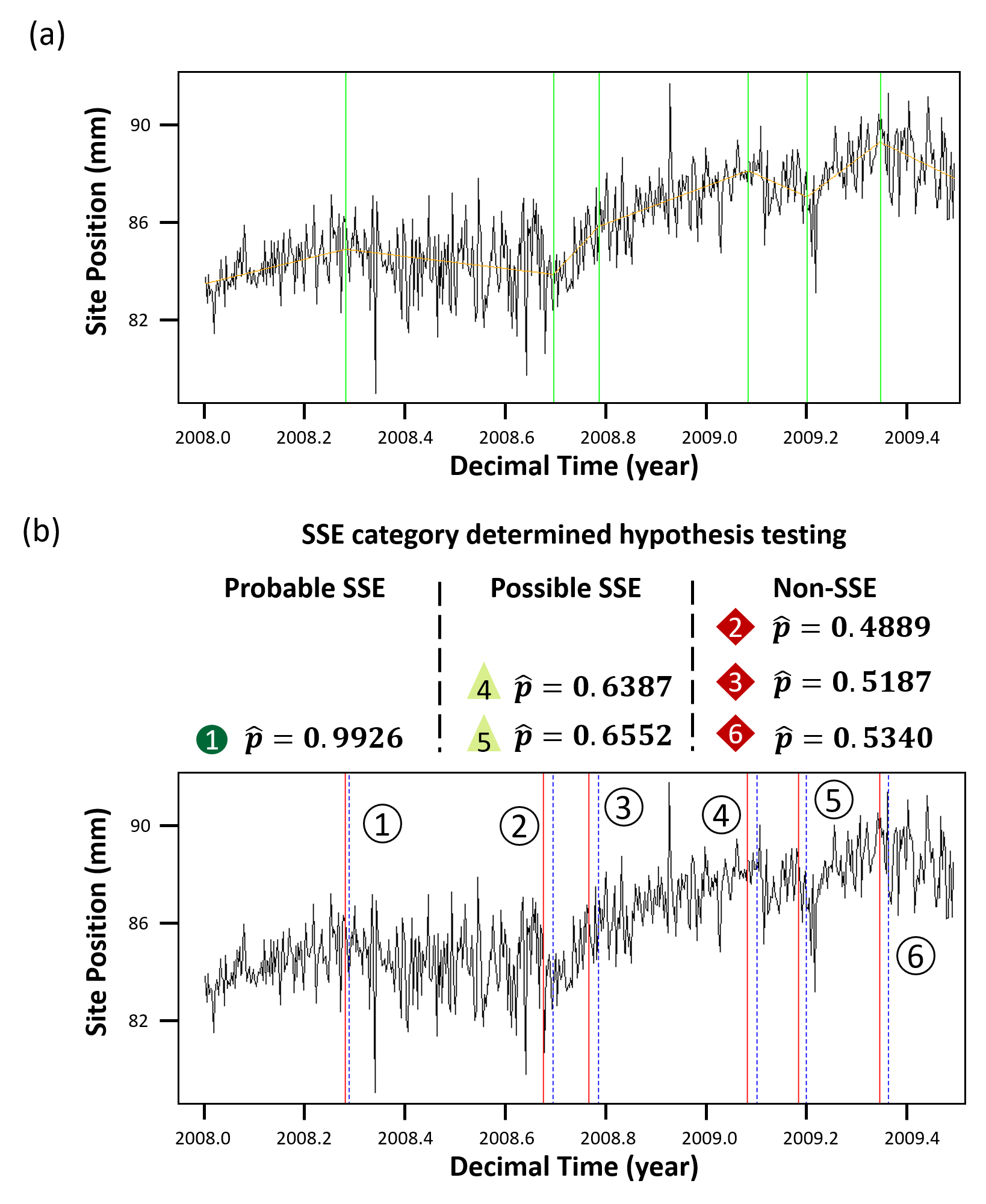}
	\caption{(a) Observed GPS data recorded by station $970828$ (see the black line) and the fitted piecewise-linear signal (see the orange line) using detected change-points by SSAID (see green lines); (b) New paired change-points of the same station $970828$ based on detected change-points in panel (a). Red lines: starting change-points; blue dotted lines: ending change-points. \minewa{The calculated probabilities of SSE occurrences $\hat{p}$ for each pair of change-points are included on the top of panel (b) and their associated SSE categories. The markers used for different SSE categories are the same as those in Fig. \ref{fig08}, and the numbers inside markers are consistent with the numbers in circles in panel (b), i.e.~the indexes of change-point pairs.}}
	\label{fig07}	
\end{figure}

We first pre-process the detected change-points to associate them,  using hypothesis testing, with the start and end times of an SSE. We refer to $\hat{N}_j$ as the number of detected change-points by SSAID at the $j$-th station, where $j$ is the station index ($j=1,\cdots,36$), which sequentially coincides with the station names on the $y$-axis of Fig. \ref{fig06} (a) from the bottom to the top. Although we could expect all $\hat{N}_j$ to be even numbers, only $13$ of them in Fig. \ref{fig06} (a) are even (see station names highlighted in red). This implies that SSAID in most stations misses some change-points associated with SSEs and/or detects spurious change-points not associated with SSEs. We also observe in multiple stations that the time difference between two neighbouring detected change-points can be in the order of months (e.g. the first and the second change-points in Fig. \ref{fig07} (a), which shows the GPS data recorded at station 970828). Such a long duration is not consistent with past studies in this region, which show that potential short-term SSEs during the period analyzed last about $7$ days \cite{Hirose2010,Obara2016,Obara2020}. Therefore, two neighbouring change-points with a large time difference cannot be paired as the start and end times of the same SSE. The above observations indicate that many single change-points were identified as potential SSEs (e.g., see green lines in Fig. \ref{fig07} (a)). 

To remedy this pathology, we create a change-point pair for each single change-point. The procedure contains the following five steps with details provided in the next few paragraphs: (1) we fit a piecewise-linear signal to the noisy SSE data (e.g. the orange line in Fig. \ref{fig07} (a)) using the detected change-points by SSAID shown in Fig. \ref{fig06} (a); (2) we calculate the slopes of each segment in the fitted model; (3) based on these slopes, we identify change-point pairs and single change-points; (4) we create several change-point pair candidates for each single change-point; and (5) \minewa{(5) we fit a piecewise-linear signal for each change-point pair candidate using the detected change-points along with the change-point pair candidate itself; (6) we calculate the Schwarz Information Criterion (SIC) value for each fitted piecewise signal (see the equation to calculate SIC in \citeA{Anastasiou2022}); (7) select the best pair candidate with the minimum SIC value for each single change-point. The SIC balances model fit and complexity, penalizing models with more parameters to avoid overfitting, which is widely used to compare different statistical models and select the best one among them} \cite{Yao1988,Anastasiou2022}. \minewa{See more details about how to pair single change-points by this pre-processing procedure in Text S8 and Figure S15 in the supplement.}

We now illustrate how to pair detected change-points based on the calculated slopes of the segments between change-points. We refer to $k_b^i$ and $k_a^i$ as the slope of the segment before and after the $i$-th detected change-point, respectively. We pair two consecutive change-points ($i$-th and $(i+1)$-th, say) as the start and end times of a unique SSE, if they simultaneously satisfy the following conditions: (1) $k_b^i$ has the same sign as the secular displacement rate; (2) the sign of $k_a^i$ is opposite to that of the secular displacement rate; (3) the time difference between the two neighbouring change-points (i.e. the duration of the SSE) is no more than a duration threshold, denoted by $D_{max}$. Here, we estimate the sign of the secular displacement rate (i.e positive or negative) at each GPS station by taking the slope of a linear model fitted to the whole noisy data. 

\minewa{All change points that have not been paired in the previous step are classified as single change points.} In the study area considered, the expected duration of an SSE is $3-7$ days \cite{Obara2020}. We found that the detected change-point location error by SSAID is at most $3$ days (see Text S2 and Fig.~S8(b) in the supplement). In the worst case, an SSE with duration $7$ days could be detected by a pair of change-points separated by up to $14$ days (assuming maximum error). Therefore, we set $D_{max}$ as $14$ days.

We then generate candidates of undetected change-points to pair with each single change-point. We first assume that each single change-point is associated with either the start or the end time of an SSE, and the duration of SSEs is $3-7$ days. This implies that the undetected change-point candidates are located in a window spanning $\pm{(3-7)}$ days around the detected single change-point. To be more specific, if the detected single change-point is the start time of an SSE, denoted by $\bar{x}_\text{cp}$, the associated change-point candidates for the undetected end time of this SSE include $\bar{x}_\text{cp}+3$, $\bar{x}_\text{cp}+4$, $\cdots$, $\bar{x}_\text{cp}+7$; conversely, if it is the end time of an SSE, the candidates for the start time are $\bar{x}_\text{cp}-7$, $\bar{x}_\text{cp}-6$, $\cdots$, $\bar{x}_\text{cp}-3$. Based on the slopes of two consecutive segments fitted in Step 2, we can determine if each single change-point is the start or the end time of an SSE. We have three possible situations: (1) if $k_b^i$ and $k_a^i$ have the same and the opposite sign as the secular displacement rate, respectively, then we regard the detected single change-point as the start time of an SSE; (2) if $k_b^i$ and $k_a^i$ have the opposite and the same sign as the secular displacement rate, respectively, then we regard the detected single change-point as the end time of an SSE; (3) in other cases, the detected single change-point can be the start time or the end time of an SSE. 

Next, we fit different piecewise-linear curves through the GPS data for every combination of change-point pair candidates. \minewa{The number of fitted piecewise-linear curves for each single change-point corresponds to the number of change-point pair candidates. If the single change-point is either the starting or ending point, there will be $5$ change-point pair candidates, resulting in $5$ piecewise-linear curves. However, if the type of the single change-point is unknown, there will be $10$ change-point pair candidates, resulting in $10$ piecewise-linear curves.} We select the piecewise-linear curve best fitted to the noisy data through the SIC. We then take the associated change-point candidate to pair with the single change-point, and obtain new paired change-points as shown in Fig. \ref{fig06} (b) and Fig. \ref{fig07} (b), in which we have two change-points for the start and end times of each potential SSE (red and blue, respectively). We denote by $\bar{N}^j=2\bar{N}^j_s$ the number of change points at each station $j$ after pairing the single change-points, where $\bar{N}^j_s$ is the number of starting change-points. In our analysis, almost all the \minewa{raw }detected change-points \minewa{by SSAID} were identified as single change-points. Note that we also imposed some manual constraints on the paired change-points to avoid the overlaps of two neighbouring pairs and discard some single change-points with obvious deviations. For example, the first detected change-point in the station $031124$ was identified as an ending change-point at the second day of the analyzed period, while we expected the starting change-point to be $3-7$ days preceding the detected ending change-point, so that we discarded this change-point.

\subsubsection{Hypothesis test}
\label{subsubsec:null_hypothesis}
As discussed in \textsection{\ref{sec:method}}, the overall trend of GPS data is a noisy linear process if no SSE occurs, while the occurrence of an SSE redirects the original trend in a different direction. Upon completion of the SSE, the trend reverses back to its previous state. As shown in Fig. \ref{fig02}, the sign of the displacement rate at the start time of an SSE is opposite to that of the secular displacement rate. The sign change of the displacement rate at the start times of SSEs constitutes the basis of the null hypothesis test, therefore the following tests are only conducted on the starting change-points. In our tests, the null hypothesis is that SSEs do not occur, and the alternative hypothesis is that SSEs occur. \minewc{Let $B$ be a random variable representing the test statistic under the null hypothesis, assumed to follow the standard Gaussian distribution.} 
Following the approach of \citeA{Yano2022}, the test statistic for testing if the $k$-th starting change-point at the $j$-th station is associated with an SSE \minewc{, i.e.~the observed value of $B$, is} set as

\begin{linenomath*}
\begin{equation}
    \bar{B}_j^k = \sgn{\left(v_0^j\right)} \frac{\bar{v}_k^j-\bar{v}_{0}^j}{\frac{1}{\bar{N}_{s}^j-1}\sqrt{\sum_{k=1}^{\bar{N}_{s}^{j}}(\bar{v}_k^j-\bar{v}_{0}^j)^2}},
\label{eq09}
\end{equation}
\end{linenomath*}

\noindent where $\sgn$ refers to the sign function; and $\bar{v}_k^j$ and $\bar{v}_{0}^j$ refer to the displacement rate at the $k$-th starting change-point and the secular displacement rate of the $j$-th station, respectively. We estimate the probability \minewc{under the null hypothesis} that SSEs do not occur at the $k$-th starting point of the $j$-th station by


\begin{linenomath*}
\begin{equation}
    p_j^k=\Prob\left(B\leq{\bar{B}_j^k}\right) = \Phi\left({\bar{B}_j^k}\right),
\label{eq10}
\end{equation}
\end{linenomath*}

\noindent where \minewc{$\Phi\left(\cdot\right)$} refers to the cumulative distribution function of the standard Gaussian distribution. 
\minewc{Here, $p_j^k$ serves as a $p$-value.} The closer \minewc{$\Phi\left({\bar{B}_j^k}\right)$} is to $0$, the more confidently we can reject the null hypothesis. To reduce Type I errors\minewa{ (i.e. false positives)}, we combine $p$-values of stations neighbouring the $j$-th station into a new single $p$-value through the harmonic mean $p$-value method \cite{Wilson2019,Yano2022}, denoted by $\hat{p}_j^k$. Finally, we quantify the confidence of occurrence of SSEs by

\begin{linenomath*}
\begin{equation}
    \Tilde{p}_j^k=1-\hat{p}_j^k.
\label{eq12}
\end{equation}
\end{linenomath*}

\noindent More details about how to calculate $\Tilde{p}_j^k$ can be found in the supplement and in \citeA{Yano2022}.

\subsubsection{Identifying SSE candidates}
\label{subsubsec:identifying_probable_SSEs}
\begin{figure}[!ht]	
	\centering
	\includegraphics[width=\textwidth]{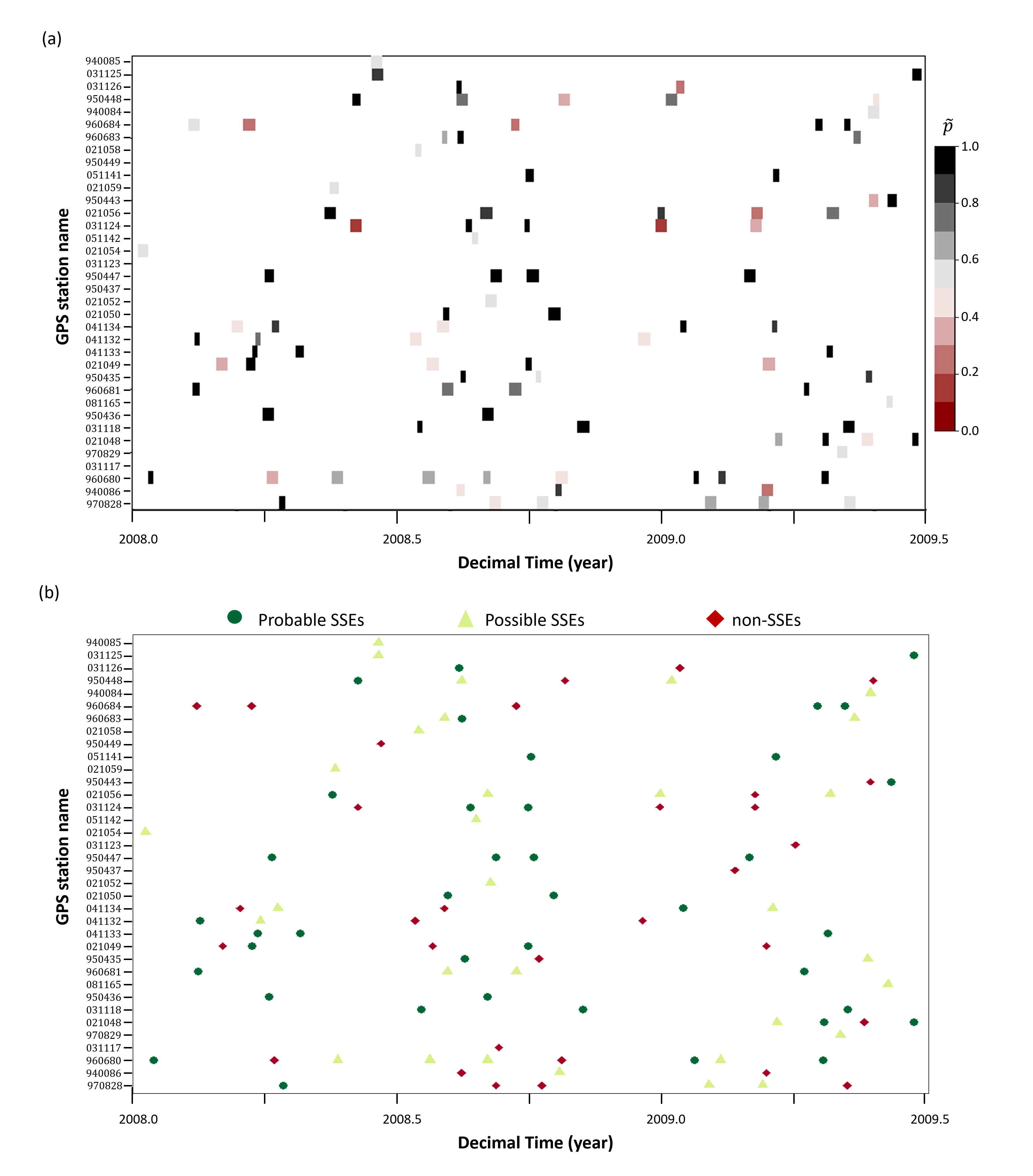}
	\caption{(a) Estimated confidence $\Tilde{p}$ of each change-point pair shown in Fig. \ref{fig06} (b). The left and the right side of each rectangle refer to the starting and the ending change-point, respectively. (b) Detected SSEs categorised as probable SSEs (green circles), possible SSEs (light green triangles) and non-SSEs (red diamonds). The location of each marker refers to the middle time of each SSE candidate. 
    }
	\label{fig08}
\end{figure}

Fig. \ref{fig08} presents the estimated probability of each detected change-point for the occurrence of an SSE by the null hypothesis test and its associated SSE category. We observe that at most stations SSAID can successfully detect SSEs with high confidence. At several stations, no such change-points are found, such as stations $021052$ and $950449$. The best detection happened at station $950447$, in which all the four detected change-points have high confidence value of $\Tilde{p}_j^k\geq{0.9}$. 

Based on the estimated $\Tilde{p}_j^k$ values, we categorize the detected change-points into probable, possible and non-SSE candidates, if $\Tilde{p}_j^k\geq{0.9}$ and $\hat{N}_a^j>1$; ${0.6}\leq{\Tilde{p}_j^k}<{0.9}$ or $\Tilde{p}_j^k\geq{0.9}$ with $\hat{N}_a^j=1$; and $\Tilde{p}_j^k<{0.6}$, respectively. \minewc{
These values were selected to be somewhat conservative in our attempt to confidently claim SSE detection.}
The introduction of $\hat{N}_a^j>1$ in the definition of probable SSE candidates is to guarantee that the detected change-points have a high confidence for the occurrence of SSEs at neighbouring stations within $30$ km simultaneously, rather than at a single station \cite{Yano2022}. Under the current classification rules, we only have a high confidence that detected change-points in the first group are associated with SSEs, and we are less confident that the other detected change-points are associated with SSEs. Fig. \ref{fig08} (b) indicates that we have identified $39$ probable SSE candidates (see green circles) and $31$ possible SSE candidates (see light green triangles) in total across all the stations. Note that some detected SSEs at different stations might be from the same SSE, indicating that the actual number of detected SSEs is likely less than the number stated above. In addition, detected change-points classified as non-SSEs still might be associated with SSEs, as other unknown non-tectonic movements or noise could affect the displacement field at the observation site so that the sign change does not significantly differ from the secular displacement rate \cite{Nishimura2013}. In the remainder of this study, we do not discuss these $2$ groups further and instead we focus on the detected change-points in the first group of probable SSE candidates. 

\subsubsection{Comparison and validation}
\label{subsubsec:comparison_and_validation}
\begin{figure}[!ht]	
	\centering
	\includegraphics[width=\textwidth]{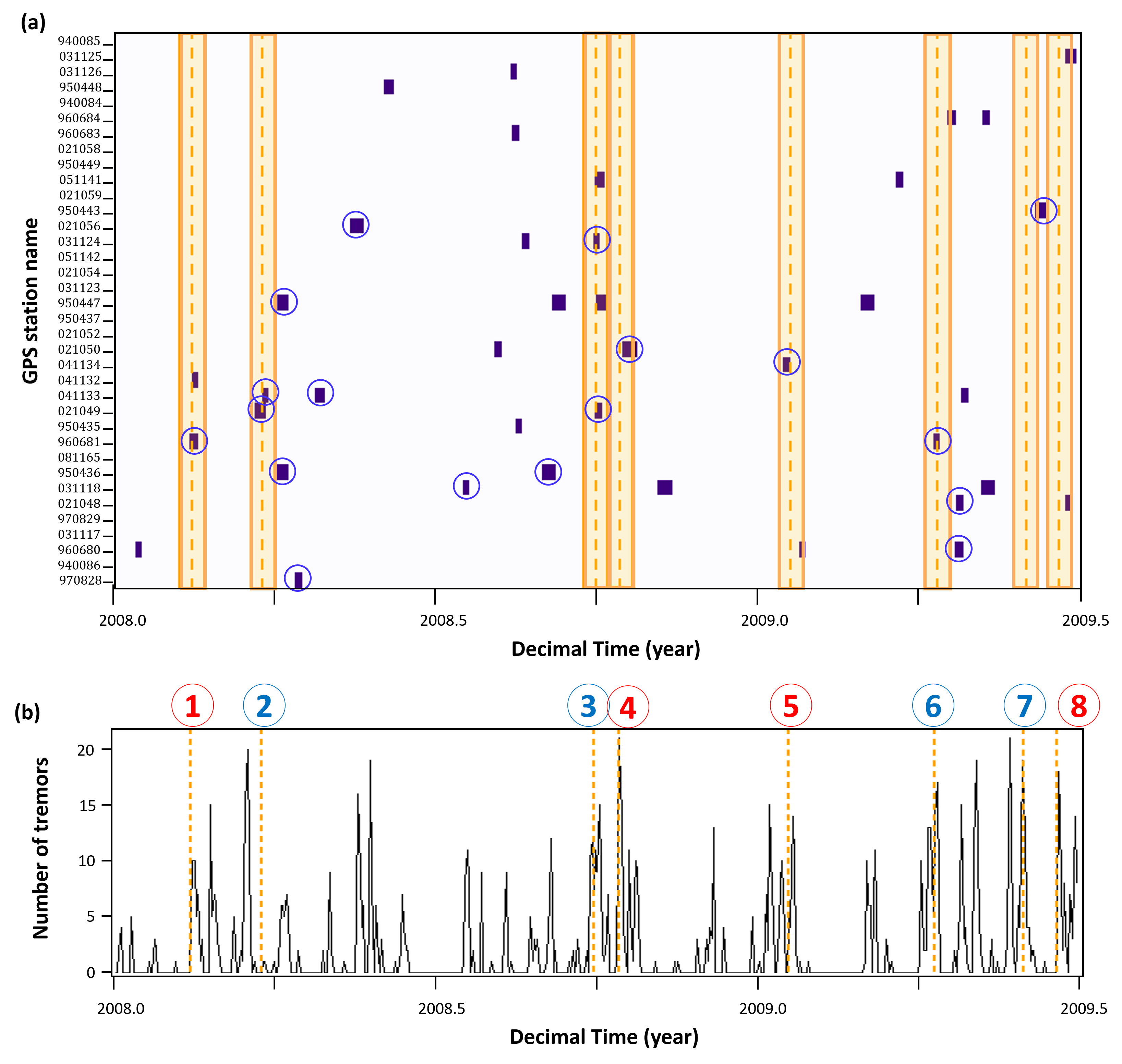}
	\caption[(a) The distribution of detected probable SSEs by SSAID and (b) the daily number of tremors in the episodic state.]{(a) The distribution of detected probable SSEs by SSAID, which are indicated by purple boxes. The left and the right sides of each purple box refer to the start and end times of an identified probable SSE by null hypothesis tests, respectively. Orange dotted lines in the middle of each shaded area refer to the occurrence times of SSEs identified by \citeA{Nishimura2013}. We assume that the start and end times of their identified SSEs are $7$ days before and after the occurrence times, respectively. Purple boxes highlighted by blue circles refer to probable SSEs identified by the fault estimation (see \textsection{\ref{subsec:fault_estimation}}). (b) The daily number of tremors in the episodic state. Numbers in circles on the top refer to the index of identified SSEs by \citeA{Nishimura2013} in Shikoku region. SSEs indicated by blue numbers are located within our research area, while those indicated by red numbers are located in the eastern Shikoku region. 
    }	
	\label{fig09}
\end{figure}

During the period analyzed in our current study, $8$ SSEs were identified in the western Shikoku region along the Bungo Channel by \citeA{Nishimura2013} \cite<see orange shaded-areas in Fig. \ref{fig09} (a); the associated SSE catalogue obtained from>[]{Kano2018}. Not only has our new method successfully detected all these $8$ SSEs in various stations \minewa{identified by \citeA<>[]{Nishimura2013}}, but SSAID is also able to detect many more previously undetected probable SSE candidates. Note that it is not expected that all the SSEs can be recorded at each GPS station, since the SNR and ground displacements caused by SSEs might greatly vary at different stations. If the SNR is too low or the ground displacement is too small at a certain station, the change-points associated with SSEs cannot be detected. 

To further verify the validity of the newly detected probable SSEs, we investigate their correlations with the tremor occurrence, since tremors often accompany SSEs \cite{Rogers2003,Obara2016,Wang2018}. An increasing daily number of tremors generally indicates that an SSE is probably occurring \cite{ITO2007}. Note that the occurrence of SSEs is not always consistent with tremor activity, which means that SSEs can also occur when no tremor activity is detected \cite{Wang2018,Kano2020,Yano2022}. In addition, not all the observed tremors are associated with the occurrence of SSEs. Based on their recurrence pattern, the tremors in the Shikoku region have been categorized into three states: episodic; weak concentration and background by \citeA{Wang2018}, among which only the tremors in the episodic state occur during SSEs. Therefore, we count the number of daily tremors in the episodic state to investigate its correlation with SSEs. As the $36$ GPS stations used in our study are concentrated in the western Shikoku region (see Fig. \ref{fig01} (b)), we only utilize the episodic tremors around these GPS stations \cite<i.e. with state index $1$-$7$ and $9$-$13$ as indicated in >[]{Wang2018}, rather than the whole observed tremor catalogue in the Shikoku region. Fig. \ref{fig09} (a) and (b) show that the identified probable SSEs are well concordant with tremor activity in the episodic states. We also notice that at its highest peaks, the number of tremors is about $20$, during the study period. \minewa{By contrast, the total number of detected probable SSEs across the $36$ GPS stations during the same period, as determined by hypothesis testing, is $39$. The number of SSEs suggested by the tremors is much lower than the identified probable SSEs. This discrepancy is reasonable because the SSE detection via hypothesis testing is on a per-station basis. In practice, the same SSE might be recorded simultaneously by different GPS stations. Therefore, SSE detection in GPS stations after hypothesis testing should be further validated by assessing spatial coherency across the regional network, as is done in the next subsection.} 



\subsection{Fault estimation}
\label{subsec:fault_estimation}
Potential SSEs are expected to bring up a systematic pattern change in the displacement field at various stations, however the above hypothesis tests fail to consider such changes in the displacement field \cite{Nishimura2013}. This can be done by estimating a fault model to describe the observed displacements \cite{Nishimura2013,Nishimura2021,Yano2022}. We use a Bayesian inversion method, i.e. the Markov chain Monte Carlo (MCMC) method with the Metropolis-Hastings algorithm \cite{Bagnardi2018,Yano2022}, to estimate a finite rectangular fault model with uniform slip for each detected probable SSE, and systematically investigate its associated displacement field. This rectangular fault model is the same as that used in \citeA{Okada1985}. Based on the processed cumulative displacement field as shown in Eq. \eqref{eq03}, the displacement field for each probable SSE candidate at various GPS stations can be simply quantified by subtracting the cumulative displacement field at the starting change-point from that at the ending change-point. These estimated daily displacement variations are used to obtain the \minewa{estimation of the fault parameters}. \minewc{A detailed exposition of the MCMC inversion method and its theoretical framework can be found \citeA{Bagnardi2018} and \citeA{Yano2022}. }

For each identified probable SSE (see purple boxes in Fig. \ref{fig09} (a)), we only use the observed displacement data of neighbouring stations \minewa{located} within a designated range as the input data of the inversion. Here, the ranges that we utilize along the dip and the strike directions are $100$ km and $150$ km, respectively, from the station where the probable SSE was identified \cite{Takagi2019}. We further rule out the data with a high percentage of invalid values (i.e. $\geq{20}\%$) during the period analyzed in our study \cite{Nishimura2021}.

Our inversion approach is divided into two stages. First, we take the approach of \citeA{Yano2022} to fully explore the source parameters while we further assume that no tensile component occurs, thus nine source parameters (length, width, depth, latitude, longitude, strike, rake, slip and dip angle) need to be determined. The initial guesses for those nine source parameters are set as follows: the latitude and the longitude of the estimated fault are set as those of the station where the probable SSE candidate was identified; the length and the width are $50$ km and $35$ km, respectively; the slip amount and the rake angle are $10$ mm and $\ang{110}$, respectively; the initial values for the strike, the dip and the depth are obtained by projecting the estimated fault model to the surface of the Philippine Sea Plate. To mitigate the effect of the initial model on the final inversion results, we further simulate $9$ realisations of the initial fault model obtained by randomly perturbing the default model described above. In total, we run the MCMC inversion $10$ times for each detected probable SSE. We then choose the output of these $10$ sets with the smallest residual as a new set of initial model parameters, and conduct a new inversion \cite{Bagnardi2018,Nishimura2021}.

In the second stage, we take the output fault models from the first stage as a new initial model, but we now follow the approach of \citeA{Nishimura2013}, which assumes that the depth, strike and dip angle of the fault model are dependent on its location to fit the surface of the Philippine Sea Plate. This means that we have $6$ free parameters instead of the previous $9$ free parameters. We then estimate a final finite fault model for each probable SSE candidate. As the slip direction of the expected SSEs in the Shikoku region should be opposite to the plate convergence direction (i.e. N$\ang{50}$W), we rule out probable SSEs candidates, for which slip directions are not between N$\ang{100}$E and N$\ang{170}$E \cite{Nishimura2013}. 

\begin{figure}[!ht]	
	\centering
	\includegraphics[width=\textwidth]{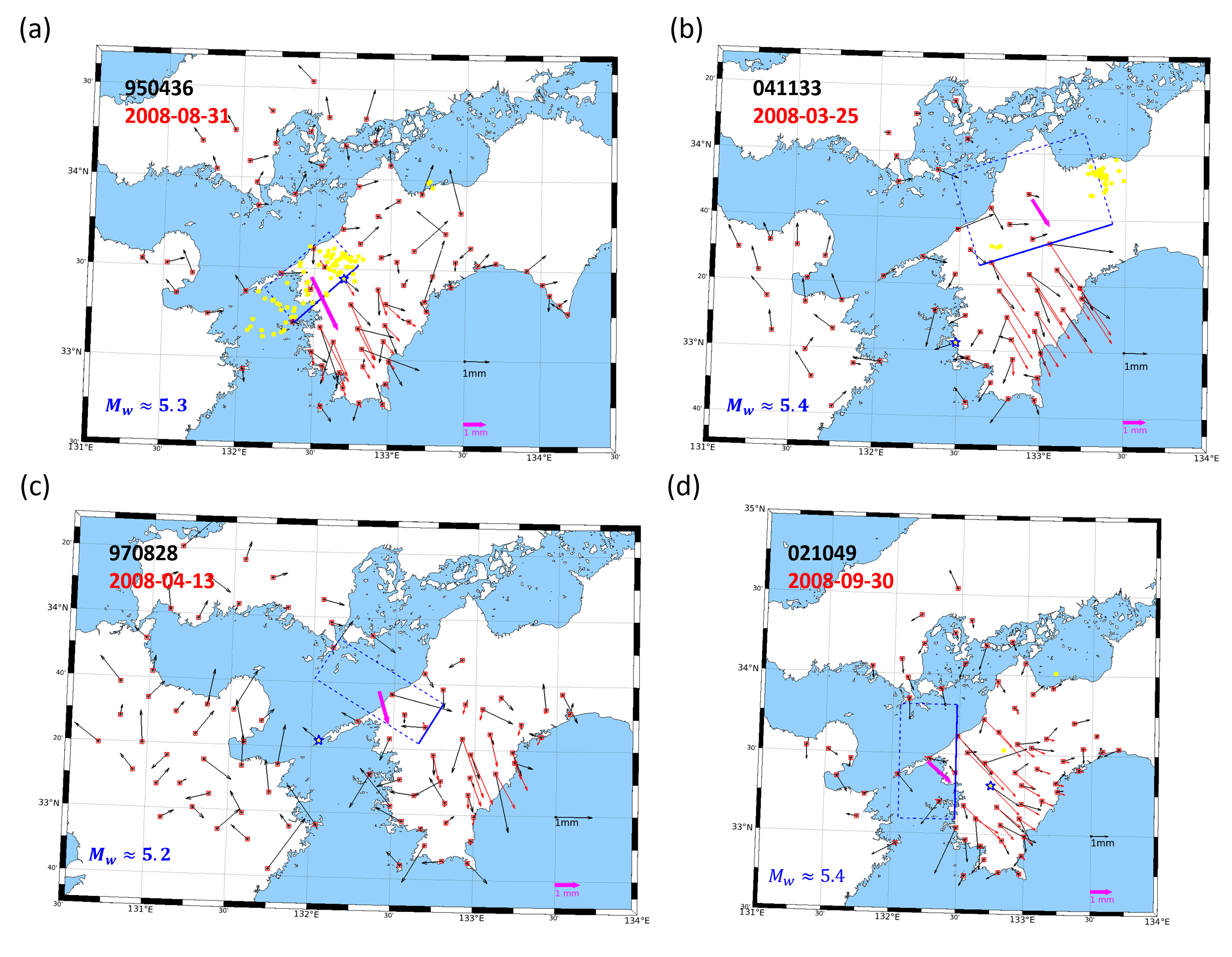}
	\caption{Representative examples of the estimated fault model for identified probable SSE candidates at the different stations: (a) station $970828$; (b) station $021049$; (c) station $950436$; (d) station $041133$. The date in red under the site name refers to the start date of this probable SSE candidate. \minewb{The quantities $M_w$ in blue color refer to the estimated moment magnitude for these identified SSEs. }The star in the map indicates the location of the station where this SSE candidate was identified. The black and the pink arrows in the right-bottom corner are the scale arrows for the observed displacement and the slip amount of the estimated model, respectively. The synthetic displacements by the displacement model of \citeA{Okada1985} have the same arrow scale as the observed ones. \minewa{Yellow} dots indicate the epicentre of tremors in the episodic state $5$ days before and after the date (see the date on the left-upper corner) when this candidate was found. The blue solid line of the rectangle refers to the top edge of the estimated fault model.\minewb{ Note that the GPS stations displayed in each panel correspond to those whose time series are utilized for fault estimation. The specific stations included may vary depending on the location of the detected change-points and their neighboring stations.}}	
	\label{fig10}
\end{figure}

We obtain $18$ potential SSEs in our current research area (see blue circles in Fig. \ref{fig09} (a)). Fig. \ref{fig10} shows representative examples of estimated fault models for four identified probable SSEs (see the other results in the supplement). These identified SSEs have an opposite slip direction to that of the plate convergence. The locations of some estimated faults coincide well with the epicenters of the tremors (see Fig. \ref{fig10} (a) and (b)), suggesting the possible occurrence of episodic tremor and slip (ETS). We also notice that no tremor activities were observed around the estimated fault model in Fig. \ref{fig10} (c) and (d), even though the estimated location is still close to the locations of known SSEs (see Fig. \ref{fig01} (a)). \minewb{We further estimate the moment magnitude ($M_w$) for all identified SSEs using the estimated fault models. $M_w$ is calculated using the formula $M_w = \frac{2}{3} (\log M_0 - 9.1)$, where $M_0 = G \times D \times S$. In this formula, $G$ represents the rigidity of the medium, $D$ the rupture surface area, and $S$ the slip of the estimated fault model\cite{Bormann2011}. We assume $G$ to be $30$ GPa. The estimated moment magnitudes of these identified SSEs range from $4.9$ to $6.1$, with most being between $5.0$ and $5.3$. This is lower than the magnitudes identified in past studies \cite{Nishimura2013}.} 

\section{Conclusions}
\label{sec:conclusions}
We developed a novel statistical method, labelled SSAID, to automatically detect short-term SSEs in GPS data. We demonstrated its effectiveness on a range of noisy simulated SSE data and illustrated its superior detection performance compared to two existing detection methods, i.e. linear regression with $\Delta{AIC}$ and $l_1$ trend filtering. We then applied SSAID to detect short-term SSEs in observed GPS data in the western Shikoku region. The results show that SSAID successfully detects multiple change-points in various GPS stations. We utilized the null hypothesis test to identify probable SSE candidates from these detected change-points, based on the sign of the displacement rate being different from that of the secular displacement rate. These SSE candidates include all known SSEs identified by \citeA{Nishimura2013} during the period analyzed, as well as previously undetected SSEs. We further estimated the parameters of a finite fault model generating the observed displacement field for each probable SSE candidate using a Bayesian inversion technique. Selecting the SSEs for which the azimuth directions of the slip vectors of the estimated fault models are opposite to that of the plate convergence, we managed to identify new SSEs in the western Shikoku region that should be added to the existing catalogue. Our results demonstrate the effectiveness of SSAID in detecting SSEs in observed GPS data. \minewa{Existing methods for detecting short-term SSEs require specifying a suitable threshold to identify the start and end change-points of SSEs. An inappropriate threshold can lead to the misestimation of the number of change-points, making detection performance heavily dependent on the threshold choice. Since different time series require different thresholds, selecting a suitable one for a group of time series is impractical. Our method, however, does not require specifying such parameters, offering greater general applicability to various time series.}

\appendix
\minewc{
\section{Methodology of SSAID}
SSAID is roughly divided into four steps, as shown in Fig. \ref{fig02_workflow}: (1) decomposing the input data into different components by singular spectrum analysis (SSA) and then reconstructing data with different noise levels; (2) adding independent Gaussian noise with various noise levels back to each reconstructed signal to generate new noisy data, some of which are in-SNL data; (3) identifying in-SNL data from the new noisy data generated in Step 2; (4) outputting the locations of estimated change-points for the input data. The pseudocode of SSAID can be found in the supplement  (see Text S5).

\subsection{Step 1: Decomposition process}
\label{subsec:dec_process}
SSA is a powerful non-parametric tool for separating underlying signals from the noise, without the need of \textit{a priori} knowledge of the underlying dynamics \cite{Ghil2002,Chen2013,Walwer2016}. However, it is not designed for detecting change-points. SSA decomposes the noisy data into different components, and then chooses some of these components in order to reconstruct the signal for the underlying true dynamics. We first use SSA to decompose $X_t$ into $M$ components, $X_t = \sum_{j=1}^{M}R_t^j$, each $R_t^j$ $(j=1,\cdots,M)$ denoting an oscillatory component. We then create $M$ sequences of data $Y_t$ by
\begin{linenomath*}
	\begin{equation}
		Y_t^k = \sum_{j=1}^{k}R_t^j \quad (k=1,\cdots,M; t=1,\cdots,T).
		\label{eqA1}
	\end{equation}
\end{linenomath*}
The components $R_t^j$ $(j=1,\cdots,M)$ are sorted in a decreasing order according to their correlation with the underlying dynamics. That is, $R_t^j$ with smaller $j$ are important components of the underlying signal, while those with larger $j$ mostly contain noise. Therefore, the noise level in $Y_t^k$ increases with $k$, such that $Y_t^M=X_t$, that is, no information is lost by this decomposition process. If the noise level in the input data is lower than its minimum SNL, the noise levels for all $Y_t^k$, $k<M$, are also lower than its minimum SNL. Even if the noise level of the input data is large enough to incorporate the SNL range, it is still possible that all the generated $Y_t^k$ do not have an SNL, since the noise level of these $Y_t^k$ is increasing with $k$ at uneven intervals. Therefore, this decomposition cannot ensure the existence of in-SNL data; this is the reason why Step 2 below needs to be implemented in the proposed method. 

\subsection{Step 2: Adding extra noise}
\label{subsec:extra_noise}
After Step 1, we construct $L$ new sequences of noisy data $Z_t^{k,s}$ for each denoised signal $Y_t^k$ $(k=1,\cdots,M)$ to ensure that some in-SNL data can be obtained as follows,
\begin{linenomath*}
	\begin{equation}
		Z_t^{k,s} = Y_t^k + a_s\omega_t \quad (k=1,\cdots,M; s=1,\cdots,L; t=1,\cdots,T),
		\label{eqA2}
	\end{equation}
\end{linenomath*}

\noindent where $\omega_t$ are independent standard, Gaussian random variables and $a_s$ is the level of added noise. If $Y_t^k$ has an SNL and $a_s$ is small enough, $Z_t^{k,s}$ should still have an SNL. Conversely, if $Y_t^k$ has a noise level lower than its minimum SNL and $a_s$ is large enough, $Z_t^{k,s}$ can have an SNL. Therefore, the level of added noise $a_s$ must vary over a sufficiently large range to ensure that some $Z_t^{k,s}$ have an SNL.

Once we obtain in-SNL data, the existing CPD methods for continuous piecewise-linear signals can be applied to detect their change-points (see the tests in the supplement), where we also showed that the percentage of successful cumulative detections $R_{sd}$ (see Eq. \eqref{eq08}) is never higher than $70-80\%$ (see Fig. $S4$ (b) in the supplement), even when the analysed signal has noise in the SNL range. Consequently, we refrain from directly applying the existing CPD methods to these new sequences of noisy data $Z_t^{k,s}$. Instead, we implement an enhancement scheme to increase the percentage of successful cumulative detection $R_{sd}$ for in-SNL data.

In this enhancement scheme, we follow the procedure below to increase $R_{sd}$ for in-SNL data. (1) We generate $Q$ realizations of the time series in Eq. \eqref{eqA2}, and we denote these by $Z_t^{k,s,m}= Y_t^k + a_s\omega_t^m$ $(m=1,\cdots,Q)$, where $\omega_t^m$ is the $m$-th realisation of the noise $\omega_t$ in Eq. \eqref{eqA2}. That is, for the same noise level $a_s$, $\omega_t$ is simulated $Q$ times. The $Q$ realisations of $Z_t^{k,s}$ are collected in a set $\textbf{\textit{G}}^{k,s}=\{Z_t^{k,s,1},\cdots,Z_t^{k,s,Q}\}$. For ease of presentation, this set $\textbf{\textit{G}}^{k,s}$ is called a group. Every realisation $Z_t^{k,s,m}$ in this group is called a member and has the same noise level as $Z_t^{k,s}$ (see Step 3 in Fig. \ref{fig02_workflow}). (2) The change-points in each $Z_t^{k,s,m}$ are detected by a chosen CPD method for continuous piecewise-linear signals. Here, we use the ID method of \citeA{Anastasiou2022} as it is the only one among the five methods examined in Text S2 in the supplement that exhibits an SNL range for all the simulated SSEs in spite of the number of change-points (see Fig. $S9$ in the supplement). (3) We determine $\hat{N}^{k,s}$, the number of estimated change-points in $Z_t^{k,s}$, and identify the locations of the estimated change-points in $Z_t^{k,s}$, stored in a vector $\textbf{\textit{U}}^{k,s}$. Further elaborations on the third procedure can be found in the subsequent two paragraphs. 

Firstly, we determine $\hat{N}^{k,s}$ by a majority voting rule based on the following results. Let $F$ denote the number of realisations in $\textbf{\textit{G}}^{k,s}$ with successful cumulative detections (see the definition of a successful cumulative detection in \textsection{\ref{sec:method}}). Let $P_s$ be the probability that a cumulative detection is successful for a given noise level $a_s$. As $Z_t^{k,s,m}(m=1,\cdots,Q)$ are independent of each other, the probability that at least half of these cumulative detections in group $\textbf{\textit{G}}^{k,s}$ are successful is
\begin{linenomath*}
	\begin{equation}
		\Prob\left(F \geq \frac{Q}{2}\right) = \sum_{q = \lceil Q/2\rceil}^{Q}\Prob\left(F = q\right) = \sum_{q = \lceil Q/2\rceil}^{Q} {Q \choose q}P_s^q(1-P_s)^{Q-q}.
		\label{eq5}
	\end{equation}
\end{linenomath*} 
\noindent If $Z_t^{k,s}$ is an in-SNL data, by the definition of the SNL in \textsection{\ref{sec:method}}, $P_s$ is over $0.5$. This gives $\mathbb{E}({F})=P_sQ\geq{Q/2}$, where $\mathbb{E}(\cdot)$ is the expectation, and hence $\Prob(F\geq{Q/2})$ will converge to $1$ if $Q$ is large enough. For example, $\Prob(F\geq{Q/2})$ is $0.9832$ if $P_s=0.6$ and $Q=100$. Thus, we can estimate the number of change-points for $Z_t^{k,s}$ by using the mode of the $\hat{N}^{k,s,m}$ values, denoted by $\hat{N}^{k,s}$$=Mo\{\hat{N}^{k,s,1},\cdots,\hat{N}^{k,s,Q}\}$, where $\hat{N}^{k,s,m}$ is the number of estimated change-points for $Z_t^{k,s,m}$ and $Mo\{\cdot\}$ denotes the mode. According to Eq. \eqref{eq5}, the probability that $\hat{N}^{k,s}$ is equal to the number of true change-points in $Z_t^{k,s}$ is close to $1$, if $Z_t^{k,s}$ is an in-SNL data. 

Secondly, we identify the locations of estimated change-points for $Z_t^{k,s}$. For ease of discussion, we call a member of $\textbf{\textit{G}}^{k,s}$ a qualified member if it satisfies $\hat{N}^{k,s,m}=$ $\hat{N}^{k,s}$. Let $\kappa$ denote the number of qualified members in group $\textbf{\textit{G}}^{k,s}$. The locations of the estimated change-points for the $j$-th qualified member are collected in a vector $\textbf{\textit{u}}^{k,s,j}$. All these $\textbf{\textit{u}}^{k,s,j}$ have the same length; this being $\hat{N}^{k,s}$. We store these vectors into a matrix

\begin{linenomath*}
	\begin{equation}
		D = 
		\left({\begin{array}{cccc} 
				\hat{\theta}_{1,1} & \hat{\theta}_{1,2} & \cdots  & \hat{\theta}_{1,\hat{N}^{k,s}} \\
				\vdots  & \vdots  & \vdots & \vdots \\
				\hat{\theta}_{j,1} & \hat{\theta}_{j,2} & \cdots & \hat{\theta}_{j,\hat{N}^{k,s}} \\
				\vdots  & \vdots  & \vdots & \vdots \\
				\hat{\theta}_{\kappa,1} & \hat{\theta}_{\kappa,2} & \cdots & \hat{\theta}_{\kappa,\hat{N}^{k,s}}\\
		\end{array}}\right),
		\label{eq6}
	\end{equation}
\end{linenomath*}

\noindent where $\hat{\theta}_{j,i}$ is the location of the $i$-th estimated change-point for the $j$-th qualified member, i.e. $\textbf{\textit{u}}^{k,s,j}=\{\hat{\theta}_{j,1},\cdots,\hat{\theta}_{j,\hat{N}^{k,s}}\}$, $j=1,\cdots,\kappa$. We take the mode of the $i$-th column in $D$ as the estimated location of the $i$-th change point for $Z_t^{k,s}$, denoted by $U_i^{k,s}=Mo\{\hat{\theta}_{1,i},\cdots,\hat{\theta}_{\kappa,i}\}$, $i=1,\cdots,\hat{N}^{k,s}$. Therefore, the estimated change-point locations for $Z_t^{k,s}$ are $\textbf{\textit{U}}^{k,s}=\{U_1^{k,s},\cdots,U_{\hat{N}^{k,s}}^{k,s}\}$. 

We confirm that the proposed majority voting rule above can significantly increase the percentage of successful cumulative detections $R_{sd}$ to $100\%$, when the input data has an SNL, by numerical tests (see Text $S3$ in the supplement).

\subsection{Step 3: Identifying in-SNL data}
\label{subsec:in-snl_identification}
After adding noise, we have generated $L\times{M}\times{Q}$ new noisy data $Z_t^{k,s,m}(k=1,\cdots,M;s=1,\cdots,L;m=1,\cdots,Q)$ (see Step 3 in Fig. \ref{fig02_workflow}) to produce in-SNL data from the input data. However, only some of these noisy data are in-SNL data. Based on the tests conducted in Text $S3$ in the supplement, we impose three conditions observed to identify in-SNL data: (1) $R_2\ge{50}\%$, (2) $\hat{N}\neq{0}$ and (3) $\Omega_3\leq{v}$. Here, $R_2\ge{50}\%$ refers to the percentage of qualified members (see the definition in \textsection{\ref{subsec:extra_noise}}, i.e. $\hat{N}^{k,s,m}=\hat{N}^{k,s}$) in a given group (i.e. $R_2=\kappa/Q$), $\hat{N}$ is the number of estimated change-points for each group by taking its mode (also see the definition of $\hat{N}$ for each group in \textsection{\ref{subsec:extra_noise}}), $\Omega_3$ is the third quartile of the RMSE calculated for each group and $v$ is a pre-defined threshold to define a successful cumulative detection (see \textsection{\ref{sec:method}, i.e. $v=3$ for these simulated SSE data}). The aim of the first condition is to locate in-SNL data. However, $R_2\ge{50}\%$ can occur when the noise level is an SNL or when it is much larger than the SNL range, for which the number of estimated change-points $\hat{N}=0$. This situation is demonstrated in Fig. $S11$ (a) (see the cyan areas) in the supplement. The second condition remedies this pathology. Finally, the third condition aims to remove groups with low accuracy. When calculating the RMSE, the locations of true change-points in the real-world data is estimated through the approach shown in Eq. \eqref{eq6}. Members of a group for which the three conditions are met are all in-SNL data. Otherwise, none of them is. For cases in which no change-points are present in the input data $X_t$, no groups have in-SNL data since $\hat{N}$$=0$, which means that SSAID will not output any change-points (i.e. $\hat{N}=0$). The quantities $R_2^{k,s}$, $\hat{N}^{k,s}$ and $\Omega_3^{k,s}$ indicated in Fig. \ref{fig02_workflow} (see Step 3) are $R_2$, $\hat{N}$ and $\Omega_3$ for group $\textbf{\textit{G}}^{k,s}$, respectively. 

\subsection{Step 4: Outputting the final change-points}
\label{subsec:change-points_output}

We now estimate the locations of change-points in the raw data $X_t$ based on the identified in-SNL data. First, we compute the mode of the distribution of detected change-points in all the identified in-SNL data as the number of estimated change-points $\hat{N}_X$ in the raw data $X_t$. If no non-zero $\hat{N}_X$ value is found, it indicates that SSAID did not detect any change-points in the input data $X_t$, and SSAID outputs no change-points without proceeding further. However, once a non-zero $\hat{N}_X$ is identified, we move to the next step.

Next, we collect the estimated change-points from the in-SNL data that have the same number of change-points as $\hat{N}_X$ into a matrix $D_f$, where each row of $D_f$ represents the locations of detected change-points for a corresponding in-SNL data sequence. Then, we generate two candidate sets of final change-points in $X_t$ by calculating both the mode and the average for each column of $D_f$. Finally, the sSIC criterion is used to select the set of change-points that best characterizes the input data. 


}

\section*{Open Research}
\textbf{Data and Code Availability Statement}
The simulated SSE data used for numerical tests in the study and the code of the newly developed method SSAID are available at Github via \url{https://github.com/yiming-otago/SSAID}, which are provided for private study and research purposes and are protected by copyright with all rights reserved unless otherwise indicated. The observed GPS data utilized in this study can be requested through Geospatial Information Authority of Japan (GSI) at \url{https://www.gsi.go.jp/ENGLISH/geonet_english.html}.


%


\acknowledgments
We are grateful to Associate Professor Ting Wang (University of Otago) for her constant support, insightful discussions, and technical assistance throughout this project. Her expertise and guidance played a crucial role in shaping the direction of this research. We would also like to thank Assistant Professor Akiko Takeo (University of Tokyo) and Professor Takuya Nishimura (Kyoto University) for kindly sharing the processed GPS data and the geometry file of the Philippine Sea Plate surface with us. We further acknowledge the helpful comments and suggestions from Professor Takuya Nishimura on the current work. Y.M. was supported by a University of Otago Doctoral Scholarship, and University of Otago Postgraduate Publishing Bursaries (Doctoral).
 
\clearpage


%
\bibliography{references.bib} 

\begin{thebibliography}{}

\bibitem [\protect \citeauthoryear {%
Agnew%
}{%
Agnew%
}{%
{\protect \APACyear {1992}}%
}]{%
Agnew1992}
\APACinsertmetastar {%
Agnew1992}%
\begin{APACrefauthors}%
Agnew, D\BPBI C.%
\end{APACrefauthors}%
\unskip\
\newblock
\APACrefYearMonthDay{1992}{}{}.
\newblock
{\BBOQ}\APACrefatitle {The time-domain behavior of power-law noises} {The
  time-domain behavior of power-law noises}.{\BBCQ}
\newblock
\APACjournalVolNumPages{Geophysical research letters}{19}{4}{333--336}.
\PrintBackRefs{\CurrentBib}

\bibitem [\protect \citeauthoryear {%
Amiri-Simkooei%
, Tiberius%
\BCBL {}\ \BBA {} Teunissen%
}{%
Amiri-Simkooei%
\ \protect \BOthers {.}}{%
{\protect \APACyear {2007}}%
}]{%
Amiri2007}
\APACinsertmetastar {%
Amiri2007}%
\begin{APACrefauthors}%
Amiri-Simkooei, A\BPBI R.%
, Tiberius, C\BPBI C.%
\BCBL {}\ \BBA {} Teunissen, P\BPBI J.%
\end{APACrefauthors}%
\unskip\
\newblock
\APACrefYearMonthDay{2007}{}{}.
\newblock
{\BBOQ}\APACrefatitle {Assessment of noise in GPS coordinate time series:
  methodology and results} {Assessment of noise in gps coordinate time series:
  methodology and results}.{\BBCQ}
\newblock
\APACjournalVolNumPages{Journal of Geophysical Research: Solid
  Earth}{112}{B7}{}.
\PrintBackRefs{\CurrentBib}

\bibitem [\protect \citeauthoryear {%
Anastasiou%
\ \BBA {} Fryzlewicz%
}{%
Anastasiou%
\ \BBA {} Fryzlewicz%
}{%
{\protect \APACyear {2022}}%
}]{%
Anastasiou2022}
\APACinsertmetastar {%
Anastasiou2022}%
\begin{APACrefauthors}%
Anastasiou, A.%
\BCBT {}\ \BBA {} Fryzlewicz, P.%
\end{APACrefauthors}%
\unskip\
\newblock
\APACrefYearMonthDay{2022}{}{}.
\newblock
{\BBOQ}\APACrefatitle {Detecting multiple generalized change-points by
  isolating single ones} {Detecting multiple generalized change-points by
  isolating single ones}.{\BBCQ}
\newblock
\APACjournalVolNumPages{Metrika}{85}{2}{141--174}.
\PrintBackRefs{\CurrentBib}

\bibitem [\protect \citeauthoryear {%
Bagnardi%
\ \BBA {} Hooper%
}{%
Bagnardi%
\ \BBA {} Hooper%
}{%
{\protect \APACyear {2018}}%
}]{%
Bagnardi2018}
\APACinsertmetastar {%
Bagnardi2018}%
\begin{APACrefauthors}%
Bagnardi, M.%
\BCBT {}\ \BBA {} Hooper, A.%
\end{APACrefauthors}%
\unskip\
\newblock
\APACrefYearMonthDay{2018}{}{}.
\newblock
{\BBOQ}\APACrefatitle {Inversion of surface deformation data for rapid
  estimates of source parameters and uncertainties: A Bayesian approach}
  {Inversion of surface deformation data for rapid estimates of source
  parameters and uncertainties: A bayesian approach}.{\BBCQ}
\newblock
\APACjournalVolNumPages{Geochemistry, Geophysics,
  Geosystems}{19}{7}{2194--2211}.
\PrintBackRefs{\CurrentBib}

\bibitem [\protect \citeauthoryear {%
Barbot%
}{%
Barbot%
}{%
{\protect \APACyear {2019}}%
}]{%
Barbot2019}
\APACinsertmetastar {%
Barbot2019}%
\begin{APACrefauthors}%
Barbot, S.%
\end{APACrefauthors}%
\unskip\
\newblock
\APACrefYearMonthDay{2019}{}{}.
\newblock
{\BBOQ}\APACrefatitle {Slow-slip, slow earthquakes, period-two cycles, full and
  partial ruptures, and deterministic chaos in a single asperity fault}
  {Slow-slip, slow earthquakes, period-two cycles, full and partial ruptures,
  and deterministic chaos in a single asperity fault}.{\BBCQ}
\newblock
\APACjournalVolNumPages{Tectonophysics}{768}{}{228171}.
\PrintBackRefs{\CurrentBib}

\bibitem [\protect \citeauthoryear {%
Bartlow%
, Wallace%
, Beavan%
, Bannister%
\BCBL {}\ \BBA {} Segall%
}{%
Bartlow%
\ \protect \BOthers {.}}{%
{\protect \APACyear {2014}}%
}]{%
Bartlow2014}
\APACinsertmetastar {%
Bartlow2014}%
\begin{APACrefauthors}%
Bartlow, N\BPBI M.%
, Wallace, L\BPBI M.%
, Beavan, R\BPBI J.%
, Bannister, S.%
\BCBL {}\ \BBA {} Segall, P.%
\end{APACrefauthors}%
\unskip\
\newblock
\APACrefYearMonthDay{2014}{}{}.
\newblock
{\BBOQ}\APACrefatitle {Time-dependent modeling of slow slip events and
  associated seismicity and tremor at the Hikurangi subduction zone, New
  Zealand} {Time-dependent modeling of slow slip events and associated
  seismicity and tremor at the hikurangi subduction zone, new zealand}.{\BBCQ}
\newblock
\APACjournalVolNumPages{Journal of Geophysical Research: Solid
  Earth}{119}{1}{734--753}.
\PrintBackRefs{\CurrentBib}

\bibitem [\protect \citeauthoryear {%
Bedford%
\ \BBA {} Bevis%
}{%
Bedford%
\ \BBA {} Bevis%
}{%
{\protect \APACyear {2018}}%
}]{%
Bedford2018}
\APACinsertmetastar {%
Bedford2018}%
\begin{APACrefauthors}%
Bedford, J.%
\BCBT {}\ \BBA {} Bevis, M.%
\end{APACrefauthors}%
\unskip\
\newblock
\APACrefYearMonthDay{2018}{}{}.
\newblock
{\BBOQ}\APACrefatitle {Greedy automatic signal decomposition and its
  application to daily GPS time series} {Greedy automatic signal decomposition
  and its application to daily gps time series}.{\BBCQ}
\newblock
\APACjournalVolNumPages{Journal of Geophysical Research: Solid
  Earth}{123}{8}{6992--7003}.
\PrintBackRefs{\CurrentBib}

\bibitem [\protect \citeauthoryear {%
Beeler%
, Roeloffs%
\BCBL {}\ \BBA {} McCausland%
}{%
Beeler%
\ \protect \BOthers {.}}{%
{\protect \APACyear {2014}}%
}]{%
Beeler2014}
\APACinsertmetastar {%
Beeler2014}%
\begin{APACrefauthors}%
Beeler, N\BPBI M.%
, Roeloffs, E.%
\BCBL {}\ \BBA {} McCausland, W.%
\end{APACrefauthors}%
\unskip\
\newblock
\APACrefYearMonthDay{2014}{}{}.
\newblock
{\BBOQ}\APACrefatitle {Re-estimated effects of deep episodic slip on the
  occurrence and probability of great earthquakes in Cascadia} {Re-estimated
  effects of deep episodic slip on the occurrence and probability of great
  earthquakes in cascadia}.{\BBCQ}
\newblock
\APACjournalVolNumPages{Bulletin of the Seismological Society of
  America}{104}{1}{128--144}.
\PrintBackRefs{\CurrentBib}

\bibitem [\protect \citeauthoryear {%
Bletery%
\ \BBA {} Nocquet%
}{%
Bletery%
\ \BBA {} Nocquet%
}{%
{\protect \APACyear {2020}}%
}]{%
Bletery2020}
\APACinsertmetastar {%
Bletery2020}%
\begin{APACrefauthors}%
Bletery, Q.%
\BCBT {}\ \BBA {} Nocquet, J\BHBI M.%
\end{APACrefauthors}%
\unskip\
\newblock
\APACrefYearMonthDay{2020}{}{}.
\newblock
{\BBOQ}\APACrefatitle {Slip bursts during coalescence of slow slip events in
  Cascadia} {Slip bursts during coalescence of slow slip events in
  cascadia}.{\BBCQ}
\newblock
\APACjournalVolNumPages{Nature communications}{11}{1}{1--6}.
\PrintBackRefs{\CurrentBib}

\bibitem [\protect \citeauthoryear {%
Bormann%
\ \BBA {} Di~Giacomo%
}{%
Bormann%
\ \BBA {} Di~Giacomo%
}{%
{\protect \APACyear {2011}}%
}]{%
Bormann2011}
\APACinsertmetastar {%
Bormann2011}%
\begin{APACrefauthors}%
Bormann, P.%
\BCBT {}\ \BBA {} Di~Giacomo, D.%
\end{APACrefauthors}%
\unskip\
\newblock
\APACrefYearMonthDay{2011}{}{}.
\newblock
{\BBOQ}\APACrefatitle {The moment magnitude M w and the energy magnitude M e:
  common roots and differences} {The moment magnitude m w and the energy
  magnitude m e: common roots and differences}.{\BBCQ}
\newblock
\APACjournalVolNumPages{Journal of Seismology}{15}{}{411--427}.
\PrintBackRefs{\CurrentBib}

\bibitem [\protect \citeauthoryear {%
Cavali{\'e}%
\ \protect \BOthers {.}}{%
Cavali{\'e}%
\ \protect \BOthers {.}}{%
{\protect \APACyear {2013}}%
}]{%
Cavalie2013}
\APACinsertmetastar {%
Cavalie2013}%
\begin{APACrefauthors}%
Cavali{\'e}, O.%
, Pathier, E.%
, Radiguet, M.%
, Vergnolle, M.%
, Cotte, N.%
, Walpersdorf, A.%
\BDBL {}Cotton, F.%
\end{APACrefauthors}%
\unskip\
\newblock
\APACrefYearMonthDay{2013}{}{}.
\newblock
{\BBOQ}\APACrefatitle {Slow slip event in the Mexican subduction zone: Evidence
  of shallower slip in the Guerrero seismic gap for the 2006 event revealed by
  the joint inversion of InSAR and GPS data} {Slow slip event in the mexican
  subduction zone: Evidence of shallower slip in the guerrero seismic gap for
  the 2006 event revealed by the joint inversion of insar and gps data}.{\BBCQ}
\newblock
\APACjournalVolNumPages{Earth and Planetary Science Letters}{367}{}{52--60}.
\PrintBackRefs{\CurrentBib}

\bibitem [\protect \citeauthoryear {%
Chen%
\ \protect \BOthers {.}}{%
Chen%
\ \protect \BOthers {.}}{%
{\protect \APACyear {2013}}%
}]{%
Chen2013}
\APACinsertmetastar {%
Chen2013}%
\begin{APACrefauthors}%
Chen, Q.%
, van Dam, T.%
, Sneeuw, N.%
, Collilieux, X.%
, Weigelt, M.%
\BCBL {}\ \BBA {} Rebischung, P.%
\end{APACrefauthors}%
\unskip\
\newblock
\APACrefYearMonthDay{2013}{}{}.
\newblock
{\BBOQ}\APACrefatitle {Singular spectrum analysis for modeling seasonal signals
  from GPS time series} {Singular spectrum analysis for modeling seasonal
  signals from gps time series}.{\BBCQ}
\newblock
\APACjournalVolNumPages{Journal of Geodynamics}{72}{}{25--35}.
\PrintBackRefs{\CurrentBib}

\bibitem [\protect \citeauthoryear {%
Crowell%
, Bock%
\BCBL {}\ \BBA {} Liu%
}{%
Crowell%
\ \protect \BOthers {.}}{%
{\protect \APACyear {2016}}%
}]{%
Crowell2016}
\APACinsertmetastar {%
Crowell2016}%
\begin{APACrefauthors}%
Crowell, B\BPBI W.%
, Bock, Y.%
\BCBL {}\ \BBA {} Liu, Z.%
\end{APACrefauthors}%
\unskip\
\newblock
\APACrefYearMonthDay{2016}{}{}.
\newblock
{\BBOQ}\APACrefatitle {Single-station automated detection of transient
  deformation in GPS time series with the relative strength index: A case study
  of Cascadian slow slip} {Single-station automated detection of transient
  deformation in gps time series with the relative strength index: A case study
  of cascadian slow slip}.{\BBCQ}
\newblock
\APACjournalVolNumPages{Journal of Geophysical Research: Solid
  Earth}{121}{12}{9077--9094}.
\PrintBackRefs{\CurrentBib}

\bibitem [\protect \citeauthoryear {%
Davis%
, Wernicke%
\BCBL {}\ \BBA {} Tamisiea%
}{%
Davis%
\ \protect \BOthers {.}}{%
{\protect \APACyear {2012}}%
}]{%
Davis2012}
\APACinsertmetastar {%
Davis2012}%
\begin{APACrefauthors}%
Davis, J\BPBI L.%
, Wernicke, B\BPBI P.%
\BCBL {}\ \BBA {} Tamisiea, M\BPBI E.%
\end{APACrefauthors}%
\unskip\
\newblock
\APACrefYearMonthDay{2012}{}{}.
\newblock
{\BBOQ}\APACrefatitle {On seasonal signals in geodetic time series} {On
  seasonal signals in geodetic time series}.{\BBCQ}
\newblock
\APACjournalVolNumPages{Journal of Geophysical Research: Solid
  Earth}{117}{B1}{}.
\PrintBackRefs{\CurrentBib}

\bibitem [\protect \citeauthoryear {%
Dmitrieva%
, Segall%
\BCBL {}\ \BBA {} DeMets%
}{%
Dmitrieva%
\ \protect \BOthers {.}}{%
{\protect \APACyear {2015}}%
}]{%
Dmitrieva2015}
\APACinsertmetastar {%
Dmitrieva2015}%
\begin{APACrefauthors}%
Dmitrieva, K.%
, Segall, P.%
\BCBL {}\ \BBA {} DeMets, C.%
\end{APACrefauthors}%
\unskip\
\newblock
\APACrefYearMonthDay{2015}{}{}.
\newblock
{\BBOQ}\APACrefatitle {Network-based estimation of time-dependent noise in GPS
  position time series} {Network-based estimation of time-dependent noise in
  gps position time series}.{\BBCQ}
\newblock
\APACjournalVolNumPages{Journal of Geodesy}{89}{6}{591--606}.
\PrintBackRefs{\CurrentBib}

\bibitem [\protect \citeauthoryear {%
Dong%
\ \protect \BOthers {.}}{%
Dong%
\ \protect \BOthers {.}}{%
{\protect \APACyear {2006}}%
}]{%
Dong2006}
\APACinsertmetastar {%
Dong2006}%
\begin{APACrefauthors}%
Dong, D.%
, Fang, P.%
, Bock, Y.%
, Webb, F.%
, Prawirodirdjo, L.%
, Kedar, S.%
\BCBL {}\ \BBA {} Jamason, P.%
\end{APACrefauthors}%
\unskip\
\newblock
\APACrefYearMonthDay{2006}{}{}.
\newblock
{\BBOQ}\APACrefatitle {Spatiotemporal filtering using principal component
  analysis and Karhunen-Loeve expansion approaches for regional GPS network
  analysis} {Spatiotemporal filtering using principal component analysis and
  karhunen-loeve expansion approaches for regional gps network
  analysis}.{\BBCQ}
\newblock
\APACjournalVolNumPages{Journal of geophysical research: solid
  earth}{111}{B3}{}.
\PrintBackRefs{\CurrentBib}

\bibitem [\protect \citeauthoryear {%
ElGharbawi%
\ \BBA {} Tamura%
}{%
ElGharbawi%
\ \BBA {} Tamura%
}{%
{\protect \APACyear {2015}}%
}]{%
Elgharbawi2015}
\APACinsertmetastar {%
Elgharbawi2015}%
\begin{APACrefauthors}%
ElGharbawi, T.%
\BCBT {}\ \BBA {} Tamura, M.%
\end{APACrefauthors}%
\unskip\
\newblock
\APACrefYearMonthDay{2015}{}{}.
\newblock
{\BBOQ}\APACrefatitle {Coseismic and postseismic deformation estimation of the
  2011 Tohoku earthquake in Kanto Region, Japan, using InSAR time series
  analysis and GPS} {Coseismic and postseismic deformation estimation of the
  2011 tohoku earthquake in kanto region, japan, using insar time series
  analysis and gps}.{\BBCQ}
\newblock
\APACjournalVolNumPages{Remote Sensing of Environment}{168}{}{374--387}.
\PrintBackRefs{\CurrentBib}

\bibitem [\protect \citeauthoryear {%
Fujita%
, Nishimura%
\BCBL {}\ \BBA {} Miyazaki%
}{%
Fujita%
\ \protect \BOthers {.}}{%
{\protect \APACyear {2019}}%
}]{%
Fujita2019}
\APACinsertmetastar {%
Fujita2019}%
\begin{APACrefauthors}%
Fujita, M.%
, Nishimura, T.%
\BCBL {}\ \BBA {} Miyazaki, S.%
\end{APACrefauthors}%
\unskip\
\newblock
\APACrefYearMonthDay{2019}{}{}.
\newblock
{\BBOQ}\APACrefatitle {Detection of small crustal deformation caused by slow
  slip events in southwest Japan using GNSS and tremor data} {Detection of
  small crustal deformation caused by slow slip events in southwest japan using
  gnss and tremor data}.{\BBCQ}
\newblock
\APACjournalVolNumPages{Earth, Planets and Space}{71}{1}{1--13}.
\PrintBackRefs{\CurrentBib}

\bibitem [\protect \citeauthoryear {%
Fukuda%
, Higuchi%
, Miyazaki%
\BCBL {}\ \BBA {} Kato%
}{%
Fukuda%
\ \protect \BOthers {.}}{%
{\protect \APACyear {2004}}%
}]{%
Fukuda2004}
\APACinsertmetastar {%
Fukuda2004}%
\begin{APACrefauthors}%
Fukuda, J.%
, Higuchi, T.%
, Miyazaki, S.%
\BCBL {}\ \BBA {} Kato, T.%
\end{APACrefauthors}%
\unskip\
\newblock
\APACrefYearMonthDay{2004}{}{}.
\newblock
{\BBOQ}\APACrefatitle {A new approach to time-dependent inversion of geodetic
  data using a Monte Carlo mixture Kalman filter} {A new approach to
  time-dependent inversion of geodetic data using a monte carlo mixture kalman
  filter}.{\BBCQ}
\newblock
\APACjournalVolNumPages{Geophysical Journal International}{159}{1}{17--39}.
\PrintBackRefs{\CurrentBib}

\bibitem [\protect \citeauthoryear {%
Fukuda%
, Miyazaki%
, Higuchi%
\BCBL {}\ \BBA {} Kato%
}{%
Fukuda%
\ \protect \BOthers {.}}{%
{\protect \APACyear {2008}}%
}]{%
Fukuda2008}
\APACinsertmetastar {%
Fukuda2008}%
\begin{APACrefauthors}%
Fukuda, J.%
, Miyazaki, S.%
, Higuchi, T.%
\BCBL {}\ \BBA {} Kato, T.%
\end{APACrefauthors}%
\unskip\
\newblock
\APACrefYearMonthDay{2008}{}{}.
\newblock
{\BBOQ}\APACrefatitle {Geodetic inversion for space—time distribution of
  fault slip with time-varying smoothing regularization} {Geodetic inversion
  for space—time distribution of fault slip with time-varying smoothing
  regularization}.{\BBCQ}
\newblock
\APACjournalVolNumPages{Geophysical Journal International}{173}{1}{25--48}.
\PrintBackRefs{\CurrentBib}

\bibitem [\protect \citeauthoryear {%
Ghil%
\ \protect \BOthers {.}}{%
Ghil%
\ \protect \BOthers {.}}{%
{\protect \APACyear {2002}}%
}]{%
Ghil2002}
\APACinsertmetastar {%
Ghil2002}%
\begin{APACrefauthors}%
Ghil, M.%
, Allen, M.%
, Dettinger, M.%
, Ide, K.%
, Kondrashov, D.%
, Mann, M.%
\BDBL {}others%
\end{APACrefauthors}%
\unskip\
\newblock
\APACrefYearMonthDay{2002}{}{}.
\newblock
{\BBOQ}\APACrefatitle {Advanced spectral methods for climatic time series}
  {Advanced spectral methods for climatic time series}.{\BBCQ}
\newblock
\APACjournalVolNumPages{Reviews of geophysics}{40}{1}{3--1}.
\PrintBackRefs{\CurrentBib}

\bibitem [\protect \citeauthoryear {%
Granat%
\ \protect \BOthers {.}}{%
Granat%
\ \protect \BOthers {.}}{%
{\protect \APACyear {2013}}%
}]{%
Granat2013}
\APACinsertmetastar {%
Granat2013}%
\begin{APACrefauthors}%
Granat, R.%
, Parker, J.%
, Kedar, S.%
, Dong, D.%
, Tang, B.%
\BCBL {}\ \BBA {} Bock, Y.%
\end{APACrefauthors}%
\unskip\
\newblock
\APACrefYearMonthDay{2013}{}{}.
\newblock
{\BBOQ}\APACrefatitle {Statistical approaches to detecting transient signals in
  GPS: Results from the 2009--2011 transient detection exercise} {Statistical
  approaches to detecting transient signals in gps: Results from the 2009--2011
  transient detection exercise}.{\BBCQ}
\newblock
\APACjournalVolNumPages{Seismological Research Letters}{84}{3}{444--454}.
\PrintBackRefs{\CurrentBib}

\bibitem [\protect \citeauthoryear {%
Hackl%
, Malservisi%
, Hugentobler%
\BCBL {}\ \BBA {} Wonnacott%
}{%
Hackl%
\ \protect \BOthers {.}}{%
{\protect \APACyear {2011}}%
}]{%
Hackl2011}
\APACinsertmetastar {%
Hackl2011}%
\begin{APACrefauthors}%
Hackl, M.%
, Malservisi, R.%
, Hugentobler, U.%
\BCBL {}\ \BBA {} Wonnacott, R.%
\end{APACrefauthors}%
\unskip\
\newblock
\APACrefYearMonthDay{2011}{}{}.
\newblock
{\BBOQ}\APACrefatitle {Estimation of velocity uncertainties from GPS time
  series: Examples from the analysis of the South African TrigNet network}
  {Estimation of velocity uncertainties from gps time series: Examples from the
  analysis of the south african trignet network}.{\BBCQ}
\newblock
\APACjournalVolNumPages{Journal of Geophysical Research: Solid
  Earth}{116}{B11}{}.
\PrintBackRefs{\CurrentBib}

\bibitem [\protect \citeauthoryear {%
Haines%
, Wallace%
\BCBL {}\ \BBA {} Dimitrova%
}{%
Haines%
\ \protect \BOthers {.}}{%
{\protect \APACyear {2019}}%
}]{%
Haines2019}
\APACinsertmetastar {%
Haines2019}%
\begin{APACrefauthors}%
Haines, J.%
, Wallace, L\BPBI M.%
\BCBL {}\ \BBA {} Dimitrova, L.%
\end{APACrefauthors}%
\unskip\
\newblock
\APACrefYearMonthDay{2019}{}{}.
\newblock
{\BBOQ}\APACrefatitle {Slow slip event detection in Cascadia using vertical
  derivatives of horizontal stress rates} {Slow slip event detection in
  cascadia using vertical derivatives of horizontal stress rates}.{\BBCQ}
\newblock
\APACjournalVolNumPages{Journal of Geophysical Research: Solid
  Earth}{124}{5}{5153--5173}.
\PrintBackRefs{\CurrentBib}

\bibitem [\protect \citeauthoryear {%
He%
\ \protect \BOthers {.}}{%
He%
\ \protect \BOthers {.}}{%
{\protect \APACyear {2017}}%
}]{%
He2017}
\APACinsertmetastar {%
He2017}%
\begin{APACrefauthors}%
He, X.%
, Montillet, J\BHBI P.%
, Fernandes, R.%
, Bos, M.%
, Yu, K.%
, Hua, X.%
\BCBL {}\ \BBA {} Jiang, W.%
\end{APACrefauthors}%
\unskip\
\newblock
\APACrefYearMonthDay{2017}{}{}.
\newblock
{\BBOQ}\APACrefatitle {Review of current GPS methodologies for producing
  accurate time series and their error sources} {Review of current gps
  methodologies for producing accurate time series and their error
  sources}.{\BBCQ}
\newblock
\APACjournalVolNumPages{Journal of Geodynamics}{106}{}{12--29}.
\PrintBackRefs{\CurrentBib}

\bibitem [\protect \citeauthoryear {%
Hirose%
, Hirahara%
, Kimata%
, Fujii%
\BCBL {}\ \BBA {} Miyazaki%
}{%
Hirose%
\ \protect \BOthers {.}}{%
{\protect \APACyear {1999}}%
}]{%
Hirose1999}
\APACinsertmetastar {%
Hirose1999}%
\begin{APACrefauthors}%
Hirose, H.%
, Hirahara, K.%
, Kimata, F.%
, Fujii, N.%
\BCBL {}\ \BBA {} Miyazaki, S.%
\end{APACrefauthors}%
\unskip\
\newblock
\APACrefYearMonthDay{1999}{}{}.
\newblock
{\BBOQ}\APACrefatitle {A slow thrust slip event following the two 1996
  Hyuganada earthquakes beneath the Bungo Channel, southwest Japan} {A slow
  thrust slip event following the two 1996 hyuganada earthquakes beneath the
  bungo channel, southwest japan}.{\BBCQ}
\newblock
\APACjournalVolNumPages{Geophysical Research Letters}{26}{21}{3237--3240}.
\PrintBackRefs{\CurrentBib}

\bibitem [\protect \citeauthoryear {%
Hirose%
, Kimura%
, Enescu%
\BCBL {}\ \BBA {} Aoi%
}{%
Hirose%
\ \protect \BOthers {.}}{%
{\protect \APACyear {2012}}%
}]{%
Hirose2012}
\APACinsertmetastar {%
Hirose2012}%
\begin{APACrefauthors}%
Hirose, H.%
, Kimura, H.%
, Enescu, B.%
\BCBL {}\ \BBA {} Aoi, S.%
\end{APACrefauthors}%
\unskip\
\newblock
\APACrefYearMonthDay{2012}{}{}.
\newblock
{\BBOQ}\APACrefatitle {Recurrent slow slip event likely hastened by the 2011
  Tohoku earthquake} {Recurrent slow slip event likely hastened by the 2011
  tohoku earthquake}.{\BBCQ}
\newblock
\APACjournalVolNumPages{Proceedings of the National Academy of
  Sciences}{109}{38}{15157--15161}.
\PrintBackRefs{\CurrentBib}

\bibitem [\protect \citeauthoryear {%
Hirose%
\ \BBA {} Kimura%
}{%
Hirose%
\ \BBA {} Kimura%
}{%
{\protect \APACyear {2020}}%
}]{%
Hirose2020}
\APACinsertmetastar {%
Hirose2020}%
\begin{APACrefauthors}%
Hirose, H.%
\BCBT {}\ \BBA {} Kimura, T.%
\end{APACrefauthors}%
\unskip\
\newblock
\APACrefYearMonthDay{2020}{}{}.
\newblock
{\BBOQ}\APACrefatitle {Slip Distributions of Short-Term Slow Slip Events in
  Shikoku, Southwest Japan, From 2001 to 2019 Based on Tilt Change
  Measurements} {Slip distributions of short-term slow slip events in shikoku,
  southwest japan, from 2001 to 2019 based on tilt change measurements}.{\BBCQ}
\newblock
\APACjournalVolNumPages{Journal of Geophysical Research: Solid
  Earth}{125}{6}{e2020JB019601}.
\PrintBackRefs{\CurrentBib}

\bibitem [\protect \citeauthoryear {%
Hirose%
\ \BBA {} Obara%
}{%
Hirose%
\ \BBA {} Obara%
}{%
{\protect \APACyear {2010}}%
}]{%
Hirose2010}
\APACinsertmetastar {%
Hirose2010}%
\begin{APACrefauthors}%
Hirose, H.%
\BCBT {}\ \BBA {} Obara, K.%
\end{APACrefauthors}%
\unskip\
\newblock
\APACrefYearMonthDay{2010}{}{}.
\newblock
{\BBOQ}\APACrefatitle {Recurrence behavior of short-term slow slip and
  correlated nonvolcanic tremor episodes in western Shikoku, southwest Japan}
  {Recurrence behavior of short-term slow slip and correlated nonvolcanic
  tremor episodes in western shikoku, southwest japan}.{\BBCQ}
\newblock
\APACjournalVolNumPages{Journal of Geophysical Research: Solid
  Earth}{115}{B6}{}.
\PrintBackRefs{\CurrentBib}

\bibitem [\protect \citeauthoryear {%
Ikari%
, Marone%
, Saffer%
\BCBL {}\ \BBA {} Kopf%
}{%
Ikari%
\ \protect \BOthers {.}}{%
{\protect \APACyear {2013}}%
}]{%
Ikari2013}
\APACinsertmetastar {%
Ikari2013}%
\begin{APACrefauthors}%
Ikari, M\BPBI J.%
, Marone, C.%
, Saffer, D\BPBI M.%
\BCBL {}\ \BBA {} Kopf, A\BPBI J.%
\end{APACrefauthors}%
\unskip\
\newblock
\APACrefYearMonthDay{2013}{}{}.
\newblock
{\BBOQ}\APACrefatitle {Slip weakening as a mechanism for slow earthquakes}
  {Slip weakening as a mechanism for slow earthquakes}.{\BBCQ}
\newblock
\APACjournalVolNumPages{Nature geoscience}{6}{6}{468--472}.
\PrintBackRefs{\CurrentBib}

\bibitem [\protect \citeauthoryear {%
Ito%
\ \protect \BOthers {.}}{%
Ito%
\ \protect \BOthers {.}}{%
{\protect \APACyear {2013}}%
}]{%
Ito2013}
\APACinsertmetastar {%
Ito2013}%
\begin{APACrefauthors}%
Ito, Y.%
, Hino, R.%
, Kido, M.%
, Fujimoto, H.%
, Osada, Y.%
, Inazu, D.%
\BDBL {}others%
\end{APACrefauthors}%
\unskip\
\newblock
\APACrefYearMonthDay{2013}{}{}.
\newblock
{\BBOQ}\APACrefatitle {Episodic slow slip events in the Japan subduction zone
  before the 2011 Tohoku-Oki earthquake} {Episodic slow slip events in the
  japan subduction zone before the 2011 tohoku-oki earthquake}.{\BBCQ}
\newblock
\APACjournalVolNumPages{Tectonophysics}{600}{}{14--26}.
\PrintBackRefs{\CurrentBib}

\bibitem [\protect \citeauthoryear {%
Ito%
, Obara%
, Shiomi%
, Sekine%
\BCBL {}\ \BBA {} Hirose%
}{%
Ito%
\ \protect \BOthers {.}}{%
{\protect \APACyear {2007}}%
}]{%
ITO2007}
\APACinsertmetastar {%
ITO2007}%
\begin{APACrefauthors}%
Ito, Y.%
, Obara, K.%
, Shiomi, K.%
, Sekine, S.%
\BCBL {}\ \BBA {} Hirose, H.%
\end{APACrefauthors}%
\unskip\
\newblock
\APACrefYearMonthDay{2007}{}{}.
\newblock
{\BBOQ}\APACrefatitle {Slow earthquakes coincident with episodic tremors and
  slow slip events} {Slow earthquakes coincident with episodic tremors and slow
  slip events}.{\BBCQ}
\newblock
\APACjournalVolNumPages{Science}{315}{5811}{503--506}.
\PrintBackRefs{\CurrentBib}

\bibitem [\protect \citeauthoryear {%
Ji%
\ \BBA {} Herring%
}{%
Ji%
\ \BBA {} Herring%
}{%
{\protect \APACyear {2013}}%
}]{%
Ji2013}
\APACinsertmetastar {%
Ji2013}%
\begin{APACrefauthors}%
Ji, K\BPBI H.%
\BCBT {}\ \BBA {} Herring, T\BPBI A.%
\end{APACrefauthors}%
\unskip\
\newblock
\APACrefYearMonthDay{2013}{}{}.
\newblock
{\BBOQ}\APACrefatitle {A method for detecting transient signals in GPS position
  time-series: smoothing and principal component analysis} {A method for
  detecting transient signals in gps position time-series: smoothing and
  principal component analysis}.{\BBCQ}
\newblock
\APACjournalVolNumPages{Geophysical Journal International}{193}{1}{171--186}.
\PrintBackRefs{\CurrentBib}

\bibitem [\protect \citeauthoryear {%
Jiang%
\ \protect \BOthers {.}}{%
Jiang%
\ \protect \BOthers {.}}{%
{\protect \APACyear {2012}}%
}]{%
Jiang2012}
\APACinsertmetastar {%
Jiang2012}%
\begin{APACrefauthors}%
Jiang, Y.%
, Wdowinski, S.%
, Dixon, T\BPBI H.%
, Hackl, M.%
, Protti, M.%
\BCBL {}\ \BBA {} Gonzalez, V.%
\end{APACrefauthors}%
\unskip\
\newblock
\APACrefYearMonthDay{2012}{}{}.
\newblock
{\BBOQ}\APACrefatitle {Slow slip events in Costa Rica detected by continuous
  GPS observations, 2002--2011} {Slow slip events in costa rica detected by
  continuous gps observations, 2002--2011}.{\BBCQ}
\newblock
\APACjournalVolNumPages{Geochemistry, Geophysics, Geosystems}{13}{4}{}.
\PrintBackRefs{\CurrentBib}

\bibitem [\protect \citeauthoryear {%
Jordan%
\ \BBA {} Jones%
}{%
Jordan%
\ \BBA {} Jones%
}{%
{\protect \APACyear {2010}}%
}]{%
Jordan2010}
\APACinsertmetastar {%
Jordan2010}%
\begin{APACrefauthors}%
Jordan, T\BPBI H.%
\BCBT {}\ \BBA {} Jones, L\BPBI M.%
\end{APACrefauthors}%
\unskip\
\newblock
\APACrefYearMonthDay{2010}{}{}.
\newblock
{\BBOQ}\APACrefatitle {Operational earthquake forecasting: Some thoughts on why
  and how} {Operational earthquake forecasting: Some thoughts on why and
  how}.{\BBCQ}
\newblock
\APACjournalVolNumPages{Seismological Research Letters}{81}{4}{571--574}.
\PrintBackRefs{\CurrentBib}

\bibitem [\protect \citeauthoryear {%
Kano%
\ \protect \BOthers {.}}{%
Kano%
\ \protect \BOthers {.}}{%
{\protect \APACyear {2018}}%
}]{%
Kano2018}
\APACinsertmetastar {%
Kano2018}%
\begin{APACrefauthors}%
Kano, M.%
, Aso, N.%
, Matsuzawa, T.%
, Ide, S.%
, Annoura, S.%
, Arai, R.%
\BDBL {}others%
\end{APACrefauthors}%
\unskip\
\newblock
\APACrefYearMonthDay{2018}{}{}.
\newblock
{\BBOQ}\APACrefatitle {Development of a slow earthquake database} {Development
  of a slow earthquake database}.{\BBCQ}
\newblock
\APACjournalVolNumPages{Seismological Research Letters}{89}{4}{1566--1575}.
\PrintBackRefs{\CurrentBib}

\bibitem [\protect \citeauthoryear {%
Kano%
\ \BBA {} Kato%
}{%
Kano%
\ \BBA {} Kato%
}{%
{\protect \APACyear {2020}}%
}]{%
Kano2020}
\APACinsertmetastar {%
Kano2020}%
\begin{APACrefauthors}%
Kano, M.%
\BCBT {}\ \BBA {} Kato, A.%
\end{APACrefauthors}%
\unskip\
\newblock
\APACrefYearMonthDay{2020}{}{}.
\newblock
{\BBOQ}\APACrefatitle {Detailed spatial slip distribution for short-term slow
  slip events along the Nankai subduction zone, southwest Japan} {Detailed
  spatial slip distribution for short-term slow slip events along the nankai
  subduction zone, southwest japan}.{\BBCQ}
\newblock
\APACjournalVolNumPages{Journal of Geophysical Research: Solid
  Earth}{125}{7}{e2020JB019613}.
\PrintBackRefs{\CurrentBib}

\bibitem [\protect \citeauthoryear {%
Lohman%
\ \BBA {} Murray%
}{%
Lohman%
\ \BBA {} Murray%
}{%
{\protect \APACyear {2013}}%
}]{%
Lohman2013}
\APACinsertmetastar {%
Lohman2013}%
\begin{APACrefauthors}%
Lohman, R\BPBI B.%
\BCBT {}\ \BBA {} Murray, J\BPBI R.%
\end{APACrefauthors}%
\unskip\
\newblock
\APACrefYearMonthDay{2013}{}{}.
\newblock
{\BBOQ}\APACrefatitle {The SCEC geodetic transient detection validation
  exercise} {The scec geodetic transient detection validation exercise}.{\BBCQ}
\newblock
\APACjournalVolNumPages{Seismological Research Letters}{84}{3}{419--425}.
\PrintBackRefs{\CurrentBib}

\bibitem [\protect \citeauthoryear {%
Mao%
, Harrison%
\BCBL {}\ \BBA {} Dixon%
}{%
Mao%
\ \protect \BOthers {.}}{%
{\protect \APACyear {1999}}%
}]{%
Mao1999}
\APACinsertmetastar {%
Mao1999}%
\begin{APACrefauthors}%
Mao, A.%
, Harrison, C\BPBI G.%
\BCBL {}\ \BBA {} Dixon, T\BPBI H.%
\end{APACrefauthors}%
\unskip\
\newblock
\APACrefYearMonthDay{1999}{}{}.
\newblock
{\BBOQ}\APACrefatitle {Noise in GPS coordinate time series} {Noise in gps
  coordinate time series}.{\BBCQ}
\newblock
\APACjournalVolNumPages{Journal of Geophysical Research: Solid
  Earth}{104}{B2}{2797--2816}.
\PrintBackRefs{\CurrentBib}

\bibitem [\protect \citeauthoryear {%
Mazzotti%
\ \BBA {} Adams%
}{%
Mazzotti%
\ \BBA {} Adams%
}{%
{\protect \APACyear {2004}}%
}]{%
Mazzotti2004}
\APACinsertmetastar {%
Mazzotti2004}%
\begin{APACrefauthors}%
Mazzotti, S.%
\BCBT {}\ \BBA {} Adams, J.%
\end{APACrefauthors}%
\unskip\
\newblock
\APACrefYearMonthDay{2004}{}{}.
\newblock
{\BBOQ}\APACrefatitle {Variability of near-term probability for the next great
  earthquake on the Cascadia subduction zone} {Variability of near-term
  probability for the next great earthquake on the cascadia subduction
  zone}.{\BBCQ}
\newblock
\APACjournalVolNumPages{Bulletin of the Seismological Society of
  America}{94}{5}{1954--1959}.
\PrintBackRefs{\CurrentBib}

\bibitem [\protect \citeauthoryear {%
McCaffrey%
}{%
McCaffrey%
}{%
{\protect \APACyear {2009}}%
}]{%
Mccaffrey2009}
\APACinsertmetastar {%
Mccaffrey2009}%
\begin{APACrefauthors}%
McCaffrey, R.%
\end{APACrefauthors}%
\unskip\
\newblock
\APACrefYearMonthDay{2009}{}{}.
\newblock
{\BBOQ}\APACrefatitle {Time-dependent inversion of three-component continuous
  GPS for steady and transient sources in northern Cascadia} {Time-dependent
  inversion of three-component continuous gps for steady and transient sources
  in northern cascadia}.{\BBCQ}
\newblock
\APACjournalVolNumPages{Geophysical Research Letters}{36}{7}{}.
\PrintBackRefs{\CurrentBib}

\bibitem [\protect \citeauthoryear {%
McGuire%
\ \BBA {} Segall%
}{%
McGuire%
\ \BBA {} Segall%
}{%
{\protect \APACyear {2003}}%
}]{%
McGuire2003}
\APACinsertmetastar {%
McGuire2003}%
\begin{APACrefauthors}%
McGuire, J\BPBI J.%
\BCBT {}\ \BBA {} Segall, P.%
\end{APACrefauthors}%
\unskip\
\newblock
\APACrefYearMonthDay{2003}{}{}.
\newblock
{\BBOQ}\APACrefatitle {Imaging of aseismic fault slip transients recorded by
  dense geodetic networks} {Imaging of aseismic fault slip transients recorded
  by dense geodetic networks}.{\BBCQ}
\newblock
\APACjournalVolNumPages{Geophysical Journal International}{155}{3}{778--788}.
\PrintBackRefs{\CurrentBib}

\bibitem [\protect \citeauthoryear {%
Melbourne%
, Szeliga%
, Miller%
\BCBL {}\ \BBA {} Santillan%
}{%
Melbourne%
\ \protect \BOthers {.}}{%
{\protect \APACyear {2005}}%
}]{%
Melbourne2005}
\APACinsertmetastar {%
Melbourne2005}%
\begin{APACrefauthors}%
Melbourne, T\BPBI I.%
, Szeliga, W\BPBI M.%
, Miller, M\BPBI M.%
\BCBL {}\ \BBA {} Santillan, V\BPBI M.%
\end{APACrefauthors}%
\unskip\
\newblock
\APACrefYearMonthDay{2005}{}{}.
\newblock
{\BBOQ}\APACrefatitle {Extent and duration of the 2003 Cascadia slow
  earthquake} {Extent and duration of the 2003 cascadia slow
  earthquake}.{\BBCQ}
\newblock
\APACjournalVolNumPages{Geophysical Research Letters}{32}{4}{}.
\PrintBackRefs{\CurrentBib}

\bibitem [\protect \citeauthoryear {%
Mitsui%
\ \BBA {} Hirahara%
}{%
Mitsui%
\ \BBA {} Hirahara%
}{%
{\protect \APACyear {2006}}%
}]{%
Mitsui2006}
\APACinsertmetastar {%
Mitsui2006}%
\begin{APACrefauthors}%
Mitsui, N.%
\BCBT {}\ \BBA {} Hirahara, K.%
\end{APACrefauthors}%
\unskip\
\newblock
\APACrefYearMonthDay{2006}{}{}.
\newblock
{\BBOQ}\APACrefatitle {Slow slip events controlled by the slab dip and its
  lateral change along a trench} {Slow slip events controlled by the slab dip
  and its lateral change along a trench}.{\BBCQ}
\newblock
\APACjournalVolNumPages{Earth and Planetary Science
  Letters}{245}{1-2}{344--358}.
\PrintBackRefs{\CurrentBib}

\bibitem [\protect \citeauthoryear {%
Miyazaki%
\ \BBA {} Heki%
}{%
Miyazaki%
\ \BBA {} Heki%
}{%
{\protect \APACyear {2001}}%
}]{%
Miyazaki2001}
\APACinsertmetastar {%
Miyazaki2001}%
\begin{APACrefauthors}%
Miyazaki, S.%
\BCBT {}\ \BBA {} Heki, K.%
\end{APACrefauthors}%
\unskip\
\newblock
\APACrefYearMonthDay{2001}{}{}.
\newblock
{\BBOQ}\APACrefatitle {Crustal velocity field of southwest Japan: Subduction
  and arc-arc collision} {Crustal velocity field of southwest japan: Subduction
  and arc-arc collision}.{\BBCQ}
\newblock
\APACjournalVolNumPages{Journal of Geophysical Research: Solid
  Earth}{106}{B3}{4305--4326}.
\PrintBackRefs{\CurrentBib}

\bibitem [\protect \citeauthoryear {%
Miyazaki%
, McGuire%
\BCBL {}\ \BBA {} Segall%
}{%
Miyazaki%
\ \protect \BOthers {.}}{%
{\protect \APACyear {2003}}%
}]{%
Miyazaki2003}
\APACinsertmetastar {%
Miyazaki2003}%
\begin{APACrefauthors}%
Miyazaki, S.%
, McGuire, J\BPBI J.%
\BCBL {}\ \BBA {} Segall, P.%
\end{APACrefauthors}%
\unskip\
\newblock
\APACrefYearMonthDay{2003}{}{}.
\newblock
{\BBOQ}\APACrefatitle {A transient subduction zone slip episode in southwest
  Japan observed by the nationwide GPS array} {A transient subduction zone slip
  episode in southwest japan observed by the nationwide gps array}.{\BBCQ}
\newblock
\APACjournalVolNumPages{Journal of Geophysical Research: Solid
  Earth}{108}{B2}{}.
\PrintBackRefs{\CurrentBib}

\bibitem [\protect \citeauthoryear {%
Nikolaidis%
}{%
Nikolaidis%
}{%
{\protect \APACyear {2002}}%
}]{%
Nikolaidis2002}
\APACinsertmetastar {%
Nikolaidis2002}%
\begin{APACrefauthors}%
Nikolaidis, R.%
\end{APACrefauthors}%
\unskip\
\newblock
\APACrefYear{2002}.
\newblock
\APACrefbtitle {Observation of geodetic and seismic deformation with the Global
  Positioning System} {Observation of geodetic and seismic deformation with the
  global positioning system}.
\newblock
\APACaddressPublisher{}{University of California, San Diego}.
\PrintBackRefs{\CurrentBib}

\bibitem [\protect \citeauthoryear {%
Nishikawa%
\ \protect \BOthers {.}}{%
Nishikawa%
\ \protect \BOthers {.}}{%
{\protect \APACyear {2019}}%
}]{%
Nishikawa2019}
\APACinsertmetastar {%
Nishikawa2019}%
\begin{APACrefauthors}%
Nishikawa, T.%
, Matsuzawa, T.%
, Ohta, K.%
, Uchida, N.%
, Nishimura, T.%
\BCBL {}\ \BBA {} Ide, S.%
\end{APACrefauthors}%
\unskip\
\newblock
\APACrefYearMonthDay{2019}{}{}.
\newblock
{\BBOQ}\APACrefatitle {The slow earthquake spectrum in the Japan Trench
  illuminated by the S-net seafloor observatories} {The slow earthquake
  spectrum in the japan trench illuminated by the s-net seafloor
  observatories}.{\BBCQ}
\newblock
\APACjournalVolNumPages{Science}{365}{6455}{808--813}.
\PrintBackRefs{\CurrentBib}

\bibitem [\protect \citeauthoryear {%
Nishimura%
}{%
Nishimura%
}{%
{\protect \APACyear {2014}}%
}]{%
Nishimura2014}
\APACinsertmetastar {%
Nishimura2014}%
\begin{APACrefauthors}%
Nishimura, T.%
\end{APACrefauthors}%
\unskip\
\newblock
\APACrefYearMonthDay{2014}{}{}.
\newblock
{\BBOQ}\APACrefatitle {Short-term slow slip events along the Ryukyu Trench,
  southwestern Japan, observed by continuous GNSS} {Short-term slow slip events
  along the ryukyu trench, southwestern japan, observed by continuous
  gnss}.{\BBCQ}
\newblock
\APACjournalVolNumPages{Progress in Earth and Planetary Science}{1}{1}{1--13}.
\PrintBackRefs{\CurrentBib}

\bibitem [\protect \citeauthoryear {%
Nishimura%
}{%
Nishimura%
}{%
{\protect \APACyear {2021}}%
}]{%
Nishimura2021}
\APACinsertmetastar {%
Nishimura2021}%
\begin{APACrefauthors}%
Nishimura, T.%
\end{APACrefauthors}%
\unskip\
\newblock
\APACrefYearMonthDay{2021}{}{}.
\newblock
{\BBOQ}\APACrefatitle {Slow Slip Events in the Kanto and Tokai Regions of
  Central Japan Detected Using Global Navigation Satellite System Data During
  1994--2020} {Slow slip events in the kanto and tokai regions of central japan
  detected using global navigation satellite system data during
  1994--2020}.{\BBCQ}
\newblock
\APACjournalVolNumPages{Geochemistry, Geophysics,
  Geosystems}{22}{2}{e2020GC009329}.
\PrintBackRefs{\CurrentBib}

\bibitem [\protect \citeauthoryear {%
Nishimura%
, Matsuzawa%
\BCBL {}\ \BBA {} Obara%
}{%
Nishimura%
\ \protect \BOthers {.}}{%
{\protect \APACyear {2013}}%
}]{%
Nishimura2013}
\APACinsertmetastar {%
Nishimura2013}%
\begin{APACrefauthors}%
Nishimura, T.%
, Matsuzawa, T.%
\BCBL {}\ \BBA {} Obara, K.%
\end{APACrefauthors}%
\unskip\
\newblock
\APACrefYearMonthDay{2013}{}{}.
\newblock
{\BBOQ}\APACrefatitle {Detection of short-term slow slip events along the
  Nankai Trough, southwest Japan, using GNSS data} {Detection of short-term
  slow slip events along the nankai trough, southwest japan, using gnss
  data}.{\BBCQ}
\newblock
\APACjournalVolNumPages{Journal of Geophysical Research: Solid
  Earth}{118}{6}{3112--3125}.
\PrintBackRefs{\CurrentBib}

\bibitem [\protect \citeauthoryear {%
Obara%
}{%
Obara%
}{%
{\protect \APACyear {2020}}%
}]{%
Obara2020}
\APACinsertmetastar {%
Obara2020}%
\begin{APACrefauthors}%
Obara, K.%
\end{APACrefauthors}%
\unskip\
\newblock
\APACrefYearMonthDay{2020}{}{}.
\newblock
{\BBOQ}\APACrefatitle {Characteristic activities of slow earthquakes in Japan}
  {Characteristic activities of slow earthquakes in japan}.{\BBCQ}
\newblock
\APACjournalVolNumPages{Proceedings of the Japan Academy, Series
  B}{96}{7}{297--315}.
\PrintBackRefs{\CurrentBib}

\bibitem [\protect \citeauthoryear {%
Obara%
\ \BBA {} Kato%
}{%
Obara%
\ \BBA {} Kato%
}{%
{\protect \APACyear {2016}}%
}]{%
Obara2016}
\APACinsertmetastar {%
Obara2016}%
\begin{APACrefauthors}%
Obara, K.%
\BCBT {}\ \BBA {} Kato, A.%
\end{APACrefauthors}%
\unskip\
\newblock
\APACrefYearMonthDay{2016}{}{}.
\newblock
{\BBOQ}\APACrefatitle {Connecting slow earthquakes to huge earthquakes}
  {Connecting slow earthquakes to huge earthquakes}.{\BBCQ}
\newblock
\APACjournalVolNumPages{Science}{353}{6296}{253--257}.
\PrintBackRefs{\CurrentBib}

\bibitem [\protect \citeauthoryear {%
Ohtani%
, McGuire%
\BCBL {}\ \BBA {} Segall%
}{%
Ohtani%
\ \protect \BOthers {.}}{%
{\protect \APACyear {2010}}%
}]{%
Ohtani2010}
\APACinsertmetastar {%
Ohtani2010}%
\begin{APACrefauthors}%
Ohtani, R.%
, McGuire, J\BPBI J.%
\BCBL {}\ \BBA {} Segall, P.%
\end{APACrefauthors}%
\unskip\
\newblock
\APACrefYearMonthDay{2010}{}{}.
\newblock
{\BBOQ}\APACrefatitle {Network strain filter: A new tool for monitoring and
  detecting transient deformation signals in GPS arrays} {Network strain
  filter: A new tool for monitoring and detecting transient deformation signals
  in gps arrays}.{\BBCQ}
\newblock
\APACjournalVolNumPages{Journal of Geophysical Research: Solid
  Earth}{115}{B12}{}.
\PrintBackRefs{\CurrentBib}

\bibitem [\protect \citeauthoryear {%
Okada%
}{%
Okada%
}{%
{\protect \APACyear {1985}}%
}]{%
Okada1985}
\APACinsertmetastar {%
Okada1985}%
\begin{APACrefauthors}%
Okada, Y.%
\end{APACrefauthors}%
\unskip\
\newblock
\APACrefYearMonthDay{1985}{}{}.
\newblock
{\BBOQ}\APACrefatitle {Surface deformation due to shear and tensile faults in a
  half-space} {Surface deformation due to shear and tensile faults in a
  half-space}.{\BBCQ}
\newblock
\APACjournalVolNumPages{Bulletin of the Seismological Society of
  America}{75}{4}{1135--1154}.
\PrintBackRefs{\CurrentBib}

\bibitem [\protect \citeauthoryear {%
Okada%
, Nishimura%
, Tabei%
, Matsushima%
\BCBL {}\ \BBA {} Hirose%
}{%
Okada%
\ \protect \BOthers {.}}{%
{\protect \APACyear {2022}}%
}]{%
Okada2022}
\APACinsertmetastar {%
Okada2022}%
\begin{APACrefauthors}%
Okada, Y.%
, Nishimura, T.%
, Tabei, T.%
, Matsushima, T.%
\BCBL {}\ \BBA {} Hirose, H.%
\end{APACrefauthors}%
\unskip\
\newblock
\APACrefYearMonthDay{2022}{}{}.
\newblock
{\BBOQ}\APACrefatitle {Development of a detection method for short-term slow
  slip events using GNSS data and its application to the Nankai subduction
  zone} {Development of a detection method for short-term slow slip events
  using gnss data and its application to the nankai subduction zone}.{\BBCQ}
\newblock
\APACjournalVolNumPages{Earth, Planets and Space}{74}{1}{1--18}.
\PrintBackRefs{\CurrentBib}

\bibitem [\protect \citeauthoryear {%
Ozawa%
, Hatano%
\BCBL {}\ \BBA {} Kame%
}{%
Ozawa%
\ \protect \BOthers {.}}{%
{\protect \APACyear {2019}}%
}]{%
Ozawa2019}
\APACinsertmetastar {%
Ozawa2019}%
\begin{APACrefauthors}%
Ozawa, S\BPBI W.%
, Hatano, T.%
\BCBL {}\ \BBA {} Kame, N.%
\end{APACrefauthors}%
\unskip\
\newblock
\APACrefYearMonthDay{2019}{}{}.
\newblock
{\BBOQ}\APACrefatitle {Longer migration and spontaneous decay of aseismic slip
  pulse caused by fault roughness} {Longer migration and spontaneous decay of
  aseismic slip pulse caused by fault roughness}.{\BBCQ}
\newblock
\APACjournalVolNumPages{Geophysical Research Letters}{46}{2}{636--643}.
\PrintBackRefs{\CurrentBib}

\bibitem [\protect \citeauthoryear {%
Radiguet%
\ \protect \BOthers {.}}{%
Radiguet%
\ \protect \BOthers {.}}{%
{\protect \APACyear {2016}}%
}]{%
Radiguet2016}
\APACinsertmetastar {%
Radiguet2016}%
\begin{APACrefauthors}%
Radiguet, M.%
, Perfettini, H.%
, Cotte, N.%
, Gualandi, A.%
, Valette, B.%
, Kostoglodov, V.%
\BDBL {}Campillo, M.%
\end{APACrefauthors}%
\unskip\
\newblock
\APACrefYearMonthDay{2016}{}{}.
\newblock
{\BBOQ}\APACrefatitle {Triggering of the 2014 Mw7. 3 Papanoa earthquake by a
  slow slip event in Guerrero, Mexico} {Triggering of the 2014 mw7. 3 papanoa
  earthquake by a slow slip event in guerrero, mexico}.{\BBCQ}
\newblock
\APACjournalVolNumPages{Nature Geoscience}{9}{11}{829--833}.
\PrintBackRefs{\CurrentBib}

\bibitem [\protect \citeauthoryear {%
Riel%
, Simons%
, Agram%
\BCBL {}\ \BBA {} Zhan%
}{%
Riel%
\ \protect \BOthers {.}}{%
{\protect \APACyear {2014}}%
}]{%
Riel2014}
\APACinsertmetastar {%
Riel2014}%
\begin{APACrefauthors}%
Riel, B.%
, Simons, M.%
, Agram, P.%
\BCBL {}\ \BBA {} Zhan, Z.%
\end{APACrefauthors}%
\unskip\
\newblock
\APACrefYearMonthDay{2014}{}{}.
\newblock
{\BBOQ}\APACrefatitle {Detecting transient signals in geodetic time series
  using sparse estimation techniques} {Detecting transient signals in geodetic
  time series using sparse estimation techniques}.{\BBCQ}
\newblock
\APACjournalVolNumPages{Journal of Geophysical Research: Solid
  Earth}{119}{6}{5140--5160}.
\PrintBackRefs{\CurrentBib}

\bibitem [\protect \citeauthoryear {%
Rogers%
\ \BBA {} Dragert%
}{%
Rogers%
\ \BBA {} Dragert%
}{%
{\protect \APACyear {2003}}%
}]{%
Rogers2003}
\APACinsertmetastar {%
Rogers2003}%
\begin{APACrefauthors}%
Rogers, G.%
\BCBT {}\ \BBA {} Dragert, H.%
\end{APACrefauthors}%
\unskip\
\newblock
\APACrefYearMonthDay{2003}{}{}.
\newblock
{\BBOQ}\APACrefatitle {Episodic tremor and slip on the Cascadia subduction
  zone: The chatter of silent slip} {Episodic tremor and slip on the cascadia
  subduction zone: The chatter of silent slip}.{\BBCQ}
\newblock
\APACjournalVolNumPages{Science}{300}{5627}{1942--1943}.
\PrintBackRefs{\CurrentBib}

\bibitem [\protect \citeauthoryear {%
Rousset%
\ \protect \BOthers {.}}{%
Rousset%
\ \protect \BOthers {.}}{%
{\protect \APACyear {2017}}%
}]{%
Rousset2017}
\APACinsertmetastar {%
Rousset2017}%
\begin{APACrefauthors}%
Rousset, B.%
, Campillo, M.%
, Lasserre, C.%
, Frank, W\BPBI B.%
, Cotte, N.%
, Walpersdorf, A.%
\BDBL {}Kostoglodov, V.%
\end{APACrefauthors}%
\unskip\
\newblock
\APACrefYearMonthDay{2017}{}{}.
\newblock
{\BBOQ}\APACrefatitle {A geodetic matched filter search for slow slip with
  application to the Mexico subduction zone} {A geodetic matched filter search
  for slow slip with application to the mexico subduction zone}.{\BBCQ}
\newblock
\APACjournalVolNumPages{Journal of Geophysical Research: Solid
  Earth}{122}{12}{10--498}.
\PrintBackRefs{\CurrentBib}

\bibitem [\protect \citeauthoryear {%
Saffer%
\ \BBA {} Wallace%
}{%
Saffer%
\ \BBA {} Wallace%
}{%
{\protect \APACyear {2015}}%
}]{%
Saffer2015}
\APACinsertmetastar {%
Saffer2015}%
\begin{APACrefauthors}%
Saffer, D\BPBI M.%
\BCBT {}\ \BBA {} Wallace, L\BPBI M.%
\end{APACrefauthors}%
\unskip\
\newblock
\APACrefYearMonthDay{2015}{}{}.
\newblock
{\BBOQ}\APACrefatitle {The frictional, hydrologic, metamorphic and thermal
  habitat of shallow slow earthquakes} {The frictional, hydrologic, metamorphic
  and thermal habitat of shallow slow earthquakes}.{\BBCQ}
\newblock
\APACjournalVolNumPages{Nature Geoscience}{8}{8}{594--600}.
\PrintBackRefs{\CurrentBib}

\bibitem [\protect \citeauthoryear {%
Segall%
, B{\"u}rgmann%
\BCBL {}\ \BBA {} Matthews%
}{%
Segall%
\ \protect \BOthers {.}}{%
{\protect \APACyear {2000}}%
}]{%
Segall2000}
\APACinsertmetastar {%
Segall2000}%
\begin{APACrefauthors}%
Segall, P.%
, B{\"u}rgmann, R.%
\BCBL {}\ \BBA {} Matthews, M.%
\end{APACrefauthors}%
\unskip\
\newblock
\APACrefYearMonthDay{2000}{}{}.
\newblock
{\BBOQ}\APACrefatitle {Time-dependent triggered afterslip following the 1989
  Loma Prieta earthquake} {Time-dependent triggered afterslip following the
  1989 loma prieta earthquake}.{\BBCQ}
\newblock
\APACjournalVolNumPages{Journal of Geophysical Research: Solid
  Earth}{105}{B3}{5615--5634}.
\PrintBackRefs{\CurrentBib}

\bibitem [\protect \citeauthoryear {%
Segall%
, Desmarais%
, Shelly%
, Miklius%
\BCBL {}\ \BBA {} Cervelli%
}{%
Segall%
\ \protect \BOthers {.}}{%
{\protect \APACyear {2006}}%
}]{%
Segall2006}
\APACinsertmetastar {%
Segall2006}%
\begin{APACrefauthors}%
Segall, P.%
, Desmarais, E\BPBI K.%
, Shelly, D.%
, Miklius, A.%
\BCBL {}\ \BBA {} Cervelli, P.%
\end{APACrefauthors}%
\unskip\
\newblock
\APACrefYearMonthDay{2006}{}{}.
\newblock
{\BBOQ}\APACrefatitle {Earthquakes triggered by silent slip events on
  K{\=\i}lauea volcano, Hawaii} {Earthquakes triggered by silent slip events on
  k{\=\i}lauea volcano, hawaii}.{\BBCQ}
\newblock
\APACjournalVolNumPages{Nature}{442}{7098}{71--74}.
\PrintBackRefs{\CurrentBib}

\bibitem [\protect \citeauthoryear {%
Segall%
\ \BBA {} Matthews%
}{%
Segall%
\ \BBA {} Matthews%
}{%
{\protect \APACyear {1997}}%
}]{%
Segall1997b}
\APACinsertmetastar {%
Segall1997b}%
\begin{APACrefauthors}%
Segall, P.%
\BCBT {}\ \BBA {} Matthews, M.%
\end{APACrefauthors}%
\unskip\
\newblock
\APACrefYearMonthDay{1997}{}{}.
\newblock
{\BBOQ}\APACrefatitle {Time dependent inversion of geodetic data} {Time
  dependent inversion of geodetic data}.{\BBCQ}
\newblock
\APACjournalVolNumPages{Journal of Geophysical Research: Solid
  Earth}{102}{B10}{22391--22409}.
\PrintBackRefs{\CurrentBib}

\bibitem [\protect \citeauthoryear {%
Smith%
\ \BBA {} Gomberg%
}{%
Smith%
\ \BBA {} Gomberg%
}{%
{\protect \APACyear {2009}}%
}]{%
Smith2009}
\APACinsertmetastar {%
Smith2009}%
\begin{APACrefauthors}%
Smith, E\BPBI F.%
\BCBT {}\ \BBA {} Gomberg, J.%
\end{APACrefauthors}%
\unskip\
\newblock
\APACrefYearMonthDay{2009}{}{}.
\newblock
{\BBOQ}\APACrefatitle {A search in strainmeter data for slow slip associated
  with triggered and ambient tremor near Parkfield, California} {A search in
  strainmeter data for slow slip associated with triggered and ambient tremor
  near parkfield, california}.{\BBCQ}
\newblock
\APACjournalVolNumPages{Journal of Geophysical Research: Solid
  Earth}{114}{B12}{}.
\PrintBackRefs{\CurrentBib}

\bibitem [\protect \citeauthoryear {%
Takagi%
, Uchida%
\BCBL {}\ \BBA {} Obara%
}{%
Takagi%
\ \protect \BOthers {.}}{%
{\protect \APACyear {2019}}%
}]{%
Takagi2019}
\APACinsertmetastar {%
Takagi2019}%
\begin{APACrefauthors}%
Takagi, R.%
, Uchida, N.%
\BCBL {}\ \BBA {} Obara, K.%
\end{APACrefauthors}%
\unskip\
\newblock
\APACrefYearMonthDay{2019}{}{}.
\newblock
{\BBOQ}\APACrefatitle {Along-strike variation and migration of long-term slow
  slip events in the western Nankai subduction zone, Japan} {Along-strike
  variation and migration of long-term slow slip events in the western nankai
  subduction zone, japan}.{\BBCQ}
\newblock
\APACjournalVolNumPages{Journal of Geophysical Research: Solid
  Earth}{124}{4}{3853--3880}.
\PrintBackRefs{\CurrentBib}

\bibitem [\protect \citeauthoryear {%
Vergnolle%
\ \protect \BOthers {.}}{%
Vergnolle%
\ \protect \BOthers {.}}{%
{\protect \APACyear {2010}}%
}]{%
Vergnolle2010}
\APACinsertmetastar {%
Vergnolle2010}%
\begin{APACrefauthors}%
Vergnolle, M.%
, Walpersdorf, A.%
, Kostoglodov, V.%
, Tregoning, P.%
, Santiago, J.%
, Cotte, N.%
\BCBL {}\ \BBA {} Franco, S.%
\end{APACrefauthors}%
\unskip\
\newblock
\APACrefYearMonthDay{2010}{}{}.
\newblock
{\BBOQ}\APACrefatitle {Slow slip events in Mexico revised from the processing
  of 11 year GPS observations} {Slow slip events in mexico revised from the
  processing of 11 year gps observations}.{\BBCQ}
\newblock
\APACjournalVolNumPages{Journal of Geophysical Research: Solid
  Earth}{115}{B8}{}.
\PrintBackRefs{\CurrentBib}

\bibitem [\protect \citeauthoryear {%
Voss%
\ \protect \BOthers {.}}{%
Voss%
\ \protect \BOthers {.}}{%
{\protect \APACyear {2018}}%
}]{%
Voss2018}
\APACinsertmetastar {%
Voss2018}%
\begin{APACrefauthors}%
Voss, N.%
, Dixon, T\BPBI H.%
, Liu, Z.%
, Malservisi, R.%
, Protti, M.%
\BCBL {}\ \BBA {} Schwartz, S.%
\end{APACrefauthors}%
\unskip\
\newblock
\APACrefYearMonthDay{2018}{}{}.
\newblock
{\BBOQ}\APACrefatitle {Do slow slip events trigger large and great megathrust
  earthquakes?} {Do slow slip events trigger large and great megathrust
  earthquakes?}{\BBCQ}
\newblock
\APACjournalVolNumPages{Science advances}{4}{10}{eaat8472}.
\PrintBackRefs{\CurrentBib}

\bibitem [\protect \citeauthoryear {%
Wallace%
}{%
Wallace%
}{%
{\protect \APACyear {2020}}%
}]{%
Wallace2020}
\APACinsertmetastar {%
Wallace2020}%
\begin{APACrefauthors}%
Wallace, L\BPBI M.%
\end{APACrefauthors}%
\unskip\
\newblock
\APACrefYearMonthDay{2020}{}{}.
\newblock
{\BBOQ}\APACrefatitle {Slow slip events in New Zealand} {Slow slip events in
  new zealand}.{\BBCQ}
\newblock
\APACjournalVolNumPages{Annual Review of Earth and Planetary
  Sciences}{48}{}{175--203}.
\PrintBackRefs{\CurrentBib}

\bibitem [\protect \citeauthoryear {%
Wallace%
\ \protect \BOthers {.}}{%
Wallace%
\ \protect \BOthers {.}}{%
{\protect \APACyear {2018}}%
}]{%
Wallace2018}
\APACinsertmetastar {%
Wallace2018}%
\begin{APACrefauthors}%
Wallace, L\BPBI M.%
, Hreinsd{\'o}ttir, S.%
, Ellis, S.%
, Hamling, I.%
, D'Anastasio, E.%
\BCBL {}\ \BBA {} Denys, P.%
\end{APACrefauthors}%
\unskip\
\newblock
\APACrefYearMonthDay{2018}{}{}.
\newblock
{\BBOQ}\APACrefatitle {Triggered slow slip and afterslip on the southern
  Hikurangi subduction zone following the Kaik{\=o}ura earthquake} {Triggered
  slow slip and afterslip on the southern hikurangi subduction zone following
  the kaik{\=o}ura earthquake}.{\BBCQ}
\newblock
\APACjournalVolNumPages{Geophysical Research Letters}{45}{10}{4710--4718}.
\PrintBackRefs{\CurrentBib}

\bibitem [\protect \citeauthoryear {%
Wallace%
\ \protect \BOthers {.}}{%
Wallace%
\ \protect \BOthers {.}}{%
{\protect \APACyear {2017}}%
}]{%
Wallace2017}
\APACinsertmetastar {%
Wallace2017}%
\begin{APACrefauthors}%
Wallace, L\BPBI M.%
, Kaneko, Y.%
, Hreinsd{\'o}ttir, S.%
, Hamling, I.%
, Peng, Z.%
, Bartlow, N.%
\BDBL {}Fry, B.%
\end{APACrefauthors}%
\unskip\
\newblock
\APACrefYearMonthDay{2017}{}{}.
\newblock
{\BBOQ}\APACrefatitle {Large-scale dynamic triggering of shallow slow slip
  enhanced by overlying sedimentary wedge} {Large-scale dynamic triggering of
  shallow slow slip enhanced by overlying sedimentary wedge}.{\BBCQ}
\newblock
\APACjournalVolNumPages{Nature Geoscience}{10}{10}{765--770}.
\PrintBackRefs{\CurrentBib}

\bibitem [\protect \citeauthoryear {%
Walwer%
, Calais%
\BCBL {}\ \BBA {} Ghil%
}{%
Walwer%
\ \protect \BOthers {.}}{%
{\protect \APACyear {2016}}%
}]{%
Walwer2016}
\APACinsertmetastar {%
Walwer2016}%
\begin{APACrefauthors}%
Walwer, D.%
, Calais, E.%
\BCBL {}\ \BBA {} Ghil, M.%
\end{APACrefauthors}%
\unskip\
\newblock
\APACrefYearMonthDay{2016}{}{}.
\newblock
{\BBOQ}\APACrefatitle {Data-adaptive detection of transient deformation in
  geodetic networks} {Data-adaptive detection of transient deformation in
  geodetic networks}.{\BBCQ}
\newblock
\APACjournalVolNumPages{Journal of Geophysical Research: Solid
  Earth}{121}{3}{2129--2152}.
\PrintBackRefs{\CurrentBib}

\bibitem [\protect \citeauthoryear {%
Wang%
, Zhuang%
, Buckby%
, Obara%
\BCBL {}\ \BBA {} Tsuruoka%
}{%
Wang%
\ \protect \BOthers {.}}{%
{\protect \APACyear {2018}}%
}]{%
Wang2018}
\APACinsertmetastar {%
Wang2018}%
\begin{APACrefauthors}%
Wang, T.%
, Zhuang, J.%
, Buckby, J.%
, Obara, K.%
\BCBL {}\ \BBA {} Tsuruoka, H.%
\end{APACrefauthors}%
\unskip\
\newblock
\APACrefYearMonthDay{2018}{}{}.
\newblock
{\BBOQ}\APACrefatitle {Identifying the Recurrence Patterns of Nonvolcanic
  Tremors Using a 2-D Hidden Markov Model With Extra Zeros} {Identifying the
  recurrence patterns of nonvolcanic tremors using a 2-d hidden markov model
  with extra zeros}.{\BBCQ}
\newblock
\APACjournalVolNumPages{Journal of Geophysical Research: Solid
  Earth}{123}{8}{6802--6825}.
\PrintBackRefs{\CurrentBib}

\bibitem [\protect \citeauthoryear {%
Wdowinski%
, Bock%
, Zhang%
, Fang%
\BCBL {}\ \BBA {} Genrich%
}{%
Wdowinski%
\ \protect \BOthers {.}}{%
{\protect \APACyear {1997}}%
}]{%
Wdowinski1997}
\APACinsertmetastar {%
Wdowinski1997}%
\begin{APACrefauthors}%
Wdowinski, S.%
, Bock, Y.%
, Zhang, J.%
, Fang, P.%
\BCBL {}\ \BBA {} Genrich, J.%
\end{APACrefauthors}%
\unskip\
\newblock
\APACrefYearMonthDay{1997}{}{}.
\newblock
{\BBOQ}\APACrefatitle {Southern California permanent GPS geodetic array:
  Spatial filtering of daily positions for estimating coseismic and postseismic
  displacements induced by the 1992 Landers earthquake} {Southern california
  permanent gps geodetic array: Spatial filtering of daily positions for
  estimating coseismic and postseismic displacements induced by the 1992
  landers earthquake}.{\BBCQ}
\newblock
\APACjournalVolNumPages{Journal of Geophysical Research: Solid
  Earth}{102}{B8}{18057--18070}.
\PrintBackRefs{\CurrentBib}

\bibitem [\protect \citeauthoryear {%
C\BPBI A.~Williams%
\ \BBA {} Wallace%
}{%
C\BPBI A.~Williams%
\ \BBA {} Wallace%
}{%
{\protect \APACyear {2015}}%
}]{%
Williams2015}
\APACinsertmetastar {%
Williams2015}%
\begin{APACrefauthors}%
Williams, C\BPBI A.%
\BCBT {}\ \BBA {} Wallace, L\BPBI M.%
\end{APACrefauthors}%
\unskip\
\newblock
\APACrefYearMonthDay{2015}{}{}.
\newblock
{\BBOQ}\APACrefatitle {Effects of material property variations on slip
  estimates for subduction interface slow-slip events} {Effects of material
  property variations on slip estimates for subduction interface slow-slip
  events}.{\BBCQ}
\newblock
\APACjournalVolNumPages{Geophysical Research Letters}{42}{4}{1113--1121}.
\PrintBackRefs{\CurrentBib}

\bibitem [\protect \citeauthoryear {%
S\BPBI D.~Williams%
\ \protect \BOthers {.}}{%
S\BPBI D.~Williams%
\ \protect \BOthers {.}}{%
{\protect \APACyear {2004}}%
}]{%
Williams2004}
\APACinsertmetastar {%
Williams2004}%
\begin{APACrefauthors}%
Williams, S\BPBI D.%
, Bock, Y.%
, Fang, P.%
, Jamason, P.%
, Nikolaidis, R\BPBI M.%
, Prawirodirdjo, L.%
\BDBL {}Johnson, D\BPBI J.%
\end{APACrefauthors}%
\unskip\
\newblock
\APACrefYearMonthDay{2004}{}{}.
\newblock
{\BBOQ}\APACrefatitle {Error analysis of continuous GPS position time series}
  {Error analysis of continuous gps position time series}.{\BBCQ}
\newblock
\APACjournalVolNumPages{Journal of Geophysical Research: Solid
  Earth}{109}{B3}{}.
\PrintBackRefs{\CurrentBib}

\bibitem [\protect \citeauthoryear {%
Wilson%
}{%
Wilson%
}{%
{\protect \APACyear {2019}}%
}]{%
Wilson2019}
\APACinsertmetastar {%
Wilson2019}%
\begin{APACrefauthors}%
Wilson, D\BPBI J.%
\end{APACrefauthors}%
\unskip\
\newblock
\APACrefYearMonthDay{2019}{}{}.
\newblock
{\BBOQ}\APACrefatitle {The harmonic mean p-value for combining dependent tests}
  {The harmonic mean p-value for combining dependent tests}.{\BBCQ}
\newblock
\APACjournalVolNumPages{Proceedings of the National Academy of
  Sciences}{116}{4}{1195--1200}.
\PrintBackRefs{\CurrentBib}

\bibitem [\protect \citeauthoryear {%
Yano%
\ \BBA {} Kano%
}{%
Yano%
\ \BBA {} Kano%
}{%
{\protect \APACyear {2022}}%
}]{%
Yano2022}
\APACinsertmetastar {%
Yano2022}%
\begin{APACrefauthors}%
Yano, K.%
\BCBT {}\ \BBA {} Kano, M.%
\end{APACrefauthors}%
\unskip\
\newblock
\APACrefYearMonthDay{2022}{}{}.
\newblock
{\BBOQ}\APACrefatitle {l1 Trend Filtering-Based Detection of Short-Term Slow
  Slip Events: Application to a GNSS Array in Southwest Japan} {l1 trend
  filtering-based detection of short-term slow slip events: Application to a
  gnss array in southwest japan}.{\BBCQ}
\newblock
\APACjournalVolNumPages{Journal of Geophysical Research: Solid
  Earth}{127}{5}{e2021JB023258}.
\PrintBackRefs{\CurrentBib}

\bibitem [\protect \citeauthoryear {%
Yao%
}{%
Yao%
}{%
{\protect \APACyear {1988}}%
}]{%
Yao1988}
\APACinsertmetastar {%
Yao1988}%
\begin{APACrefauthors}%
Yao, Y\BHBI C.%
\end{APACrefauthors}%
\unskip\
\newblock
\APACrefYearMonthDay{1988}{}{}.
\newblock
{\BBOQ}\APACrefatitle {Estimating the number of change-points via
  Schwarz'criterion} {Estimating the number of change-points via
  schwarz'criterion}.{\BBCQ}
\newblock
\APACjournalVolNumPages{Statistics \& Probability Letters}{6}{3}{181--189}.
\PrintBackRefs{\CurrentBib}

\bibitem [\protect \citeauthoryear {%
Yarai%
\ \BBA {} Ozawa%
}{%
Yarai%
\ \BBA {} Ozawa%
}{%
{\protect \APACyear {2013}}%
}]{%
Yarai2013}
\APACinsertmetastar {%
Yarai2013}%
\begin{APACrefauthors}%
Yarai, H.%
\BCBT {}\ \BBA {} Ozawa, S.%
\end{APACrefauthors}%
\unskip\
\newblock
\APACrefYearMonthDay{2013}{}{}.
\newblock
{\BBOQ}\APACrefatitle {Quasi-periodic slow slip events in the afterslip area of
  the 1996 Hyuga-nada earthquakes, Japan} {Quasi-periodic slow slip events in
  the afterslip area of the 1996 hyuga-nada earthquakes, japan}.{\BBCQ}
\newblock
\APACjournalVolNumPages{Journal of Geophysical Research: Solid
  Earth}{118}{5}{2512--2527}.
\PrintBackRefs{\CurrentBib}

\bibitem [\protect \citeauthoryear {%
Zhang%
\ \protect \BOthers {.}}{%
Zhang%
\ \protect \BOthers {.}}{%
{\protect \APACyear {1997}}%
}]{%
Zhang1997}
\APACinsertmetastar {%
Zhang1997}%
\begin{APACrefauthors}%
Zhang, J.%
, Bock, Y.%
, Johnson, H.%
, Fang, P.%
, Williams, S.%
, Genrich, J.%
\BDBL {}Behr, J.%
\end{APACrefauthors}%
\unskip\
\newblock
\APACrefYearMonthDay{1997}{}{}.
\newblock
{\BBOQ}\APACrefatitle {Southern California Permanent GPS Geodetic Array: Error
  analysis of daily position estimates and site velocities} {Southern
  california permanent gps geodetic array: Error analysis of daily position
  estimates and site velocities}.{\BBCQ}
\newblock
\APACjournalVolNumPages{Journal of geophysical research: solid
  earth}{102}{B8}{18035--18055}.
\PrintBackRefs{\CurrentBib}

\end{thebibliography}
%




%
%
%
%
%

\clearpage

{\noindent \LARGE \bf Supporting Information for ``Automated Detection of Short-term Slow Slip Events in Southwest Japan"} \par
\vspace{0.5em}

\noindent Yiming Ma\textsuperscript{1,3}, Andreas Anastasiou\textsuperscript{2}, Fabien Montiel\textsuperscript{1} \par
\vspace{0.2em}
\noindent\noindent \textsuperscript{1}Department of Mathematics and Statistics, University of Otago\par
\noindent \textsuperscript{2}Department of Mathematics and Statistics, University of Cyprus\par
\noindent \textsuperscript{3}Department of Mathematical Sciences, Auckland University of Technology\par
\vspace{1em}

\newcounter{texts}
\renewcommand{\thetexts}{S\arabic{texts}} 

\newcounter{Seq}
\renewcommand{\theSeq}{S\arabic{Seq}}  

\newcounter{Sfig}
\renewcommand{\theSfig}{S\arabic{Sfig}} 

\newcounter{Stable}
\renewcommand{\theStable}{S\arabic{Stable}} 

\section*{Contents of the supporting file}
\indent
\hangindent=3.5em
\hangafter=1
Texts S1, S2, S3, S4, S5, S6, S7, S8, S9, S10, S11.
\vspace{0.5em}

\indent
\hangindent=3.5em
\hangafter=1
Figures S1, S2, S3, S4, S5, S6, S7, S8, S9, S10, S11, S12, S13, S14, S15, S16, S17, S18, S19, S20, S21, S22, S23, S24, S25, S26, S27, S28, S29, S30, S31, S32, S33, S34, S35, S36, S37, S38, S39, S40, S41.\\

\section*{Introduction}
\indent
\hangindent=3.5em
\hangafter=1
Text S1 introduces a deterministic model to simulate SSEs.

\indent
\hangindent=3.5em
\hangafter=1
Text S2 presents the results of applying different change-point detection methods for piecewise-linear signal to simulated SSE data.

\indent
\hangindent=3.5em
\hangafter=1
\minewc{Text S3 verifies the validity of the proposed scheme in Step 2 of SSAID to improve the successful percentage $R_{sd}$ for in-SNL data by using numerical tests.}

\indent
\hangindent=3.5em
\hangafter=1
\minewc{Text S4 investigates the impacts of different factors on the performance of SSAID.}

\indent
\hangindent=3.5em
\hangafter=1
\minewa{Text S5 presents the pseudocode for SSAID.}

\indent
\hangindent=3.5em
\hangafter=1
Text \minewc{S6} elaborates how to calculate $\Tilde{p}_j^k$, the confidence of occurrence of SSEs.

\indent
\hangindent=3.5em
\hangafter=1
Text \minewc{S7}  presents the histograms of the detected change-points for all the simulated noisy SSE data from all the different seeds and noise levels by various detection methods. 

\indent
\hangindent=3.5em
\hangafter=1
\minewa{Text S8 illustrates how to pair single change-points using our pre-processing procedure.}

\indent
\hangindent=3.5em
\hangafter=1
\minewa{Text S9 demonstrates the validity of pre-processing and hypothesis testing on identifying the probable SSEs.}

\indent
\hangindent=3.5em
\hangafter=1
\minewb{Text S10 presents observed GPS time series at different GPS stations and their neighbouring GPS stations.}

\indent
\hangindent=3.5em
\hangafter=1
Text \minewc{S11} presents the results of the other $14$ identified SSEs \minewb{with estimated moment magnitudes $M_{w}$.}

\vspace{0.5em}

\indent
\hangindent=3.5em
\hangafter=1
Figure S1 shows the configurations of the modified model and the slip rate history.

\indent
\hangindent=3.5em
\hangafter=1
\minewc{Figure S2 shows examples of observed and simulated SSE data.}

\indent
\hangindent=3.5em
\hangafter=1
\minewc{Figure S3 shows simulated pure SSE data with different numbers of SSEs in a one-year period}

\indent
\hangindent=3.5em
\hangafter=1
\minewc{Figure S4 shows the average of the errors between the number of estimated change-points and the number of true change-points by different change-point detection methods for piecewise-linear signals and the percentage of successful cumulative detection $R_{sd}$ versus the white noise level.}

\indent
\hangindent=3.5em
\hangafter=1
\minewc{Figure S5 shows the difference between the number of estimated change-points and the number of true change-points by different piecewise-linear CPD methods applied to the simulated SSE data with 10 change-points: (a) CPOP; (b) DPSEG.}

\indent
\hangindent=3.5em
\hangafter=1
\minewc{Figure S6 shows the difference between the number of estimated change-points and the number of true change-points by different piecewise-linear CPD methods applied to the simulated SSE data with 10 change-points: (a) ID; (b) NOT.}

\indent
\hangindent=3.5em
\hangafter=1
\minewc{Figure S7 shows the difference between the number of estimated change-points and the number of true change-points by different piecewise-linear CPD methods applied to the simulated SSE data with 10 change-points: (a) SEGMENTED-AIC; (b) SEGMENTED-BIC.}

\indent
\hangindent=3.5em
\hangafter=1
\minewc{Figure S8 shows the successful cumulative detection percentage $R_{sd}$ as a function of noise levels against different threshold $v$ values and the quartile distributions of RMSE as a function of noise levels for ID.}

\indent
\hangindent=3.5em
\hangafter=1
\minewc{Figure S9 shows suitable noise levels (SNLs) of different CPD methods for piecewise-linear signals as a function of the number of change-points.}

\indent
\hangindent=3.5em
\hangafter=1
\minewc{Figure S10 shows the number of estimated change-points for each group by taking the mode of the number of estimated change-points in its members and the percentage $R_{sd}$ of successful cumulative detections (or the percentage $R_1$ of detections as a function of noise levels.}

\indent
\hangindent=3.5em
\hangafter=1
\minewc{Figure S11 shows the percentage $R_2$ of the qualified members and the third quartile $\Omega_3$ for each group.}

\indent
\hangindent=3.5em
\hangafter=1
\minewc{Figure S12 shows the successful cumulative detection percentage $R_{sd}$ for each white noise level as a function of the number of realisations $Q$ and $L$ values.}

\indent
\hangindent=3.5em
\hangafter=1
\minewc{Figure S13 shows An example showing how the value of the Mallows' $C_p$ changes with the hyperparameter $\lambda$ for a noisy time series. The minimum value is highlighted by the red vertical line.}

\indent
\hangindent=3.5em
\hangafter=1
Figure \minewb{S14} shows histograms of detected change-points in all the synthetic data by SSAID and $l_1$ trend filtering.

\indent
\hangindent=3.5em
\hangafter=1
Figure \minewb{S15} shows histograms of detected change-points in all the synthetic data by the linear regression with $\Delta{AIC}$ using different thresholds.

\indent
\hangindent=3.5em
\hangafter=1
\minewa{Figure S16 shows a simulated noisy time series with change-points detected by SSAID, and its associated SIC values for different change-point candidates to pair the single change-point.}

\indent
\hangindent=3.5em
\hangafter=1
\minewa{Figure S17 shows the deployment of multiple GPS stations and the simulated noisy time series across these stations.}

\indent
\hangindent=3.5em
\hangafter=1
\minewa{Figure S18 shows the histogram of the calculated detection confidence $\hat{p}$ for each change-point pair in the numerical tests, validating the proposed pre-processing and hypothesis testing for identifying probable SSEs from SSAID detection results. }

\indent
\hangindent=3.5em
\hangafter=1
\minewa{Figure S19 shows observed time series at three GPS stations ($021052$; $950449$; $950447$) and their estimated change-points by SSAID plus single change-point pairing.}

\indent
\hangindent=3.5em
\hangafter=1
\minewb{Figure S20 shows locations of four reference GPS stations ($950436$; $041133$; $970828$; $021049$) and their neighboring GPS stations.}

\indent
\hangindent=3.5em
\hangafter=1
\minewb{Figure S21 shows observed time series at the reference GPS station $950436$ and its neighbouring GPS stations.}

\indent
\hangindent=3.5em
\hangafter=1
\minewb{Figure S22 shows observed time series at different neighbouring GPS stations of station $950436$.}

\indent
\hangindent=3.5em
\hangafter=1
\minewb{Figure S23 shows observed time series at the reference GPS station $041133$ and its neighbouring GPS stations.}

\indent
\hangindent=3.5em
\hangafter=1
\minewb{Figure S24 shows observed time series at different neighbouring GPS stations of station $041133$.}

\indent
\hangindent=3.5em
\hangafter=1
\minewb{Figure S25 shows observed time series at the reference GPS station $970828$ and its neighbouring GPS station $940086$.}

\indent
\hangindent=3.5em
\hangafter=1
\minewb{Figure S26 shows observed time series at the reference GPS station $021049$ and its neighbouring GPS stations.}

\indent
\hangindent=3.5em
\hangafter=1
\minewb{Figure S27 shows observed time series at different neighbouring GPS stations of station $021049$.}

\indent
\hangindent=3.5em
\hangafter=1
Figure \minewb{S28} shows the estimated fault model of an identified probable SSE candidate at the station $021049$\minewb{ with estimated moment magnitude $M_w$}.

\indent
\hangindent=3.5em
\hangafter=1
Figure \minewb{S29}  shows the estimated fault model of an identified probable SSE candidate at the station $950447$\minewb{ with estimated moment magnitude $M_w$}.

\indent
\hangindent=3.5em
\hangafter=1
Figure \minewb{S30} shows the estimated fault model of an identified probable SSE candidate at the station $041133$\minewb{ with estimated moment magnitude $M_w$}.

\indent
\hangindent=3.5em
\hangafter=1
Figure \minewb{S31} shows the estimated fault model of an identified probable SSE candidate at the station $031118$\minewb{ with estimated moment magnitude $M_w$}.

\indent
\hangindent=3.5em
\hangafter=1
Figure \minewb{S32} shows the the estimated fault model of an identified probable SSE candidate at the station $960681$\minewb{ with estimated moment magnitude $M_w$}.

\indent
\hangindent=3.5em
\hangafter=1
Figure \minewb{S33} shows the estimated fault model of an identified probable SSE candidate at the station $960681$\minewb{ with estimated moment magnitude $M_w$}.

\indent
\hangindent=3.5em
\hangafter=1
Figure \minewb{S34} shows the estimated fault model of an identified probable SSE candidate at the station $021050$\minewb{ with estimated moment magnitude $M_w$}.

\indent
\hangindent=3.5em
\hangafter=1
Figure \minewb{S35} shows the estimated fault model of an identified probable SSE candidate at the station $031124$\minewb{ with estimated moment magnitude $M_w$}.

\indent
\hangindent=3.5em
\hangafter=1
Figure \minewb{S36} shows the estimated fault model of an identified probable SSE candidate at the station $960680$\minewb{ with estimated moment magnitude $M_w$}.

\indent
\hangindent=3.5em
\hangafter=1
Figure \minewb{S37} shows the estimated fault model of an identified probable SSE candidate at the station $950436$\minewb{ with estimated moment magnitude $M_w$}.

\indent
\hangindent=3.5em
\hangafter=1
Figure \minewb{S38} shows the estimated fault model of an identified probable SSE candidate at the station $9041134$\minewb{ with estimated moment magnitude $M_w$}.

\indent
\hangindent=3.5em
\hangafter=1
Figure \minewb{S39} shows the estimated fault model of an identified probable SSE candidate at the station $021056$\minewb{ with estimated moment magnitude $M_w$}.

\indent
\hangindent=3.5em
\hangafter=1
Figure \minewb{S40} shows the estimated fault model of an identified probable SSE candidate at the station $950443$\minewb{ with estimated moment magnitude $M_w$}.

\indent
\hangindent=3.5em
\hangafter=1
Figure \minewb{S41} shows the estimated fault model of an identified probable SSE candidate at the station $021048$\minewb{ with estimated moment magnitude $M_w$}.

\refstepcounter{texts}
\section*{Text \thetexts. The deterministic fault model to simulate SSEs}
\begin{figure}[htbp]	
	\centering
	\includegraphics[width=\textwidth]{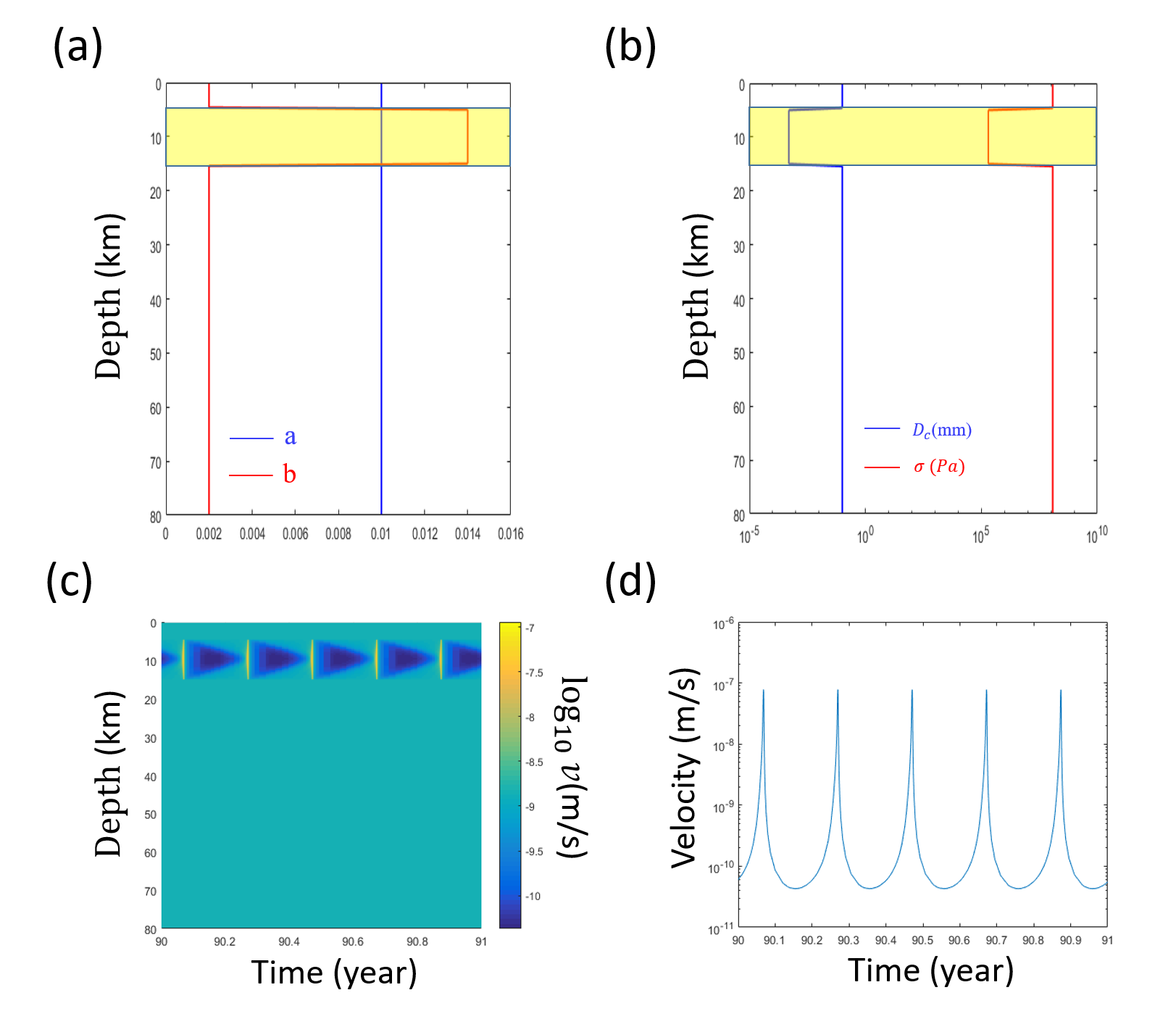}
	 \begin{minipage}{0.9\textwidth}
		\textbf{Figure \theSfig.} The spatial distribution of constitutive parameters along the depth direction in the modified reference model: (a) $a$ and $b$; (b) $D_c$ and $\sigma$. The light-yellow area refers to the VW transition zone. Slip rate history of (c) all the subfaults of the modified reference model over a 10-year period; and (d) the subfault at the middle point of the VW transition zone over a one-year period.
	\end{minipage}
	\label{fig_appen_modified_model_00}	
\end{figure}

\noindent In this section, we introduce a simplified deterministic fault model, which can spontaneously reproduce recurrent SSEs with a short duration of about a week, i.e. short-term SSEs. As shown in Fig. \ref{fig_appen_modified_model_00} (a), this model is composed of three sections, assuming that the velocity-weakening transition zone is embedded into two velocity-strengthening sections. The distributions of constitutive parameters (i.e. $\sigma$, $D_c$, $a$ and $b$) in the rate- and state-dependent friction (RSF) law are shown in Fig. \ref{fig_appen_modified_model_00} (a) and (b). The length along the strike direction and the width along the depth direction of the model are $500$ km and $80$ km, respectively. The slab angle is $\ang{15}$. We take $\Delta{w_0}=0.4/\sin{(\ang{15})}$ as its grid size and we then have $N=200$ subfaults along the dip direction. The slip rate history of the whole new modified fault model over a period of $10$ years (i.e. from the $90$-th to $100$-th year) is shown in Fig. \ref{fig_appen_modified_model_00} (c), and a one-year slip rate history of the subfault at the middle point of the VW transition is presented at Fig. \ref{fig_appen_modified_model_00} (d). We can see that the recurrent SSEs with short durations can spontaneously arise in the current model.

\refstepcounter{texts}
\section*{Text \thetexts. Piecewise-linear detection methods applied to SSE data}
\begin{figure}[htbp]	
	\centering
	\includegraphics[width=\textwidth]{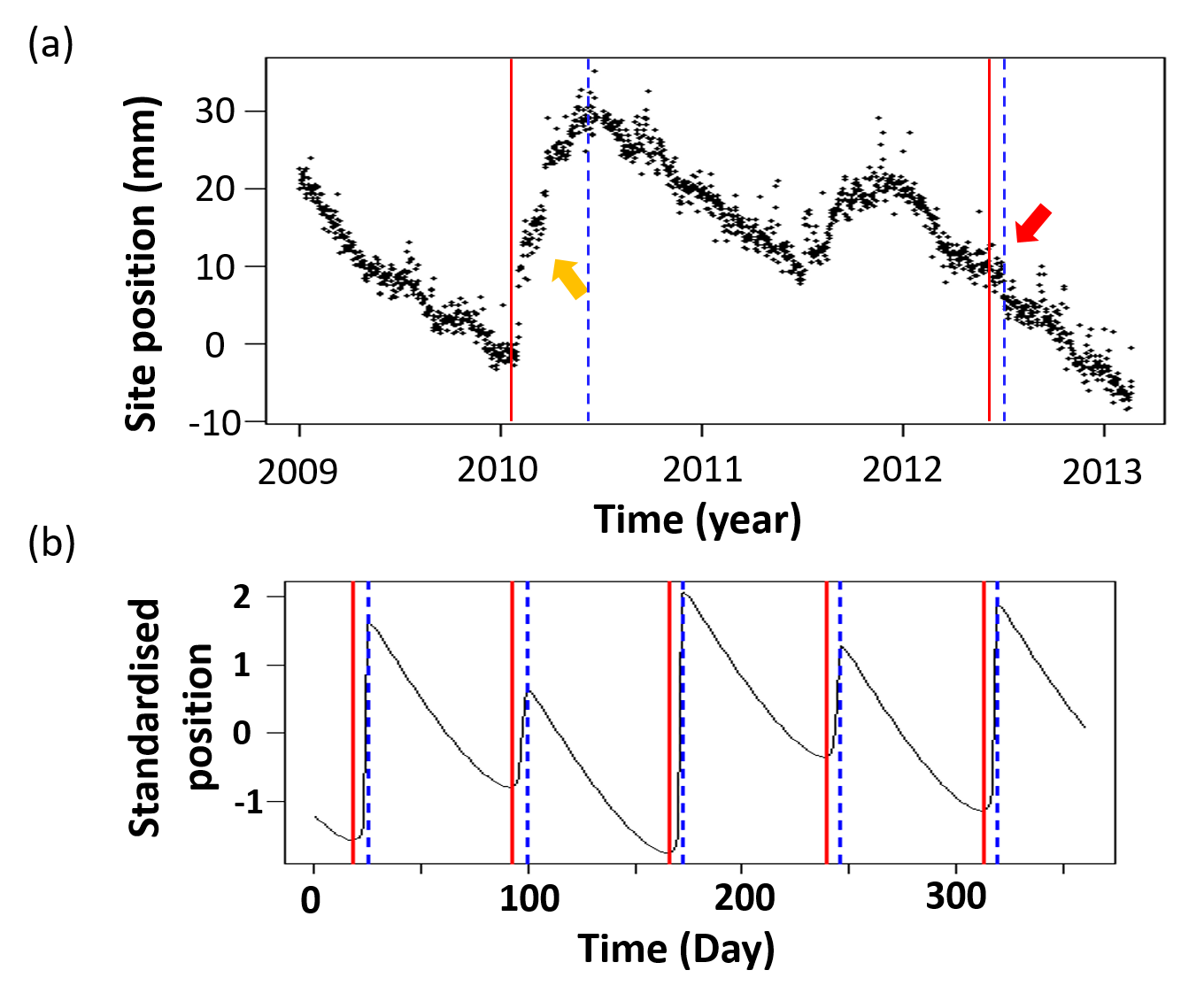}
	\begin{minipage}{0.9\textwidth}
		\textbf{Figure \theSfig.} (a) Observed SSE data recorded by the east component of a GPS station (MAHI), in Hikurangi subduction zone, New Zealand. Two arrows indicate two SSEs examples with different amplitude jumps (Wallace, 2020); (b) Simulated SSE data by a geophysical process-based model (see Text S1) with $5$ SSEs in a one-year period. There are $10$ change-points in these data indicated by both red and blue lines. Red vertical lines: the start times of SSEs; blue dotted vertical lines: the end times of SSEs.
	\end{minipage}
	\label{fig02_obs_sim_SSE}	
\end{figure}

Both the observational and simulated noisy SSE data, shown in Fig. \ref{fig02_obs_sim_SSE}, appear to have a piecewise-linear structure, even though the pure SSE signal (Fig. \ref{fig02_obs_sim_SSE} (b)) does not. Therefore, the existing change-point detection (CPD) methods for continuous piecewise-linear signals might be useful in detecting SSEs in GPS data. Therefore, in this section, we aim to quantify the impact of model misspecification of existing change-point methods designed for piecewise-linear signals, when being applied to detect change-points in simulated SSE data, in which the underlying signal has a continuous piecewise-non-linear structure, but the exact form is unknown. We evaluate the accuracy of these methods with respect to both the estimated number of change points and the estimated change-point locations. The new method of SSAID introduced in the main text uses the findings of this section to automatically detect the start and the end times of SSEs in GPS data.

\begin{figure}[htbp]	
	\centering
	\includegraphics[width=\textwidth]{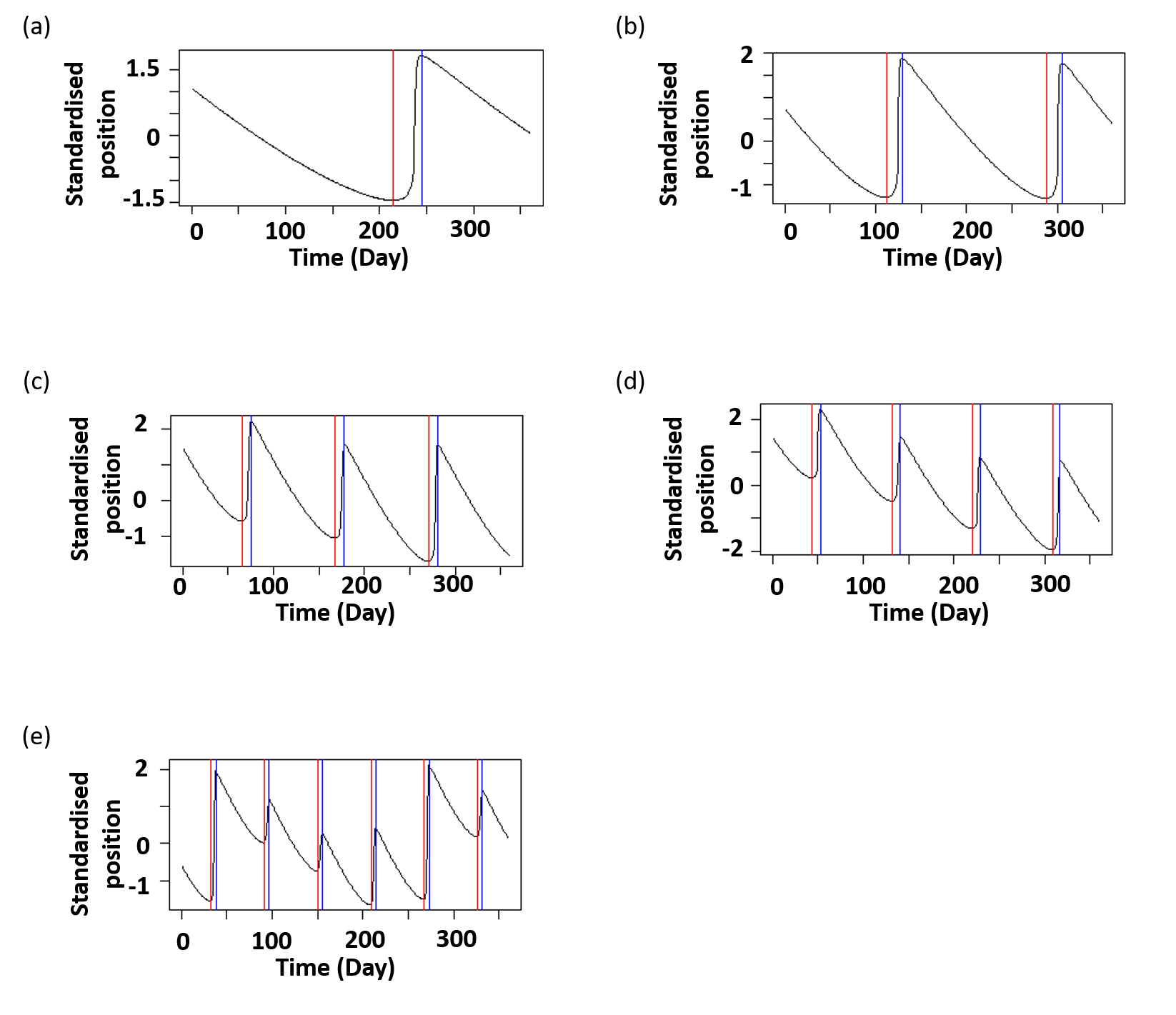}
	\begin{minipage}{0.9\textwidth}
		\textbf{Figure \theSfig.} Simulated pure SSE data with different numbers of SSEs in a one-year period: (a) $1$; (b) $2$; (c) $3$; (d) $4$; (e) $6$. Red vertical lines: the start time of an SSE; blue vertical line: the end time of an SSE.
	\end{minipage}
	\label{fig03_sim_SSE_diff_cp}	
\end{figure}

We first simulate GPS data that contain SSEs using the deterministic geophysical model introduced in Text S1. Fig. \ref{fig02_obs_sim_SSE} (b) shows a simulated signal with $5$ SSEs, in a one-year period. The recurring periodic pattern is consistent with direct SSE observations from GPS data. By changing the model parameters, we also simulated other SSE signals with different numbers of SSEs per year (see Fig. \ref{fig03_sim_SSE_diff_cp}). In these simulated signals, we define the start of an SSE when the slip velocity becomes $20\%$ higher than the plate velocity, and the end of an SSE when the slip velocity becomes lower than $1.2$ times the plate velocity. The plate velocity refers to the slip velocity of the subducting plate in our model. 

We construct the noisy simulated data $X_t$ using $X_t = f_t + C_{wn}\times\epsilon_t, \quad (t=1,\cdots,T)$ (the same formula as the main text, see Eq. (4) therein), where $T$ is the length of the data sequence, and $f_t$ is the simulated SSE signal (generated by the deterministic geophysical model in Text S1), standardised through the Z-score normalisation for ease of comparison. Note that the assumptions of the noise model are consistent with those of the five existing CPD methods for piecewise-linear signals discussed below.

We now test the performance of five well-established CPD algorithms for piecewise-linear signals on the simulated signal with $5$ SSEs (i.e. $10$ change-points; see Fig. \ref{fig02_obs_sim_SSE} (b)), which have a duration of approximately one week each: the Narrowest-Over-Threshold (NOT) algorithm (Baranowski et al., 2019), the Continuous-piecewise-linear Pruned Optimal Partitioning (CPOP) algorithm (Fearnhead et al., 2019), the Piecewise Linear Segmentation by Dynamic Programming algorithm (DPSEG) (\text{Machn\'e} and Stadler, 2020), the Fit Regression Models with Breaken-Line Relationships algorithm (i.e. known as SEGMENTED) (Muggeo, 2003, 2008) and the Isolate-Detect (ID) method (Anastasiou and Fryzlewicz, 2022). In the following tests, we choose the default values for all the parameters in the five methods, in which we test two different information criterion for SEGMENTED to estimate the number of change-points of the segmented relationship, i.e., Akaike's information criterion (AIC) and Bayesian information criterion (BIC).

\begin{figure}[htbp]	
	\centering
	\includegraphics[width=0.8\textwidth]{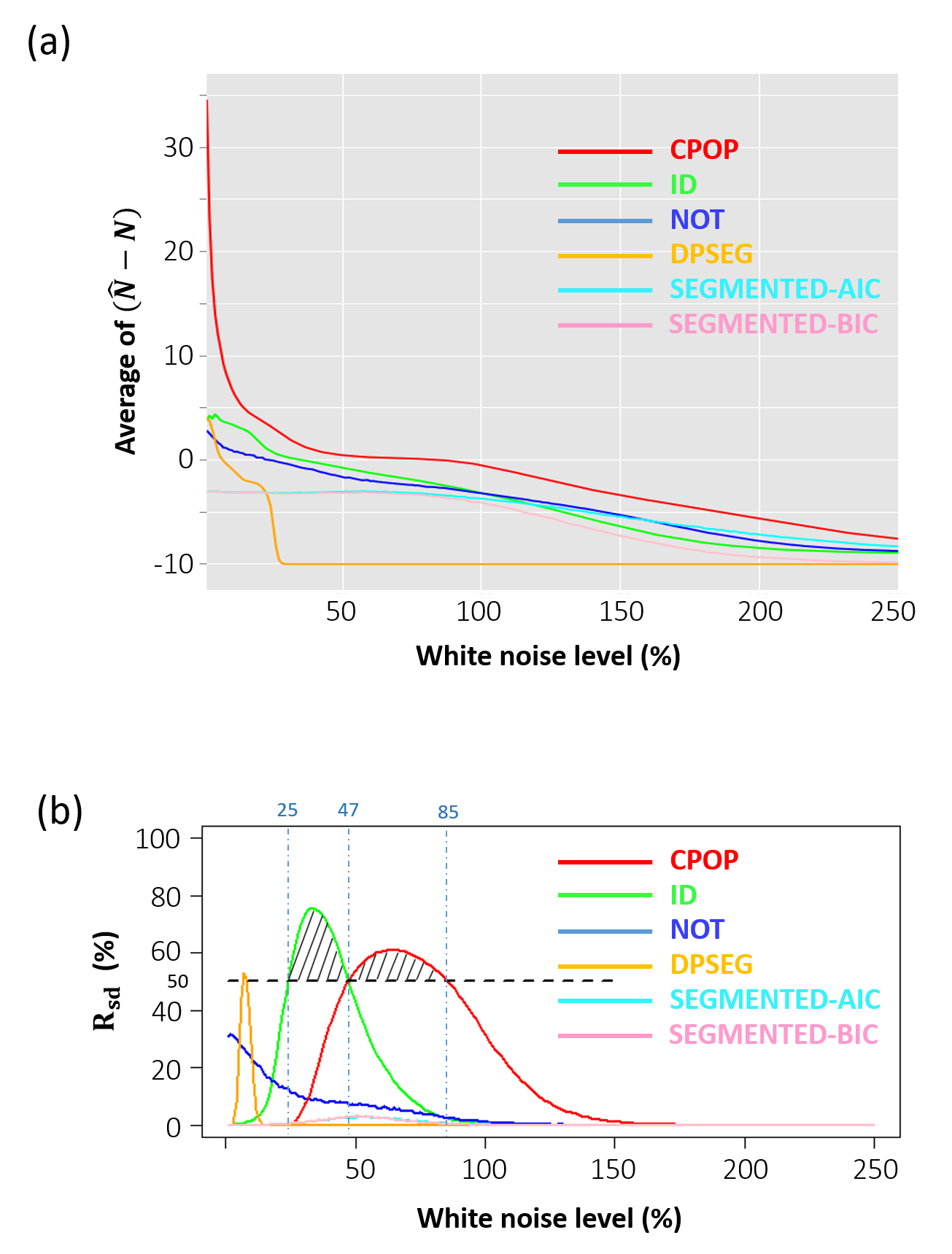}
	\begin{minipage}{0.9\textwidth}
		\textbf{Figure \theSfig.} (a) The average of the errors between the number of estimated change-points $\hat{N}$ and the number of true change-points $N=10$ for different detection methods. The number of estimated change-points, $\hat{N}$, by each CPD method is shown in Figs. $S5$-$S7$, respectively. (b) The percentage of successful cumulative detection, $R_{sd}$ (see Eq. (5) in the main text), out of $10,000$ realisations versus the white noise level $\sigma$. The shaded areas indicate $R_{sd}\geq{50\%}$.
	\end{minipage}
	\label{fig04_ave_err_N_diff_CPD}	
\end{figure}

\begin{figure}[htbp]	
	\centering
	\includegraphics[width=\textwidth]{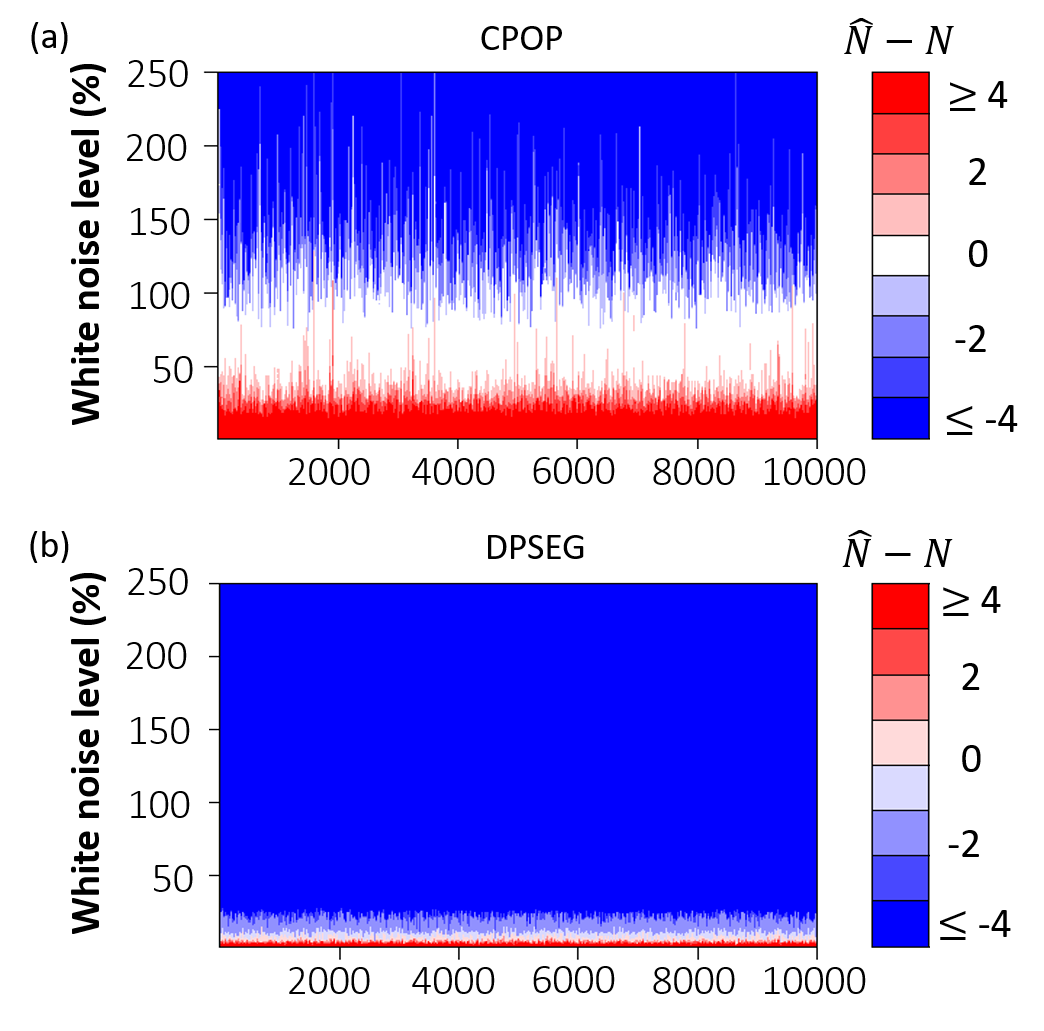}
	\begin{minipage}{0.9\textwidth}
		\textbf{Figure \theSfig.} The difference between the number of estimated change-points $\hat{N}$ and the number of true change-points $N$ for different piecewise-linear CPD methods applied to the simulated SSE data with $10$ change-points: (a) CPOP; (b) DPSEG. The underlying simulated SSE is as shown Fig. \ref{fig02_obs_sim_SSE} (b) in the main context.
	\end{minipage}
	\label{fig05_cpop_dpseg}	
\end{figure}

We carry out $10,000$ simulations for each noise level, with noise levels $\sigma=1\%$, $2\%$, $\cdots$, $250\%$. Fig. \ref{fig04_ave_err_N_diff_CPD} (a) shows that the average error of $\hat{N}-N$ increases with the noise level for all the tested methods and ultimately converges to $-10$ at very high noise levels. This is consistent with our expectations, as any piecewise CPD method would yield no change detection for signals with a high variance, while exhibiting a spurious increase in the number of detected change-points for signals unsuited to their model assumptions when the variance approaches zero. This is an ordinary outcome resulting from the continuous nature of the change-point detection due to model misspecification, the number of estimated change-points ranging from none to a high number. We also observe that a majority of the findings by the method of SEGMENTED consistently underestimated the actual number of change-points regardless of the selection criterion. This implies that not all of the tested methods are able to accurately ascertain the correct number of change-points, with the estimated number of change-points exhibiting an upper limit. Despite the success of the other four methods (i.e., CPOP, ID, NOT and DPSEG) in accurately determining the correct number within a certain range of noise levels with minimal errors, CPOP stands out by demonstrating an average error that reaches a plateau near zero for noise levels ranging from approximately $50\%$ to $95\%$. However, it is insufficient to solely rely on the average of $\hat{N}-N$ to quantify the performance of each CPD method. We conduct further analysis on the number of estimated change-points, $\hat{N}$, for each tested CPD method Figs. \ref{fig05_cpop_dpseg}-\ref{fig07_segmented}. We notice that the performance of NOT is not satisfactory for the data as NOT overestimates or underestimates the number of true change-points depending on the noise level $C_{wn}$, while DPSEG, CPOP and ID can consistently detect the number of true change-points for some noise levels in a certain range. While the correct estimation of the number of true change-points is important to acknowledge for quantifying the detection performance of these CPD methods, it is also crucial to highlight that the accuracy of the locations of the estimated change-points has not yet been taken into account, highlighting an additional important factor to consider.

We now compare the five methods with respect to the percentage of successful cumulative detections ($R_{sd}$; see its definition in Eq. (5) in the main text) for each noise level, which considers the location of estimated change-points. When defining a successful cumulative detection, we need to specify a threshold value $v$ for the calculated RMSE between the estimated change-point locations and the true locations. In general, the larger the threshold values $v$, the higher the successful percentage $R_{sd}$, but we need to control the threshold to be small enough so that the detection error is acceptable. We show the effect of different threshold values $v$ on $R_{sd}$ (see \ref{fig07_v} (a)). In our simulated noiseless signal, the duration of each SSE is about $7$ days, and the corresponding recurrence time is $74$ days. An RMSE of 3 days (i.e. $v$=3) is an acceptable error for detecting such SSEs (Holtkamp and Brudzinski, 2010; Nishimura et al., 2013). We also observe that the CPD accuracy both with respect to the estimated number and the estimated change-point locations decreases with the noise level (see \ref{fig07_v} (b)).

Fig. \ref{fig04_ave_err_N_diff_CPD} (b) shows that, for noise levels in a certain range, the DPSEG, ID and CPOP methods are able to detect all SSEs successfully in over $50\%$ of the simulations, which have the correct number of estimated change-points and high accuracy regarding their locations. We refer to this noise level range as the suitable noise level (SNL) range. For the simulated signal containing five SSEs, the SNL range for DPSEG, ID and CPOP are $7-8\%$, $25 - 47\%$ and $48 - 85\%$, respectively. There is no SNL range for the NOT and SEGMENTED methods. To investigate the influence of the number of change-points on the SNL range, we replace the underlying signal with other simulated signals shown (see Fig. \ref{fig03_sim_SSE_diff_cp}), which have different SSE duration. The results for these simulated signals are shown in Fig. \ref{fig08_SNLs}. We observe that the SNL range depends on the number of true change-points. The NOT method has a much broader SNL range than the CPOP and ID methods when the number of change-points is $8$ or less, while no SNL range exists for NOT when the number of change-points exceeds $8$. This suggests that NOT does not seem to be suitable for detecting SSEs with short durations. In contrast, SNL ranges for the CPOP method only occur when the number of change-points is $6$ or more, while the SNL ranges for the DPSEG and SEGMENTED methods are barely observed or moderately narrow if present. Interestingly, the ID method is the only one among the five methods that exhibits an SNL range for all the simulated SSEs. However, its extent varies depending on the signal. We also observe that the values for SNLs generally decrease as $N$ increases. When an SSE has a longer duration, the difference between the piecewise-non-linear shape of the SSE signal and a piecewise-linear signal becomes larger. It is sensible that more noise is needed in such cases to cover up the difference between the signals' actual structure and that of a continuous piecewise-linear signal.

We have observed that an SNL range can be found for accurate detection of change-points in complex piecewise signals such as SSEs using existing CPD methods for continuous piecewise-linear signals. Among the algorithms considered, the ID method seems to have the best behaviour overall when a range of different signals is considered with the number of change-points ranging from $2$ to $12$. However, the SNL is not consistent for different methods and signal types. Since the noise level and the underlying SSE signal in real-world GPS data are not known, the five CPD methods considered here cannot be directly employed to consistently detect SSEs. Note that despite the existence of numerous other change-point detection methods for piecewise-linear signals in the literature (Cho and Kirch, 2021; Yu, 2020), our focus is not to explore all of them. Among the five examined methods, CPOP, ID, and NOT have already been shown in an extensive simulation study carried out in Anastasiou and Fryzlewicz (2022), to perform very well in terms of accuracy regarding both the estimated number and locations of change-points in continuous piecewise-linear signals. Motivated by widening SNL ranges, we aim to develop a new algorithm based on the ID method to detect change-points in continuous signals with continuous piecewise structures while the exact form is unknown such as the form governing the behaviour of SSEs.

We conducted further analysis of the numerical test results and identified two quantities, $R_2$ and $\Omega_3$ (definied in Step 3 of SSAID in the appendix ), which can be used to identify in-SNL data (see Fig. \ref{fig10_R2_Omega3}).

\refstepcounter{texts}
\section*{Text \thetexts. Numerical tests to verify the validity of the proposed improvement scheme in Step 2 of SSAID}
\begin{figure}[htbp]	
	\centering
	\includegraphics[width=\textwidth]{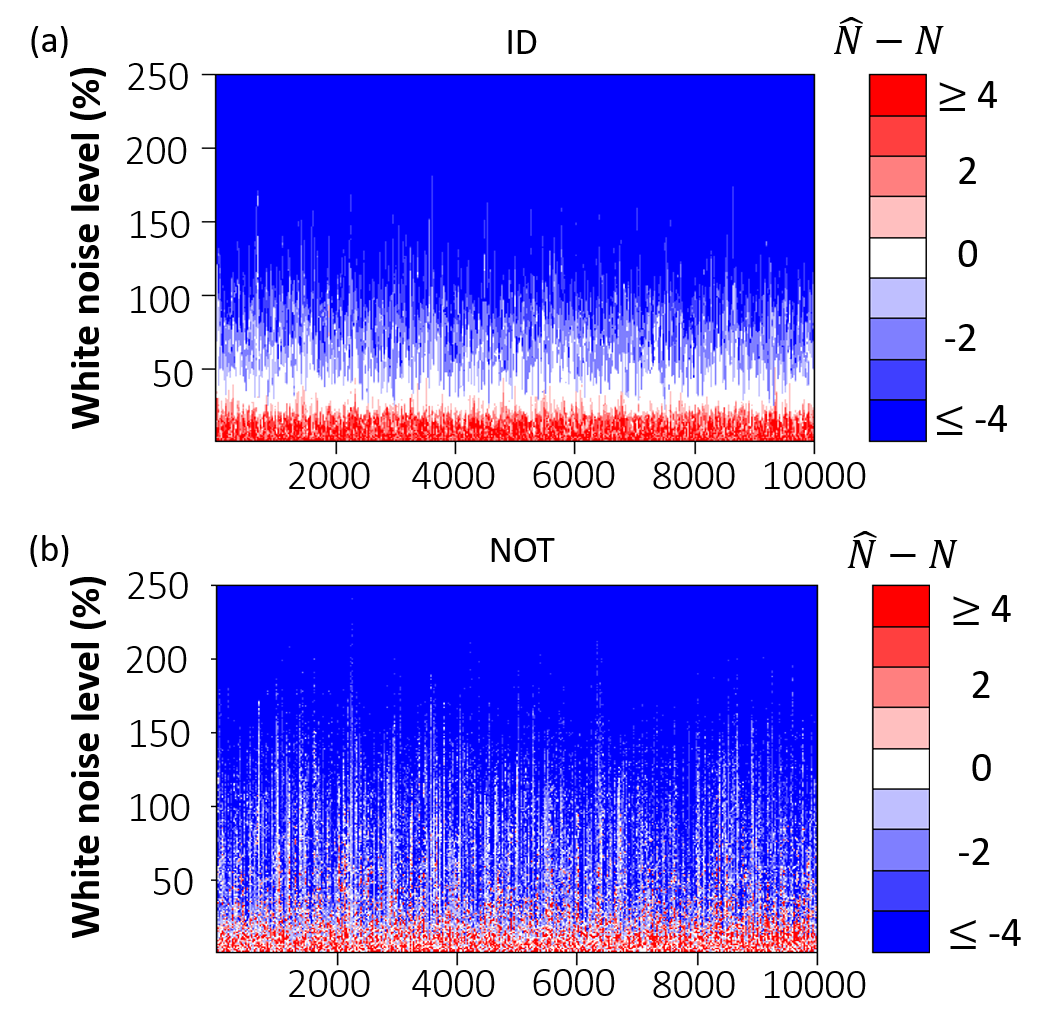}
	\begin{minipage}{0.9\textwidth}
		\textbf{Figure \theSfig.} The same as Fig. \ref{fig05_cpop_dpseg} but different CPD methods: (a) ID; (b) NOT.
	\end{minipage}
	\label{fig06_id_not}	
\end{figure}

\begin{figure}[htbp]	
	\centering
	\includegraphics[width=\textwidth]{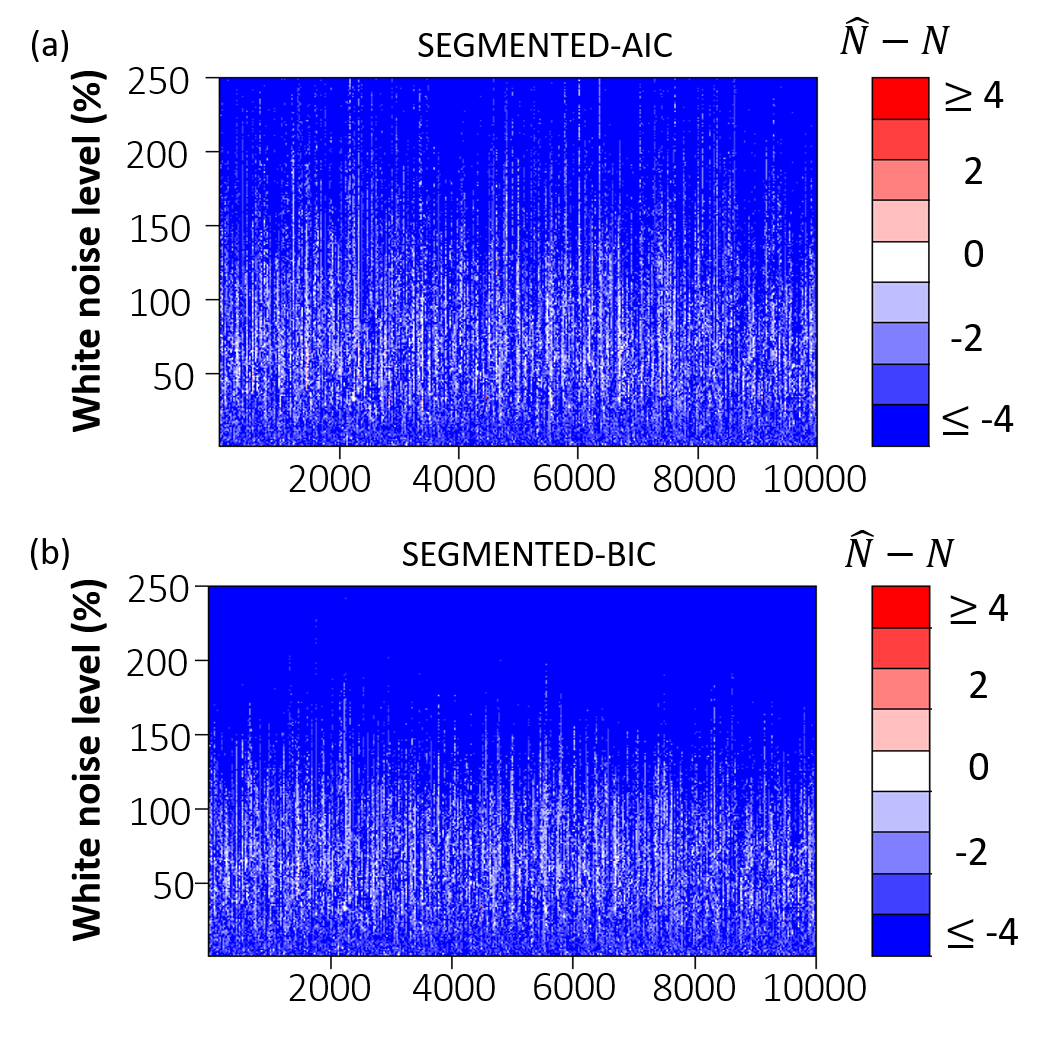}
	\begin{minipage}{0.9\textwidth}
		\textbf{Figure \theSfig.} The same as Fig. \ref{fig05_cpop_dpseg} but different CPD methods: (a) SEGMENTED-AIC; (b) SEGMENTED-BIC.
	\end{minipage}
	\label{fig07_segmented}	
\end{figure}

\begin{figure}[htbp]	
	\centering
	\includegraphics[width=0.9\textwidth]{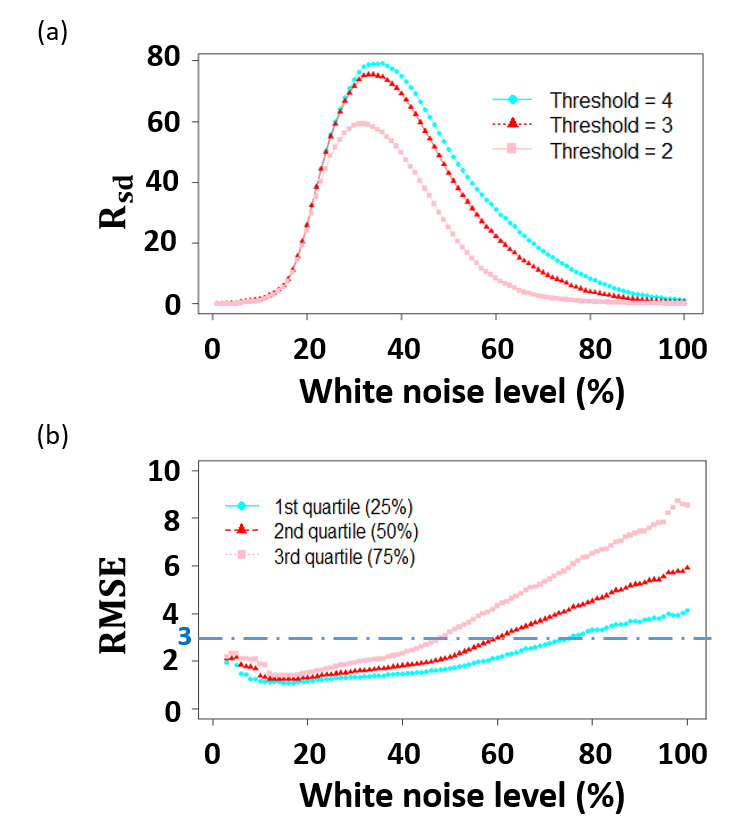}
	\begin{minipage}{0.9\textwidth}
		\textbf{Figure \theSfig.} (a) The successful cumulative detection percentage $R_{sd}$ as a function of noise levels against different threshold $v$ values; (b) The quartile distributions of RMSE as a function of noise levels for ID, the values of which are picked from these $10,000$ calculated RMSE values. The dotted horizontal blue line indicates the threshold that we use in our following tests.
	\end{minipage}
	\label{fig07_v}	
\end{figure}

\begin{figure}[htbp]	
	\centering
	\includegraphics[width=0.9\textwidth]{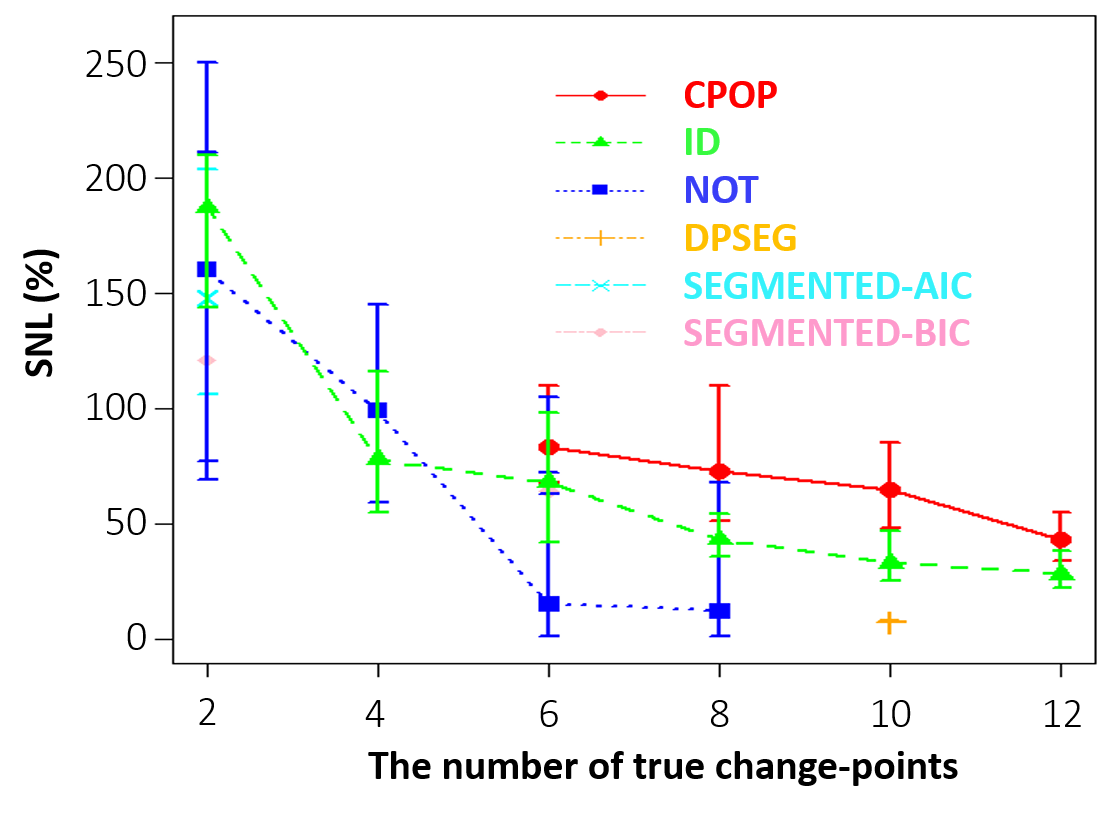}
	\begin{minipage}{0.9\textwidth}
		\textbf{Figure \theSfig.} Suitable noise levels (SNLs, see vertical intervals) of different methods as a function of the number of change-points. The indicators (i.e. red circle, green triangle, blue square, orange plus, cyan cross and pink diamond) refer to the noise level at which the percentage of successful cumulative detection is the highest. 
	\end{minipage}
	\label{fig08_SNLs}	
\end{figure}

We verify the validity of the scheme to improve the successful percentage $R_{sd}$, by conducting tests similar to those of Fig. \ref{fig06_id_not} (a) , but now using our proposed improvement scheme (see Step 2 in the Section 3 and the appendix of the main text) and exploring a slightly narrower range of noise levels. For each noisy data $X_t$ in Eq. (4) of the main text, we generate $Q$ realisations by simulating different noise models, i.e.

\begin{figure}[htbp]		
	\centering
	\includegraphics[width=0.9\textwidth]{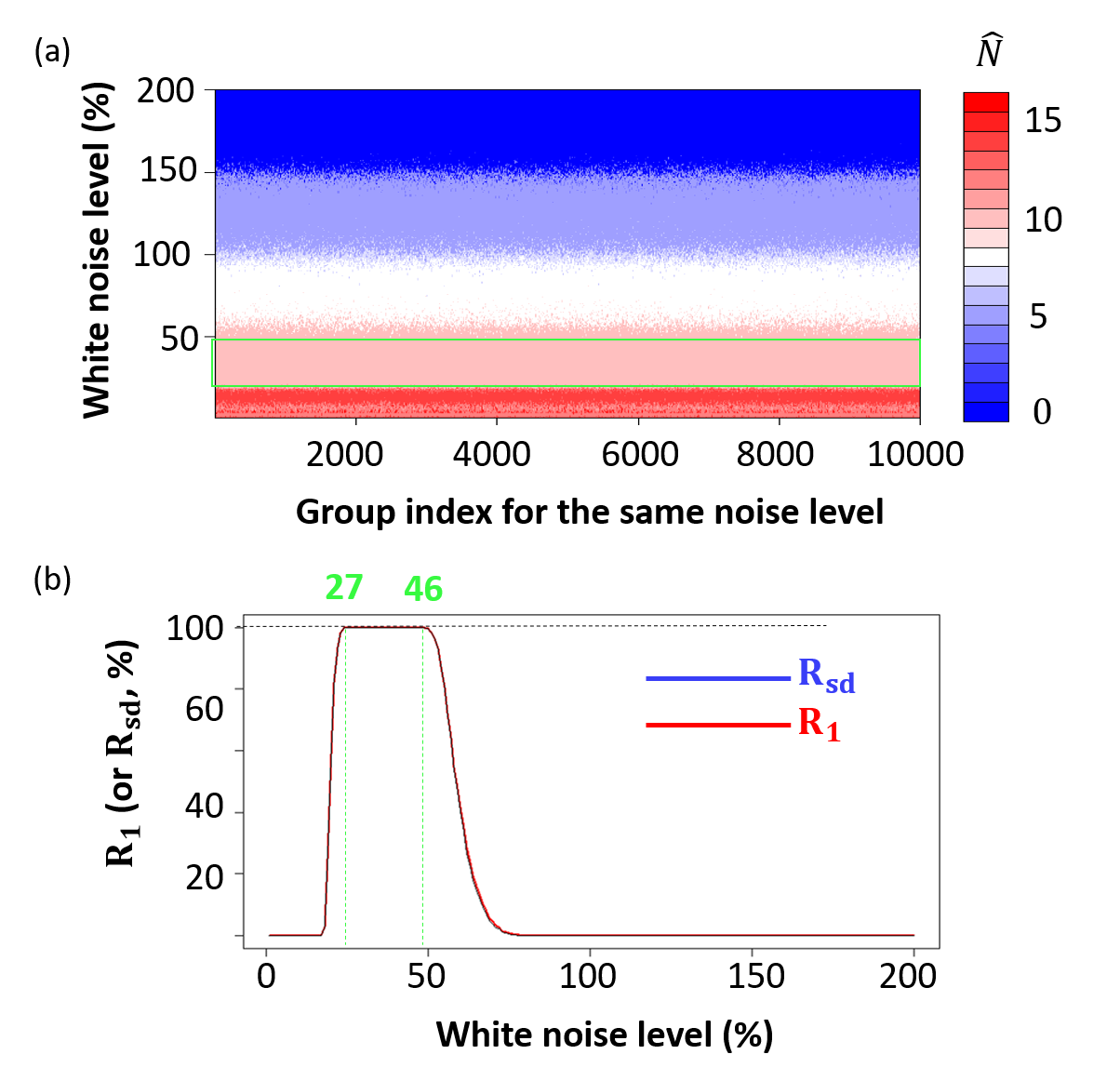}
	\begin{minipage}{0.9\textwidth}
		\textbf{Figure \theSfig.} (a) The number of estimated change-points for each group $\hat{N}$ by taking the mode of the number of estimated change-points in its members (i.e. $\hat{N}=Mo\{\hat{N}^1,\cdots,\hat{N}^Q\}$; see Text S3). The true number of change-points is $10$, indicated by the light pink. The SNL range is outlined by the green box. (b) The percentage $R_{sd}$ of successful cumulative detections (or the percentage $R_1$ of detections which satisfy $\hat{N}$$=N=10$; see the definitions for $R_{sd}$ and $R_1$ in Eq. $(5)$ in the main text) as a function of noise levels, among $10,000$ groups. The numbers $27$ and $46$ in green indicate the lower and upper limits of the noise level range where $R_{sd}=100\%$, respectively. This is consistent with the range of identified SNLs in Fig. \ref{fig04_ave_err_N_diff_CPD} (b).
	\end{minipage}
	\label{fig09_verify_majority}	
\end{figure}

\begin{figure}[htbp]		
	\centering
	\includegraphics[width=\textwidth]{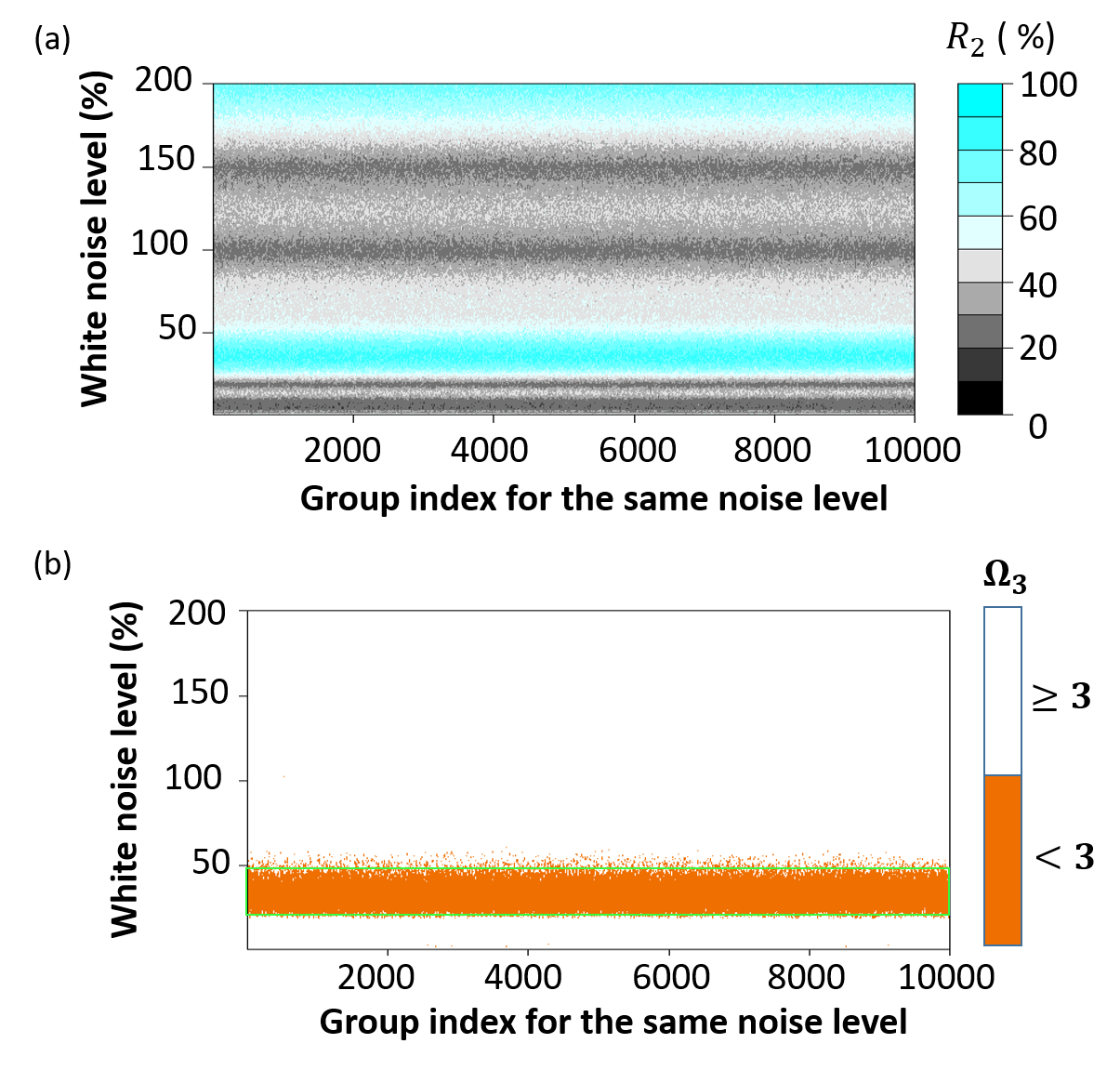}
	\begin{minipage}{0.9\textwidth}
		\textbf{Figure \theSfig.} (a) The percentage $R_2$ of the qualified members (see more details in Appendix A3 in the main text, i.e. $R_2=\kappa/Q$) for each group. (b) The third quartile $\Omega_3$ of RMSE for each group, in which we approximate the real change-points by the approach shown in Eq. (A4) in the appendix.
	\end{minipage}
	\label{fig10_R2_Omega3}	
\end{figure}

\refstepcounter{Seq} 
\begin{align}
	X_t^m = f_t + C_{wn}\times{\epsilon_t^m}, \quad (m=1,\cdots,Q;t=1,\cdots,T).
	\tag{\theSeq}
	\label{eqA0}
\end{align}

\noindent where $\epsilon_t^m$ is the $m$-th realisation of $\epsilon_t$, $C_{wn}$ changes from $1\%$ to $200\%$, with increments of $1\%$, and the underlying signal $f_t$ is kept unchanged. This set of $\{X_t^1, \cdots, X_t^Q\}$ is a group for $X_t$, the same as that mentioned before (i.e. $\textbf{\textit{G}}^{k,s}$; also see Step 3 in Fig. $3$ of the main text). Following our approach in Text S2, $10,000$ groups are randomly generated for each noise level $C_{wn}$. Each of these groups contains $Q$ realisations of the white noise. We estimate change-points for each group by using the mode as discussed above. More specifically, we apply the ID method to detect the change-points in each realisation $X_t^m$, and take the mode of $\hat{N}^m$ values as the number of estimated change-points for each group of $X_t$, denoted by $\hat{N}=Mo\{\hat{N}^1,\cdots,\hat{N}^Q\}$, where $\hat{N}^m$ is the number of estimated change-points for the $m$-th realisation $X_t^m$. We then take the approach shown in Eq. (A4) in the appendix to determine the locations of change-points for each group of $X_t$. We choose $Q=100$ here. Fig. \ref{fig09_verify_majority} (a) and (b) confirm that our majority voting rule can significantly increase the successful percentage $R_{sd}$ to $100\%$, when the input data has an SNL (see the level range outlined by the green numbers on the top of Fig. \ref{fig09_verify_majority} (b)). We observe that $R_{sd}$ overlaps with $R_1$ (see Eq. (5) in the main text), which means that the performance of the majority voting rule only depends on the noise level $C_{wn}$. 

\refstepcounter{texts}
\section*{Text \thetexts. Factors affecting the performance of SSAID}
Three key parameters may affect the performance of SSAID: the number of decomposed components $M$ in SSA, the number of realisations $Q$, and the highest level $L$ of added Gaussian noise (see Eqs. (A2)-(A3) in the appendix). The first parameter $M$ comes from the well-developed SSA algorithm, and has been widely discussed in the literature (e.g. Ghil et al., 2002; Walwer et al., 2016). Based on these studies, $M=100$ is a reasonable choice. We will mainly focus on the selection of $L$ and $Q$. We seek to guarantee the existence of in-SNL data among the $Z_t^{k,s,m}$ (see Step 3 in Fig. $3$ in the main text) while mitigating computing time. We can see from Eq. (A3) in the appendix that the percentage of successful detections $R_{sd}$ increases with $Q$ if $P_s$ is fixed.

We first generate a range of simulated SSE data $X_t$ in the form of Eq. (4) in the main text, in which $f_t$ is the pure SSE signal shown in Fig. \ref{fig02_obs_sim_SSE} (b), and the noise level $C_{wn}$ varies from $0$ to $100\%$, with increments of $1\%$. We create 100 data sequences of independent standard Gaussian random variables $\epsilon_t$ $( t=1,2,\ldots, T)$. In total, we have $100\times{101}$ noisy time series $X_t$ $(t=1,2,\ldots,T)$.

\begin{figure}[htbp]
	\centering
	\includegraphics[width=0.85\textwidth]{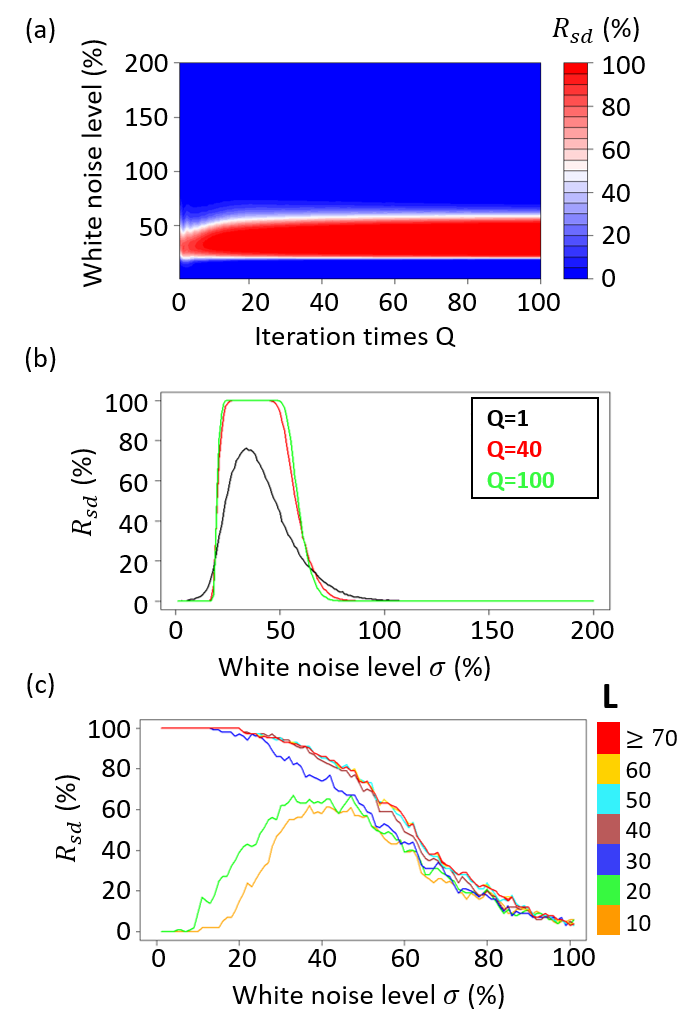}
	\begin{minipage}{0.9\textwidth}
		\textbf{Figure \theSfig.} (a) The successful percentage $R_{sd}$ for each white noise level as a function of the number of realisations $Q$. (b) and (c) $R_{sd}$ as a function of the noise level $C_{wn}$ for several $Q$ and $L$ values, respectively.
	\end{minipage}
	\label{fig11_factors}	
\end{figure}

We now apply SSAID to these noisy simulated SSE data using different $Q$ and $L$ values. Fig. \ref{fig11_factors} (a) and (b) show that the SNL range varies little for $Q\ge30$ and $R_{sd}$ reaches 100\% for in-SNL data. To ensure convergence, we take $Q=50$ in our subsequent tests. Fig. \ref{fig11_factors} (c) shows that the dependence of $R_{sd}$ on $C_{wn}$ converges rapidly with $L$ for $L\ge30$. We choose $L=80$ in our tests of current simulated SSE data, which is large enough to guarantee the existence of in-SNL data. 

\refstepcounter{texts}
\section*{Text \thetexts. The pseudocode for SSAID}
The SSAID pseudocode is divided into two tables (see Tables \ref{table_SSAID_part1} and \ref{table_SSAID_part2}).

\refstepcounter{Stable}
\begin{table}[htbp]
	\begin{minipage}{0.8\textwidth}
		\textbf{Table \theStable.} Pseudocode of SSAID (part 1).
	\end{minipage}
	\label{table_SSAID_part1}
	\renewcommand{\arraystretch}{0.7}
	\begin{tabular}{p{\dimexpr\textwidth-2\tabcolsep-\arrayrulewidth\relax}}
		\hline
		Results: Estimated change-points for the input data $X_t$ \\
		\hline
		\textbf{Step 1} (Decomposition process): Obtain denoised data with different noise levels $Y_t^{k}\quad(k=1,\cdots,M)$\\
		$R_t^j \leftarrow $ The $j$-th decomposed component of $X_t$ by SSA;\\
		$Y_t^k \leftarrow \sum_{j=1}^k{R_t^j}$;\\
		\hline
		\hline
		\textbf{Step 2} (Adding noise in the way of the $R_{sd}$ improvement scheme): Generate a range of new noisy data $Z_t^{k,s,m}\quad{(k=1,\cdots,M; s=1,\cdots,L; m=1,\cdots,Q)}$ to guarantee the existence of in-SNL data;\\		
		$Z_t^{k,s,m} \leftarrow Y_t^k + a_s\omega_t^m $;\\
		\hline
		\hline
		\textbf{Step 3} (The main part): Identifying in-SNL data among all the $Z_t^{k,s,m}$ group-by-group, through three condition: (1)$\hat{N}\neq{0}$; (2)$R_2\geq{50\%}$; (3) $\Omega_3\leq{v}$;\\ 
		\textbf{for} \textit{\underline{k=1:M}} \textbf{do}\\
		\quad\textbf{for} \textit{\underline{s=1:L}} \textbf{do}\\
		\qquad	Determine $\hat{N}^{k,s}$ for the group of $\textbf{\textit{G}}^{k,s}=\{Z_t^{k,s,1},\cdots,Z_t^{k,s,Q}\}$;\\
		\qquad\textbf{for} \textit{\underline{m=1:Q}} \textbf{do}\\
		\qquad\quad	$\hat{N}^{k,s,m} \leftarrow $ The number of estimated change-points for $Z_t^{k,s,m}$ by ID;\\
		\qquad\textbf{end}\\
		\qquad $\hat{N}^{k,s}\leftarrow Mo\{\hat{N}^{k,s,1},\cdots,\hat{N}^{k,s,Q}\}$;\\
		\qquad Condition 1:\\
		\qquad\textbf{if} \textit{\underline{$\hat{N}^{k,s}$=0}} \textbf{then}\\
		\qquad\quad All the members in $\textbf{\textit{G}}^{k,s}$ are marked as `\textbf{NOT in-SNL data}';\\
		\qquad\quad $\hat{N}^{k,s,m}(m=1,\cdots,Q)\leftarrow 0$;\\
		\qquad\textbf{else}\\
		\qquad\quad $\kappa \leftarrow$ The frequency of the mode $\hat{N}^{k,s}$ amongst $\{\hat{N}^{k,s,1},\cdots,\hat{N}^{k,s,Q}\}$;\\
		\qquad\quad $R_2^{k,s} \leftarrow \kappa/Q$;\\
		\qquad\quad Condition 2:\\ 
		\qquad\quad \textbf{if} \textit{\underline{$R_2^{k,s}<0.5$}} \textbf{then}\\
		\qquad\qquad All the members in the current group $\textbf{\textit{G}}^{k,s}$ are marked as `\textbf{NOT in-SNL data}';\\
		\qquad\qquad $\hat{N}^{k,s,m}(m=1,\cdots,Q)\leftarrow 0$;\\
		\qquad\quad\textbf{else}\\
		\qquad\qquad $D \leftarrow $ Generate a matrix shown in Eq. (6) which has a size of ($\kappa,\hat{H}^{k,s}$);\\
		\qquad\qquad $U^{k,s} \leftarrow$ The mode of each column in $D$;\\
		\qquad\qquad Condition 3:\\ 
		\qquad\qquad \shortstack[l]{$\Omega_3 \leftarrow $ The third quartile (75\%) of the RMSE for each group by assuming $U^{k,s}$ as the real\\ change-points for $Z_t^{k,s}$};\\
		\qquad\qquad\textbf{if} \textit{\underline{$\Omega_3>v$}} \textbf{then}\\
		\qquad\qquad\quad All the members in the current group $\textbf{\textit{G}}^{k,s}$ are marked as `\textbf{NOT in-SNL data}';\\ 
		\qquad\qquad\quad $\hat{N}^{k,s,m}(m=1,\cdots,Q)\leftarrow 0$;\\
		\qquad\qquad\textbf{end}\\
		\qquad\quad\textbf{end}\\
		\qquad\textbf{end}\\
		\quad\textbf{end}\\
		\quad \shortstack[l]{Output the number of change-points for $Y_t^{k}$: $\hat{N}_{tmp}\leftarrow Mo\{\hat{N}^{k,s,m}|s=1,\cdots,L; m=1,\cdots,Q;$\\ $\hat{N}^{k,s,m}\neq{0}\}$};\\
		\quad\textbf{if}\textit{\underline{$N_{tmp}$}} exists \textbf{then}\\
		\qquad $\hat{N}^{k}\leftarrow \hat{N}_{tmp}$\\
		\quad\textbf{else}\\
		\qquad{$\hat{N}^{k}\leftarrow 0$}\\
		\quad\textbf{end}\\
		\textbf{end}\\
		\hline
		\hline
		\textbf{Step 4} \shortstack[l]{Output the final estimated locations of change-points for $X_t$\\ (Continued on Table \ref{table_SSAID_part2}.)}
		
	\end{tabular}
\end{table}

\refstepcounter{Stable}
\begin{table}[htbp]
	\begin{minipage}{0.8\textwidth}
		\textbf{Table \theStable.} Pseudocode of SSAID (part 2-continued from Table \ref{table_SSAID_part1})
	\end{minipage}
	\label{table_SSAID_part2}
	\renewcommand{\arraystretch}{0.7}
	\begin{tabular}{p{\dimexpr\textwidth-2\tabcolsep-\arrayrulewidth\relax}}
		\hline
		\hline
		\textbf{Step 4} Output the final estimated locations of change-points for $X_t$.\\
		Determine the number of change-points  $\hat{N}_X$ for $X_t$:\\
		$\hat{N}_{tmp}\gets Mo\{\hat{N}^{k}|k=1\cdots,M; \hat{N}^{k}\neq{0}\}$;\\
		\textbf{if} \textit{{$N_{tmp}$}} does NOT exist \textbf{then}\\
		\quad $\hat{N}_{X}\gets 0$;\\
		\quad SSAID does not detect any change-points in $X_t$;\\
		\quad EXIT without output;\\
		\textbf{else}\\
		\quad $\hat{N}_{X}\gets \hat{N}_{tmp}$;\\
		\quad Collect all the groups which are not marked as `\textbf{NOT in-SNL data}';\\ 
		\quad \shortstack[l]{Pick up all the members with $\hat{N}^{k,s,m}=\hat{N}_X$ from in-SNL data, and then store their detected change-points\\ into a new matrix $D_f$, which is similar to the matrix $D$ in Eq. (A4), but with a different size;}\\
		\quad \shortstack[l]{Calculate the mode and the average of each column in $D_f$ to generate two candidate sets of final \\change-points in $X_t$;}\\
		\quad $U\gets$ the set of change-points with a smaller SIC value, which is the final output of SSAID.\\
		\textbf{end}\\
		\hline
		\hline			
	\end{tabular}
\end{table}

\refstepcounter{texts}
\section*{Text \thetexts. Details about hypothesis test}
In this section, we elaborate more details about how to calculate $\Tilde{p}_j^k$. We calculate the displacement rate at the $k$-th starting change-point, i.e. $\bar{v}_k^j$ in Eq. (8) of the main text, by taking the slope of the fitted linear model to the noisy data between the $k$-th starting and ending change-points. It takes three steps to estimate $\bar{v}_{0}^j$: (1) we consider the noisy SSE data as a piecewise-linear signal with $2\bar{N}_s^j$ knots; (2) we calculate the slope of each segment in the modelled piecewise-linear signal; and (3) we select the slopes which have the same sign as the secular linear process, and take their average as the estimated secular displacement rate. 

It is possible that the expected $B_j^k$ values that reject the null hypothesis depend on the sign of the secular displacement rate. If the secular displacement rate has a positive sign, at the start time of an SSE, it changes to a negative sign (see Fig. $7$(a) of the main text). This indicates that negative $B_j^k$ values are expected at the start times of SSEs. If, on the other hand, the secular displacement rate has a negative sign, positive $B_j^k$ values are expected at the start times of SSEs. Therefore, we introduce the term of the sign function in Eq. $(8)$ to make both cases have the same expected $\bar{B}_j^k$ values (i.e. negative). Under the null hypothesis, $\bar{B}_j^k$ follows the standard Gaussian distribution (Yano \& Kano, 2022). Therefore, we estimate the probability that SSEs do not occur at the $k$-th starting point of the $j$-th station by Eq. $(9)$ shown in the main text.

To reduce Type I errors, we combine $p$-values of stations neighbouring the $j$-th station into a new single $p$-value through the harmonic mean $p$-value method (Wilson, 2019; Yano \& Kano, 2022), that is

\refstepcounter{Seq} 
\begin{equation}
	\hat{p}_j^k=\frac{1}{\sum_{g=1}^{\hat{N}_a^j}{(1/{\mathring{p}_{j,g}^k}})},
	\tag{\theSeq}
	\label{eq11}
\end{equation}
\noindent where $\hat{N}_a^j$ is the number of stations neighbouring the $j$-th station, $g$ is the neighbouring station index, and $\mathring{p}_{j,g}^k$ refers to the $p$-value calculated via Eq. $(9)$ of the main text for the $g$-th station neighbouring the $j$-th station, which quantifies the probability that an SSE does not occur at the $k$-th starting change-point of the $j$-th station. Here, we refer to stations within a designated distance, denoted by $D_{\eta}$, from the $j$-th station as neighbouring stations of the $j$-th station. When selecting $D_{\eta}$, we need to guarantee that the time differences of the same detected SSE between the stations (i.e. the $j$-th station and its neighbouring stations) should be negligible. We have already indicated that SSAID can bear an error of at most $3$ days in Section $4$ of the main text, which means that the time difference should be at most $3$ days. Since the average distance between stations in GEONET is about $20$ km (Takagi et al., 2019) and the typical along-strike propagation velocity of ETS in our research area is $10-20$ km/day (Dragert et al., 2001; Obara, 2002; Obara, 2020), we take $D_{\eta}=30$ km in our following hypothesis tests, i.e. the same as that taken by Yano \& Kano (2022).

Calculating $\mathring{p}_{j,g}^k$ in Eq. \eqref{eq11} requires three steps: (1) we estimate the secular displacement rate $\bar{v}_0^{j,g}$ at the $g$-th neighbouring station of station $j$, by using the same approach as before; (2) we also take the slope of the fitted linear model to the noisy data at the $g$-th neighbouring station of the $j$-th station to estimate its displacement rate $\mathring{v}_{k}^{j,g}$ at the $k$-th starting change-point of the $j$-th station; (3) we utilize Eqs. $(8)$ and $(9)$ of the main text to quantify $\mathring{p}_{j,g}^k$. Note that in the second step, the period used to calculate $\mathring{v}_{k}^{j,g}$ is between the $k$-th starting and the $k$-th ending change-point of the $j$-th station, rather than its own change-points. This is because of the assumption that an SSE should be recorded at the same time by both the $j$-th station and its neighbouring stations (see the explanations for choosing $D_{\eta}$ in the last paragraph). Since the $j$-th station and its neighbouring stations are distributed in a nearby region, they should have similar $p$-values. If the $k$-th starting change-point at the $j$-th station is associated with an SSE, it is expected to have a small $\hat{p}_j^k$, so that we have high confidence to reject the null hypothesis. It is clear from Eq. \eqref{eq11} that $\mathring{p}_{j,g}^k$ cannot be zero. If there exists a $\mathring{p}_{j,g}^k=0$, we manually set the associated $\hat{p}_j^k$ as $0$ as we have a high probability to reject the null hypothesis. 

Finally, we can obtain the confidence of the occurrence of SSEs $\Tilde{p}_j^k$ via Eq. $(10)$ in the main text.Note that when only one pair of change-points are identified (i.e. $\hat{N}_s^j=1$), we cannot calculate $\bar{B}_j^k$ via Eq. $(8)$ in main text and conduct the following hypothesis test instead. We assume that $\Tilde{p}_j^k=0.6$ if the sign of the displacement rate at the starting change-point is opposite to that of the secular displacement rate, otherwise $\Tilde{p}_j^k=0$. The selection of these two specific values (i.e. $0.6$ and $0$) is simply set for ease of discussion, based on the SSE categories defined in section $5.1.3$.

\refstepcounter{texts}
\section*{Text \thetexts. The histograms of detected change-points by different methods}
In this section, we present the histograms of the detected change-points for all the simulated noisy SSE data from all the different seeds and noise levels by various detection methods (see \textsection{4} of the main text), including SSAID, l1 trend filtering, and the linear regression with $\Delta${AIC}, utilizing different thresholds in Figs. \ref{fig04_SSAID_L1ft} and \ref{fig04_linear_AIC_diff_thres}. We can see that most SSAID detections tend to converge to accurate locations with minimal errors, demonstrating its superior detection performance. In contrast, l1 trend filtering, despite exhibiting similar behaviors, suffers from a higher number of false detections and larger errors. The results of linear regression with $\Delta${AIC} also highlight the significant influence of the chosen threshold on the detection success. When the threshold is set at a low value, the majority of detections miss the true locations, although some successful cummulative detections do occur. Conversely, raising the threshold increases the percentage of detections that correctly identify the true change-points but also introduces a higher number of false detections.

\refstepcounter{texts}
\section*{Text \thetexts. Illustration on how to pair single change-points using our pre-processing procedure}
In this section, we illustrate the pairing of single change-points in simulated GPS time series using our proposed pre-processing procedure via the Schwarz Information Criterion (SIC). We generate a noisy time series, similar to those in Section $4.3$, comprising both white and colored noise at a $20\%$ noise level. To simplify the presentation, we reduce the time series length to $400$ days, containing $10$ true change-points. The pairing procedure remains consistent regardless of the time series length. We apply SSAID to detect change-points, resulting in the detections of $11$ change-points, with $10$ correctly detected and one false change-point. The false change-point is regarded as a single change-point. 

We now follow the pre-processing procedure from Section $5.1.1$ to pair the single change-point by creating an additional change-point. First, we calculate $k_b$ and $k_a$ to determine if the change-point is a starting or ending change-point. For our simulated time series, a starting change-point is indicated by $k_b<0$ and $k_a>0$, while an ending change-point is indicated by $k_b>0$ and $k_a<0$. For a starting change-point, we search for the paired change-point within $3-7$ days after the starting point. For an ending change-point, we search within $3-7$ days before the ending point. If the type of change-point is unclear, we search within $3-7$ days both before and after the single change-point. The search range of $3-7$ days deviated from the single change-point is based on prior information from past studies regarding the expected duration of short-term SSEs in the research area.

As shown in Fig. \ref{fig_appen_showing_how_to_pair_single_cp} (a), the current single change-point has $k_b<0$ and $k_a<0$, making it unclear whether it is a starting or ending point. Thus, the paired change-point will be within $3-7$ days both before and after the single change-point. Specifically, with the detected single change-point at day $\bar{x}_{cp}=312$, the search range includes days $305$, $306$, $307$, $308$, $309$, $315$, $316$, $317$, $318$, $319$. For each candidate change-point, we fit a piecewise-linear signal to the noisy time series and calculate its associated SIC value. We then have $10$ fitted signals, each with an SIC value. The SIC value helps evaluate how well the model fits the data, with smaller values indicating better fits (Anastasiou \& Fryzlewic, 2021). The candidate with the minimum SIC value within the search range is chosen as the best paired change-point (see Fig.\ref{fig_appen_showing_how_to_pair_single_cp} (c)). In this case, day $315$ is selected to pair the single change-point. Once paired, the earlier change-point is the starting point for a potential SSE, and the later one is the ending point. Note that this pre-processing procedure may not always correctly identify the missing paired change-points, but their association with true SSEs will be verified by subsequent hypothesis testing and fault estimation. The primary purpose of this pre-processing is to satisfy the prerequisite of having paired change-points (a starting and an ending point) for each potential SSE necessary for the subsequent analyses.

\refstepcounter{texts}
\section*{Text \thetexts. Numerical tests for pre-processing and hypothesis testing}
In this section, we conduct extensive numerical tests to verify the validity of our proposed processing chain for identifying the probable SSEs from SSAID detection results. From the numerical tests in Section 4.3 of the main text, which investigate the influence of color noise on SSAID's detection performance, we find that SSAID achieves nearly $100\%$ successful cummulative detection when the white noise level is below $25\%$ and the color noise level is below $20\%$.
	
In our current tests, we simulate $100$ noisy time series. For these sequences, the white noise level is fixed at $20\%$, while the color noise is varied $100$ times, each time with a fixed noise level of $20\%$. We employ SSAID to detect their change-points. Out of the $100$ detection scenarios, we excluded $6$ time series where SSAID's detection was unsuccessful. Consequently, we have $94$ simulated noisy time series in which SSAID successfully identifies all $20$ true change-points with an acceptable error of no more than $3$ days.  For each of the $94$ detection results, we randomly removed $2-5$ correct change-points from the $20$ correctly detected change-points and randomly added $1-3$ false change-points, resulting in $3-8$ single change-points per detection result.
	
When calculating the detection confidence $\hat{p}$ for each change-point pair, we need to utilize information from neighboring stations. To achieve this, we simulate four additional noisy time series for each test scenario, recorded by two stations on either side of the current station (see Fig. \ref{fig_example_of_time_series_at_multiple_stations} (a)). Consequently, each test scenario includes five distinct time series, each recorded by a different GPS station (see Fig. \ref{fig_example_of_time_series_at_multiple_stations} (b)-(f)). These GPS stations are regularly spaced at $15$ km intervals along the fault's strike, recording the same SSEs. The noise at the additional stations is also assumed to be a combination of white noise and flicker noise. The noise levels at the neighboring stations are randomly assigned, which means that the characteristics may vary significantly even between adjacent stations.
	
We now apply our proposed pre-processing chain to these generated detection results. We first pair the single change-points using our pre-processing scheme and then calculate the corresponding detection confidence $\hat{p}$ for each paired change-point using the hypothesis testing. In all the $94$ tested scenarios, we generate $1449$ pairs of change-points, with $354$ pairs having $\hat{p}\geq{0.9}$, $205$ pairs having ${0.6}\leq{\hat{p}}<{0.9}$, and $890$ pairs having $\hat{p}<{0.6}$. We categorize them into three groups based on the calculated $\hat{p}$: probable SSEs, possible SSEs, and non-SSEs. We then calculate the percentage of correct detections (detections with an error of no more than $3$ days from the true change-points) and false detections (detections with an error greater than $3$ days from the true change-points) for each category. 
	
As shown in Fig. \ref{fig_appen_verifying_tests}, the results indicate that in the probable SSEs category, $335$ out of $354$ pairs (about $94.6\%$) are correct change-points, while in the non-SSEs category, only $40$ out of $890$ pairs (about $4.5\%$) are correct change-points. In the possible SSEs category, $130$ out of $205$ pairs (about $63.4\%$) are correct detections, with the remaining $75$ pairs (about $36.6\%$) being false detections. This indicates that a high confidence $\hat{p}$ effectively identifies change-points likely resulting from true SSEs, while a low confidence $\hat{p}$ identifies those unlikely to be true SSEs. However, a medium confidence $\hat{p}$ is unreliable for identifying correct detections. Notably, all three categories contain both correct and false detections, so we cannot rely solely on the confidence score to identify accurate change-points. Change-points in the second and third categories may still originate from true SSEs.
	
These tests confirm that the proposed pre-processing combined with the hypothesis testing procedure can help identify change-points most likely originating from true SSEs. We will use fault estimation to further rule out false detections from the identified probable SSEs. Fault estimation has been widely applied to different observed data and its validity has been verified in past studies (see Bagnardi \& Hooper (2018); Yano \& Kano (2022) for details).

\refstepcounter{texts}
\section*{Text \thetexts. Observed GPS time series at different GPS stations and their neighbouring GPS stations}
In this section, we first present some representative time series observed at different GPS stations. Fig. \ref{fig_observed_time_series_at_three_sites} shows the three time series recorded at GPS stations $021052$, $950449$, $950447$ and their estimated change-points by SSAID plus single change-points pairing. 

We now present the four representative time series recorded by GPS stations $950436$, $041133$, $970828$ and $021049$, corresponding to the fault estimation results depicted in Fig. $12$ in the main text. Fig. \ref{fig_map_showing_four_stations_and_neighbouring_stations} shows locations of these four reference GPS stations and their neighboring GPS stations. A neighboring GPS station refer to a GPS station if its distance from its reference GPS station is no more than 3 km. The time series observed at GPS station $065$, indicated in red, is only available from early 2009 and will not be displayed.

We now present the four representative time series recorded by GPS stations $950436$, $041133$, $970828$ and $021049$, corresponding to the fault estimation results depicted in Fig. $12$ in the main text. Figure \ref{fig_map_showing_four_stations_and_neighbouring_stations} shows the locations of these four reference GPS stations and their neighboring GPS stations. A neighboring GPS station is defined as one that is within $30$ km of its reference GPS station. The time series observed at GPS station $081175$, indicated in red, is only available from early 2009 and will not be displayed. All the observed time series on these GPS stations are shown in Fig. \ref{fig_appen_time_series_01} to \ref{fig_appen_time_series_07}, and the estimated change-points by SSAID plus single change-point pairing for each reference GPS station are indicated by red vertical lines (starting points) and blue dotted lines (ending change-points).

\refstepcounter{texts}
\section*{Text \thetexts. The fault estimation results of other identified SSEs}
In Section $5.2$ of the main text, we indicated that $18$ SSEs were identified by the fault estimation using the probable SSE candidates, while only 4 representative results were included. In this section, we present the results of the other $14$ identified SSEs.

\refstepcounter{Sfig}

\refstepcounter{Sfig}

\refstepcounter{Sfig}

\refstepcounter{Sfig}

\refstepcounter{Sfig}

\refstepcounter{Sfig}

\refstepcounter{Sfig}

\refstepcounter{Sfig}

\refstepcounter{Sfig}

\refstepcounter{Sfig}

\refstepcounter{Sfig}

\refstepcounter{Sfig}

\refstepcounter{Sfig}
\begin{figure}[htbp]	
	\centering
	\includegraphics[width=0.9\textwidth]{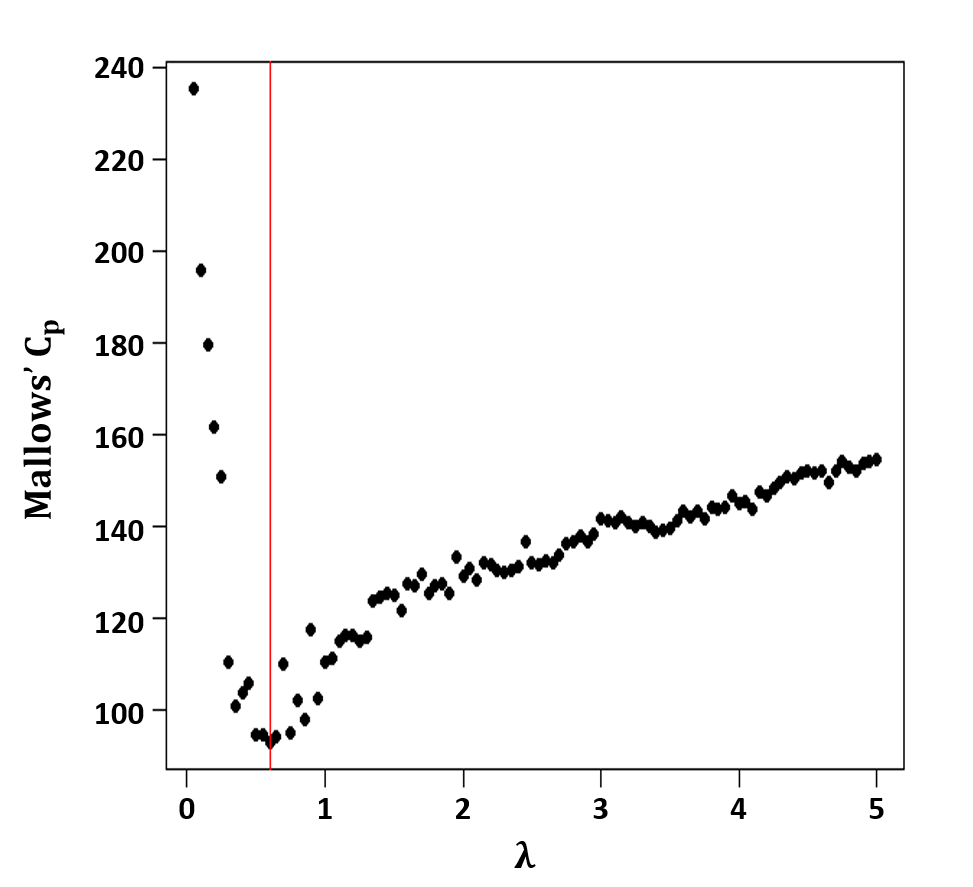}
	\begin{minipage}{0.9\textwidth}
		\textbf{Figure \theSfig.} An example showing how the value of the Mallows' $C_p$ changes with the hyperparameter $\lambda$ for a noisy time series. The minimum value is highlighted by the red vertical line.
	\end{minipage}
	\label{fig_MallowCP_lambda}	
\end{figure}

\refstepcounter{Sfig}
\begin{figure}[htbp]	
	\centering
	\includegraphics[width=0.9\textwidth]{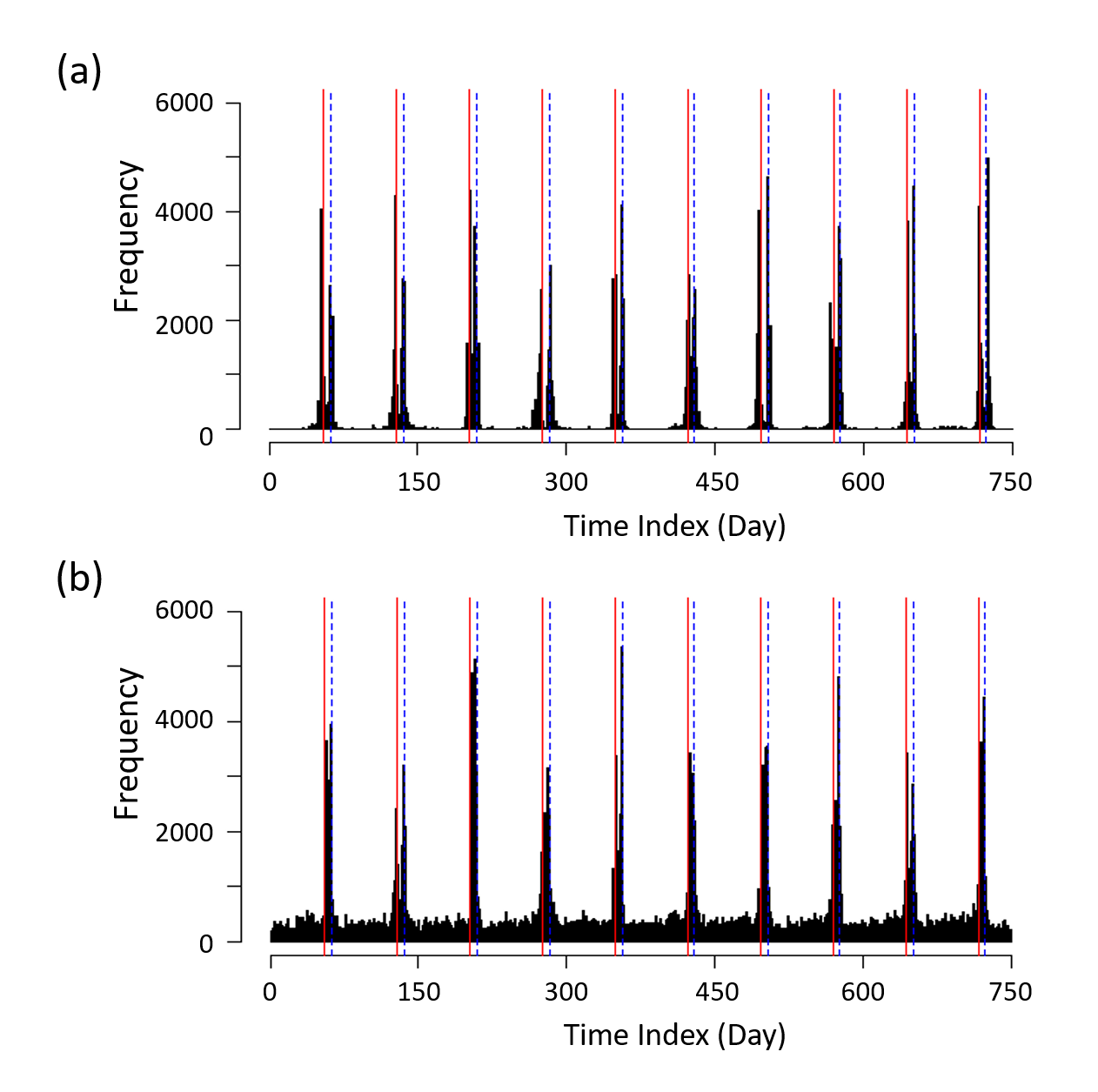}
	\begin{minipage}{0.9\textwidth}
		\textbf{Figure \theSfig.} Histogram of detected change-points in all the synthetic data in \textsection{4} of the main text by different methods: (a) SSAID; (b) $l_1$ trend filtering. Vertical red lines: start times of simulated SSEs; vertical blue dashed lines: end times of simulated SSEs.
	\end{minipage}
	\label{fig04_SSAID_L1ft}	
\end{figure}

\refstepcounter{Sfig}
\begin{figure}[htbp]	
	\centering
	\includegraphics[width=0.78\textwidth]{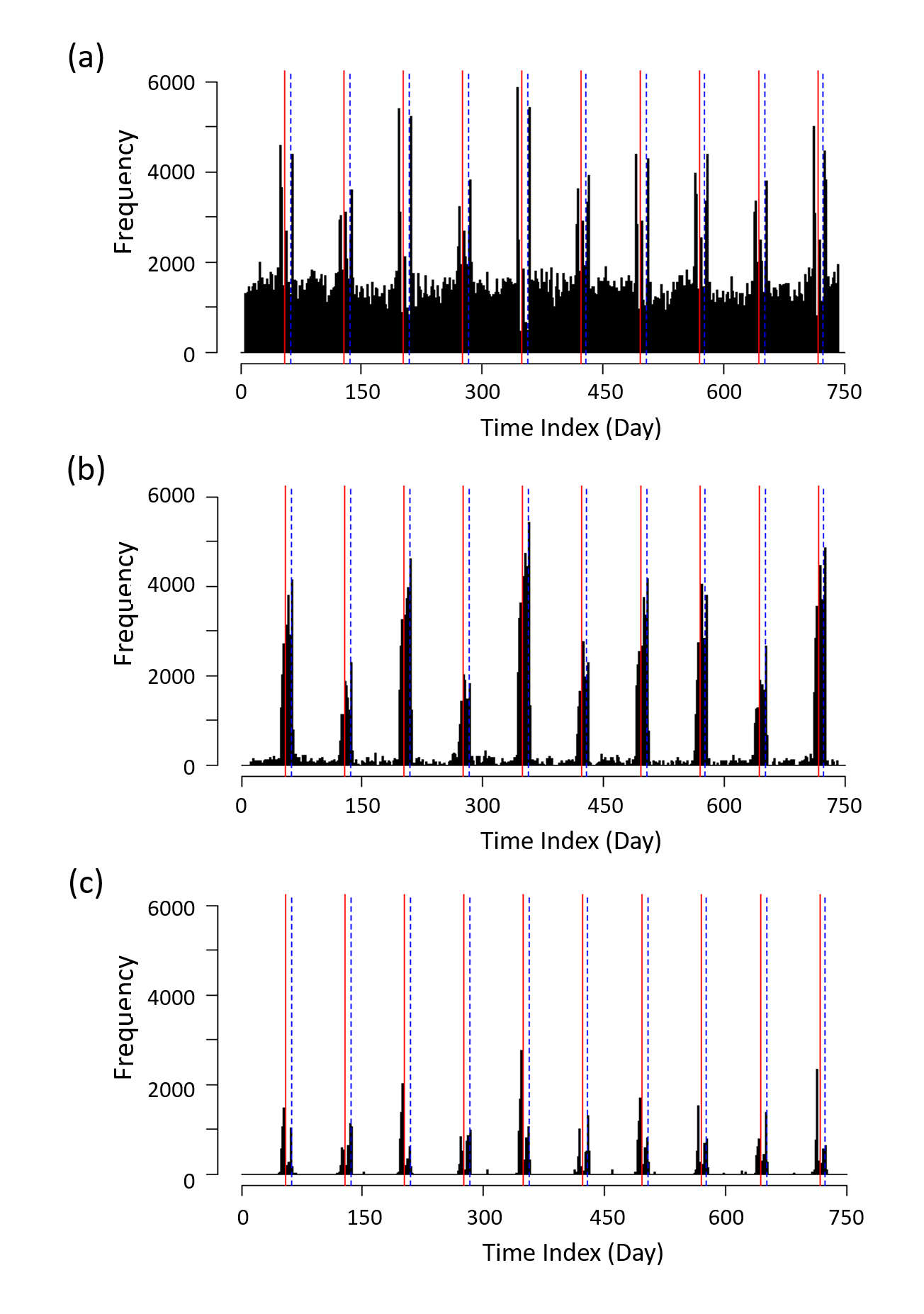}
	\begin{minipage}{0.9\textwidth}
		\textbf{Figure \theSfig.} The same histograms as Fig. \ref{fig04_SSAID_L1ft} but for the linear regression with $\Delta{AIC}$ by using different thresholds: (a) a high threshold ($\zeta$=-10); (d) a medium threshold ($\zeta$=-20); (e) a low threshold ($\zeta$=-30). The sliding window is $15$ days. Vertical red lines: start times of simulated SSEs; vertical blue dashed lines: end times of simulated SSEs.
	\end{minipage}
	\label{fig04_linear_AIC_diff_thres}	
\end{figure}

\refstepcounter{Sfig}
\begin{figure}[htbp]	
	\centering
	\includegraphics[width=0.65\textwidth]{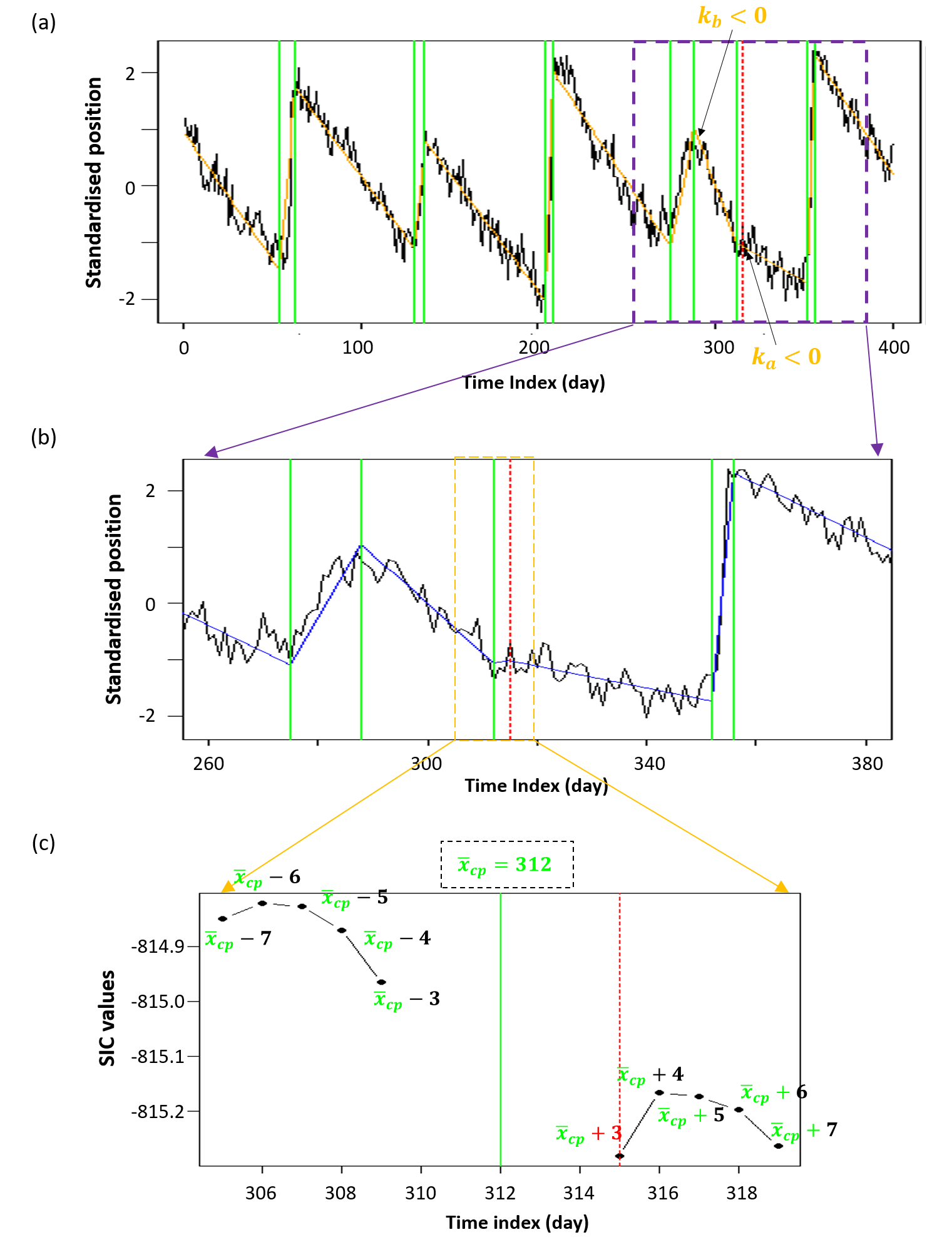}
	\begin{minipage}{0.9\textwidth}
		\textbf{Figure \theSfig.} (a) Simulated noisy time series with change-points detected by SSAID, marked by green lines. SSAID correctly identifies $10$ change-points and one false change-point at day $312$. The orange line represents a piecewise linear fit based on the $11$ detected change-points. Slopes $k_b$ and $k_a$ correspond to the linear segments before and after the single change-point. (b) Zoom-in view of the region in panel (a) within the purple dotted box. The red dotted line highlights the selected change-point to pair the false change-point. The blue line shows the piecewise linear fit using all detected change-points and the paired change-point. (c) SIC values for change-point candidates to pair the false change-point at day $\bar{x}_{cp}=312$. The search range includes days $305$, $306$, $307$, $308$, $309$, $315$, $316$, $317$, $318$, $319$. The selected change-point, day $315$, is marked by the red dotted line.
	\end{minipage}
	\label{fig_appen_showing_how_to_pair_single_cp}	
\end{figure}

\refstepcounter{Sfig}
\begin{figure}[htbp]	
	\centering
	\includegraphics[width=\textwidth]{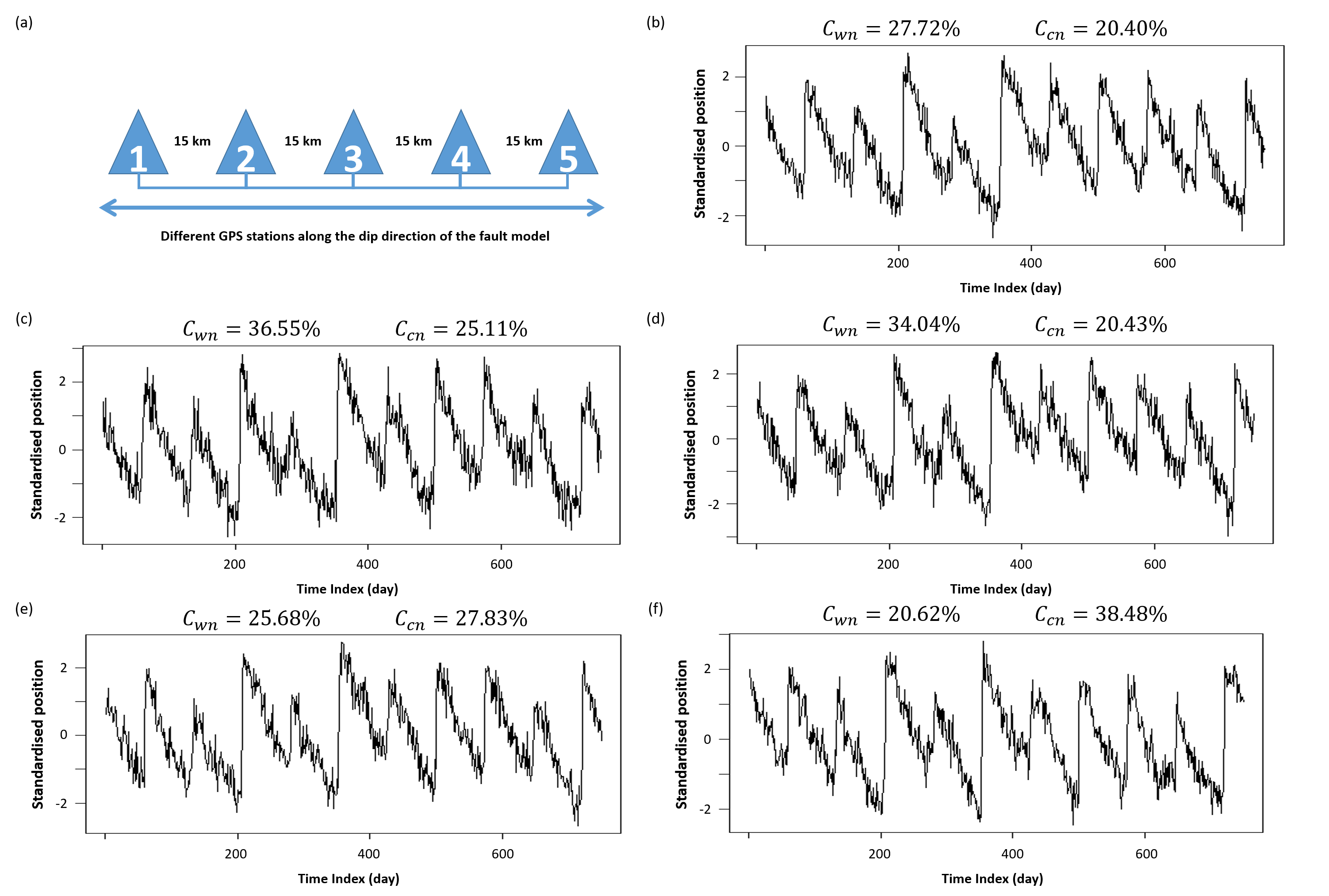}
	\begin{minipage}{0.9\textwidth}
		\textbf{Figure \theSfig.} (a) The deployment of five GPS stations along the dip direction for simulating noisy time series across multiple stations; (b)-(g) an example of simulated noisy time series recorded at the five GPS stations with different white noise levels $C_{wn}$ and color noise levels $C_{cn}$, outlined in each panel.
	\end{minipage}
	\label{fig_example_of_time_series_at_multiple_stations}	
\end{figure}

\refstepcounter{Sfig}
\begin{figure}[htbp]	
	\centering
	\includegraphics[width=\textwidth]{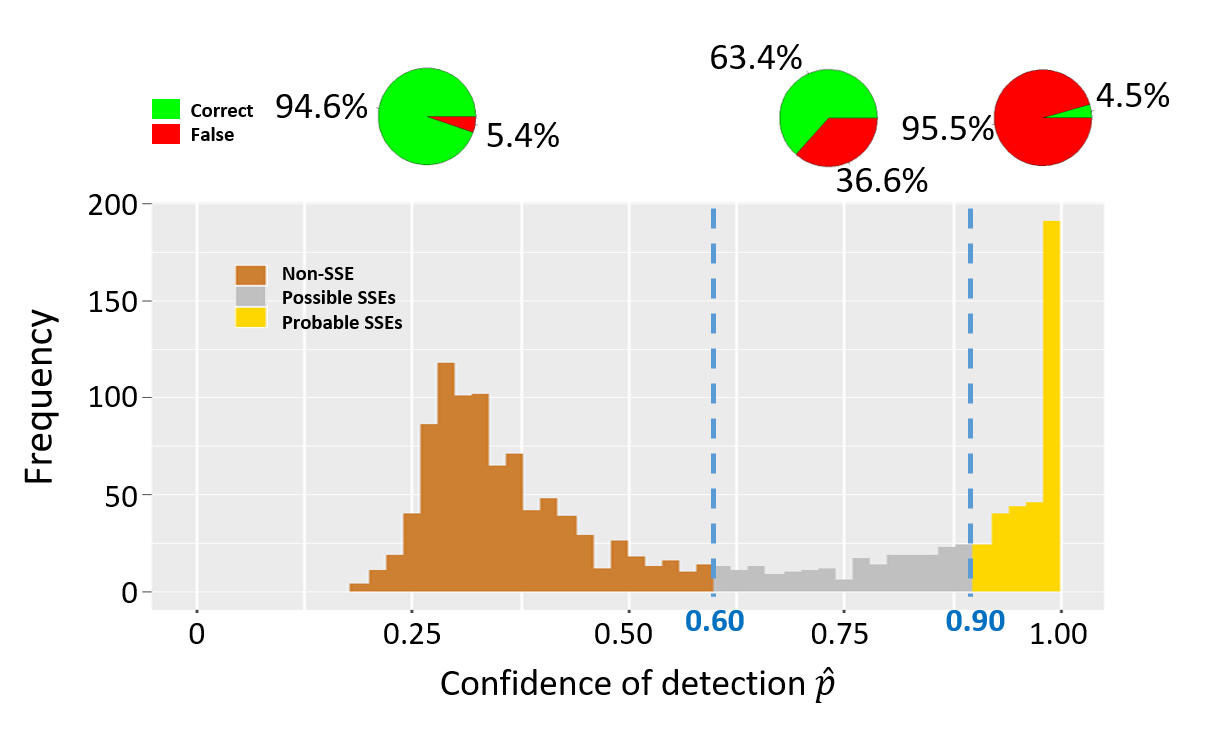}
	\begin{minipage}{0.9\textwidth}
		\textbf{Figure \theSfig.} Histogram of the calculated detection confidence $\hat{p}$ for each change-point pair in the numerical tests, validating the proposed pre-processing and hypothesis testing for identifying probable SSEs from  SSAID detection results. The three categories in dark brown, grey, and yellow, divided by the two blue dotted vertical lines, represent non-SSEs ($\hat{p}<0.6$), possible SSEs (${0.6}\leq{\hat{p}}<0.9$), and probable SSEs ($\hat{p}\geq{0.9}$), respectively. The pie charts above each category show the percentages of correct (green) and false (red) detections. A correct detection has an error of no more than $3$ days from the true change-points, while a false detection exceeds this error threshold.
	\end{minipage}
	\label{fig_appen_verifying_tests}	
\end{figure}

\refstepcounter{Sfig}
\begin{figure}[htbp]	
	\centering
	\includegraphics[width=0.9\textwidth]{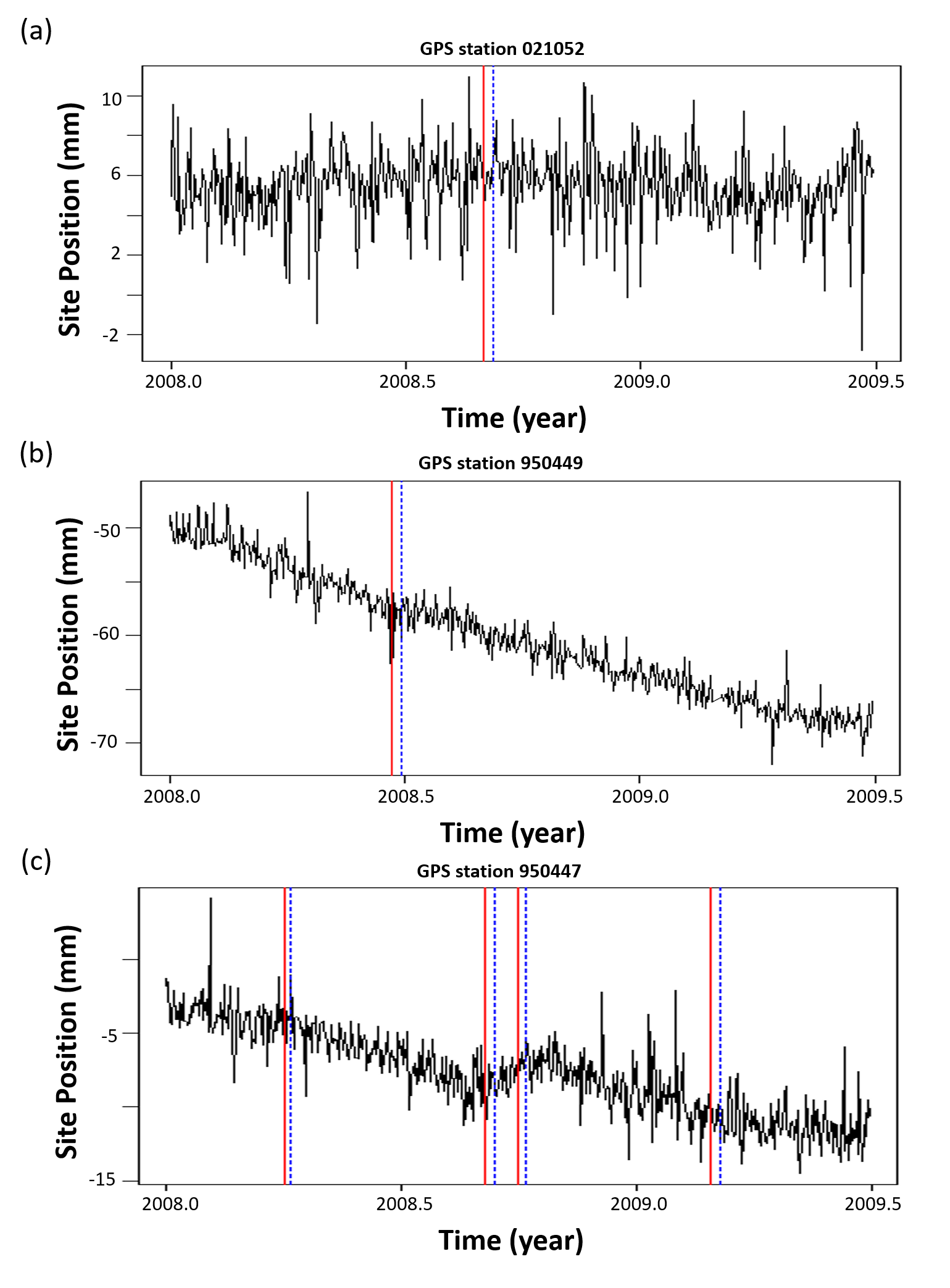}
	\begin{minipage}{0.9\textwidth}
		\textbf{Figure \theSfig.} Observed time series at different stations and their estimated change-points by SSAID plus single change-point pairing: (a) $021052$; (b) $950449$; (c) $950447$. Red vertical lines: starting change-points; blue dotted vertical lines: ending change-points.
	\end{minipage}
	\label{fig_observed_time_series_at_three_sites}	
\end{figure}

\refstepcounter{Sfig}
\begin{figure}[htbp]	
	\centering
	\includegraphics[width=0.9\textwidth]{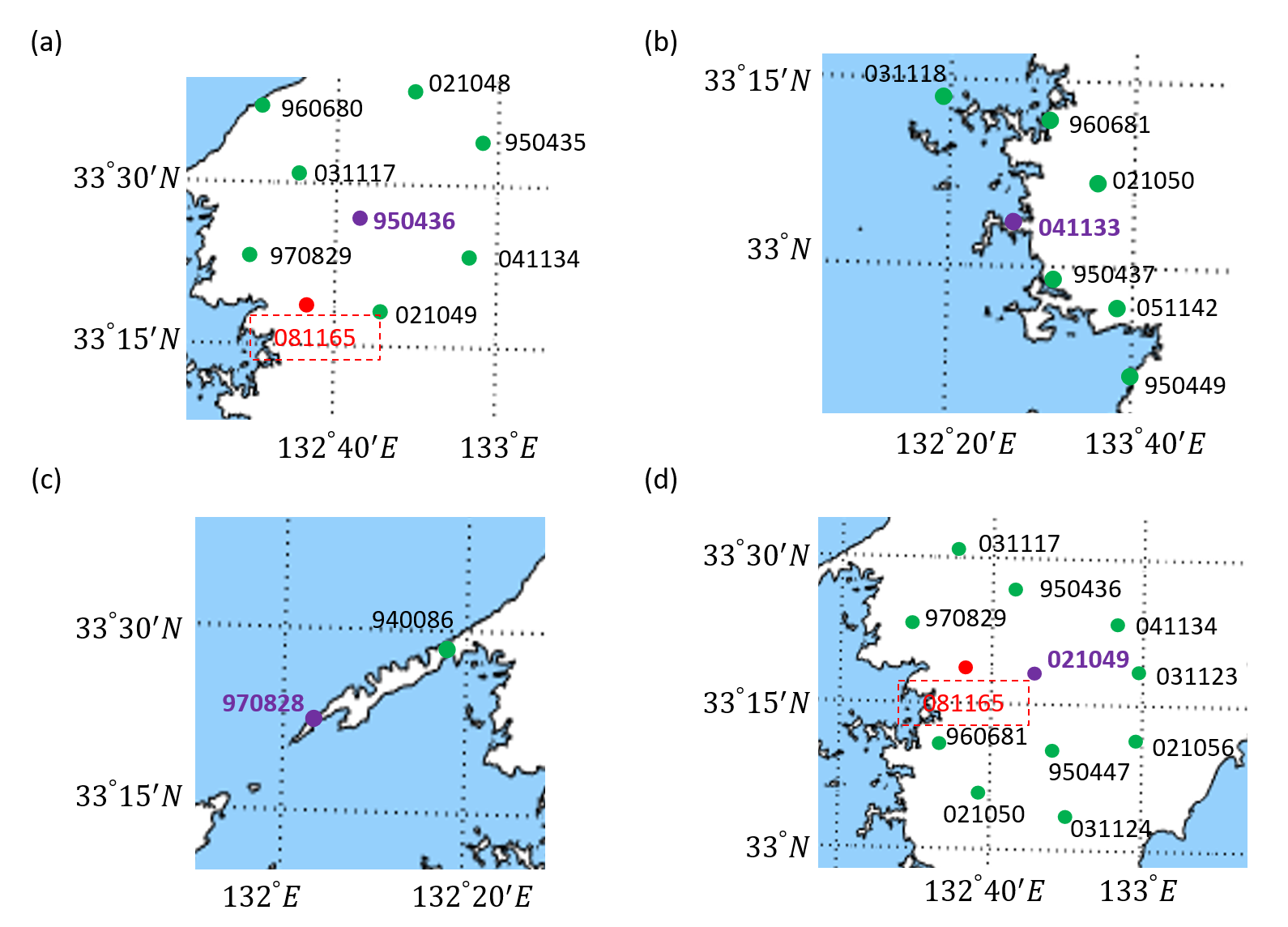}
	\begin{minipage}{0.9\textwidth}
		\textbf{Figure \theSfig.} Locations of four reference GPS stations and their neighboring GPS stations: (a) $950436$; (b) $041133$; (c) $970828$; (d) $021049$. The reference GPS stations, indicated in purple, correspond to the four GPS stations shown in Fig. \minewc{$13$} in the main text for fault estimation. The distance between each reference GPS station and its neighboring GPS stations is no more than $30$ km. The time series observed at GPS station $81165$, indicated in red, is only available from early 2009 and will not be displayed.
	\end{minipage}
	\label{fig_map_showing_four_stations_and_neighbouring_stations}	
\end{figure}

\refstepcounter{Sfig}
\begin{figure}[htbp]	
	\centering
	\includegraphics[width=0.85\textwidth]{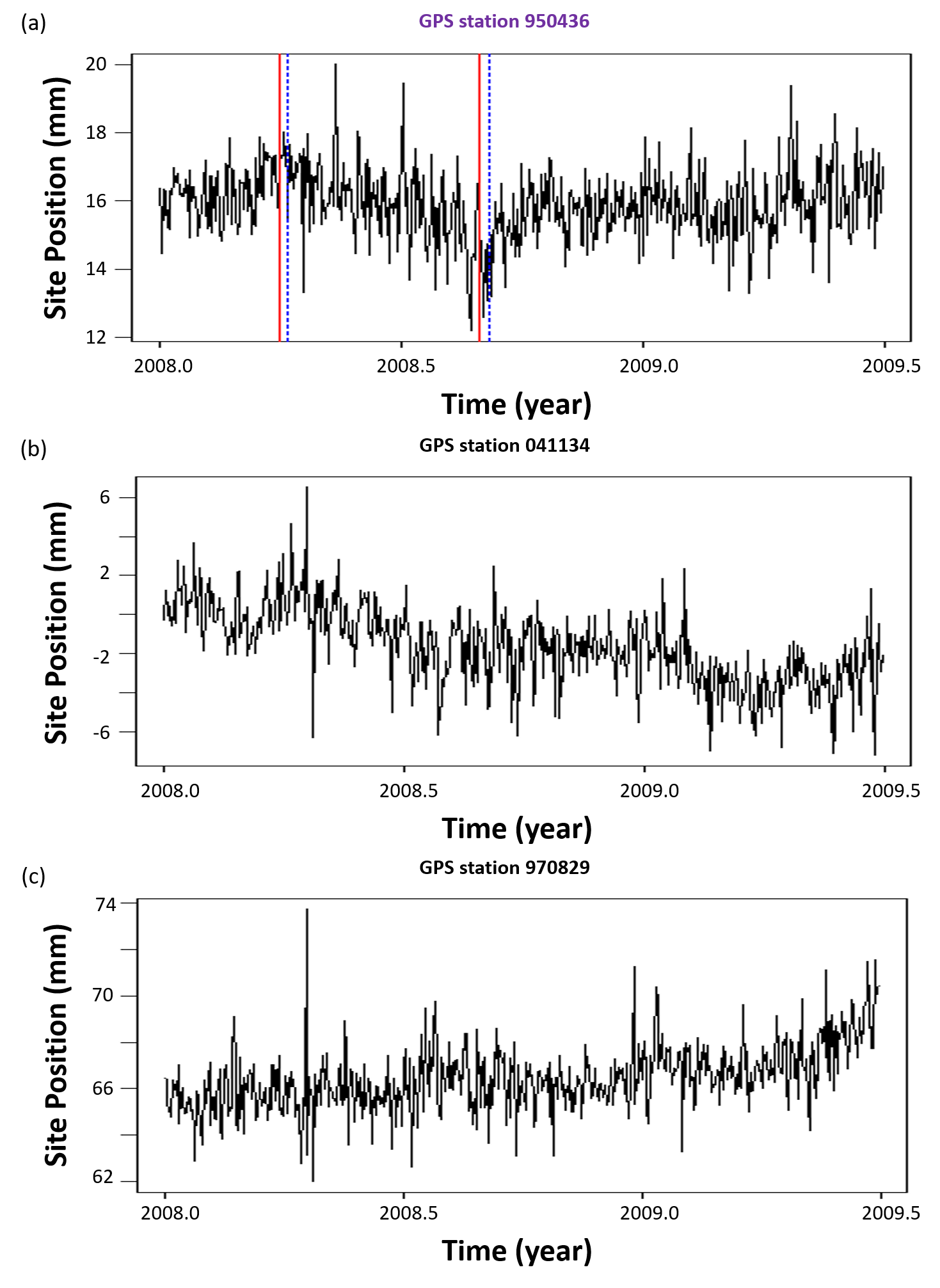}
	\begin{minipage}{0.9\textwidth}
		\textbf{Figure \theSfig.} Observed time series at the reference GPS station $950436$ (see panel (a)) and its neighbouring GPS stations, which include $041134$ (see panel (b)), $970829$ (see panel (c)), $950435$ (see Fig. \ref{fig_appen_time_series_02} (a)), $021048$ (see Fig. \ref{fig_appen_time_series_02} (b)), $960680$ (see Fig. \ref{fig_appen_time_series_02} (c)), $021049$ (see Fig. \ref{fig_appen_time_series_06} (a)), $031117$ (see Fig. \ref{fig_appen_time_series_07} (c)).  Red vertical lines: starting change-points; blue dotted vertical lines: ending change-points. 
	\end{minipage}
	\label{fig_appen_time_series_01}	
\end{figure}

\refstepcounter{Sfig}
\begin{figure}[htbp]	
	\centering
	\includegraphics[width=0.85\textwidth]{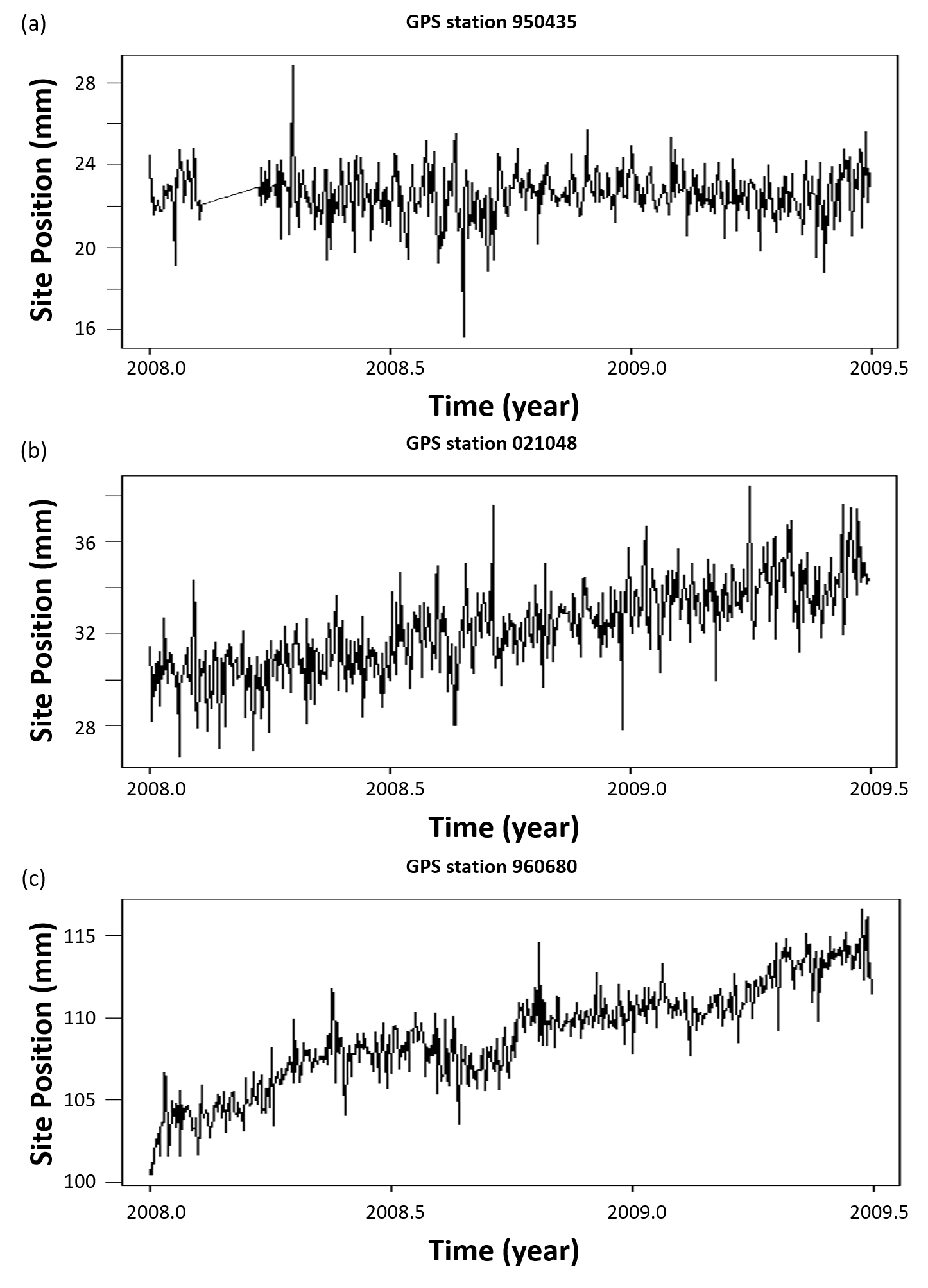}
	\begin{minipage}{0.9\textwidth}
		\textbf{Figure \theSfig.} Observed time series at different neighbouring GPS stations of station $950436$.
	\end{minipage}
	\label{fig_appen_time_series_02}	
\end{figure}

\refstepcounter{Sfig}
\begin{figure}[htbp]	
	\centering
	\includegraphics[width=0.85\textwidth]{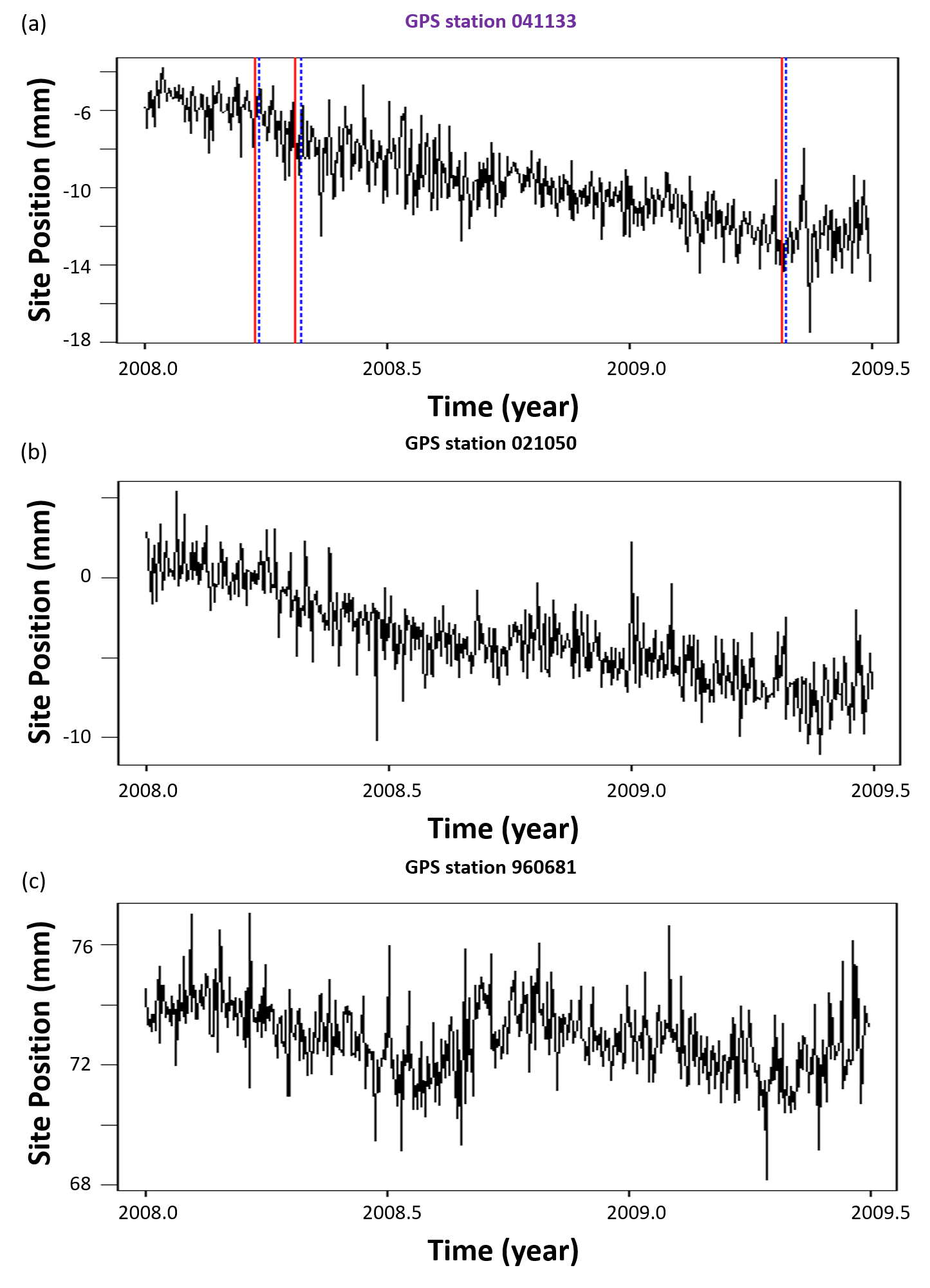}
	\begin{minipage}{0.9\textwidth}
		\textbf{Figure \theSfig.} Observed time series at the reference GPS station $041133$ (see panel (a)) and its neighbouring GPS stations, which include $021050$ (see panel (b)), $960681$ (see panel (c)), $031118$ (see Fig. \ref{fig_appen_time_series_04} (a)), $051142$ (see Fig. \ref{fig_appen_time_series_04} (b)), $950437$ (see Fig. \ref{fig_appen_time_series_04} (c)), $950449$ (see Fig. \ref{fig_observed_time_series_at_three_sites} (b)).  Red vertical lines: starting change-points; blue dotted vertical lines: ending change-points. 
	\end{minipage}
	\label{fig_appen_time_series_03}	
\end{figure}

\refstepcounter{Sfig}
\begin{figure}[htbp]	
	\centering
	\includegraphics[width=0.85\textwidth]{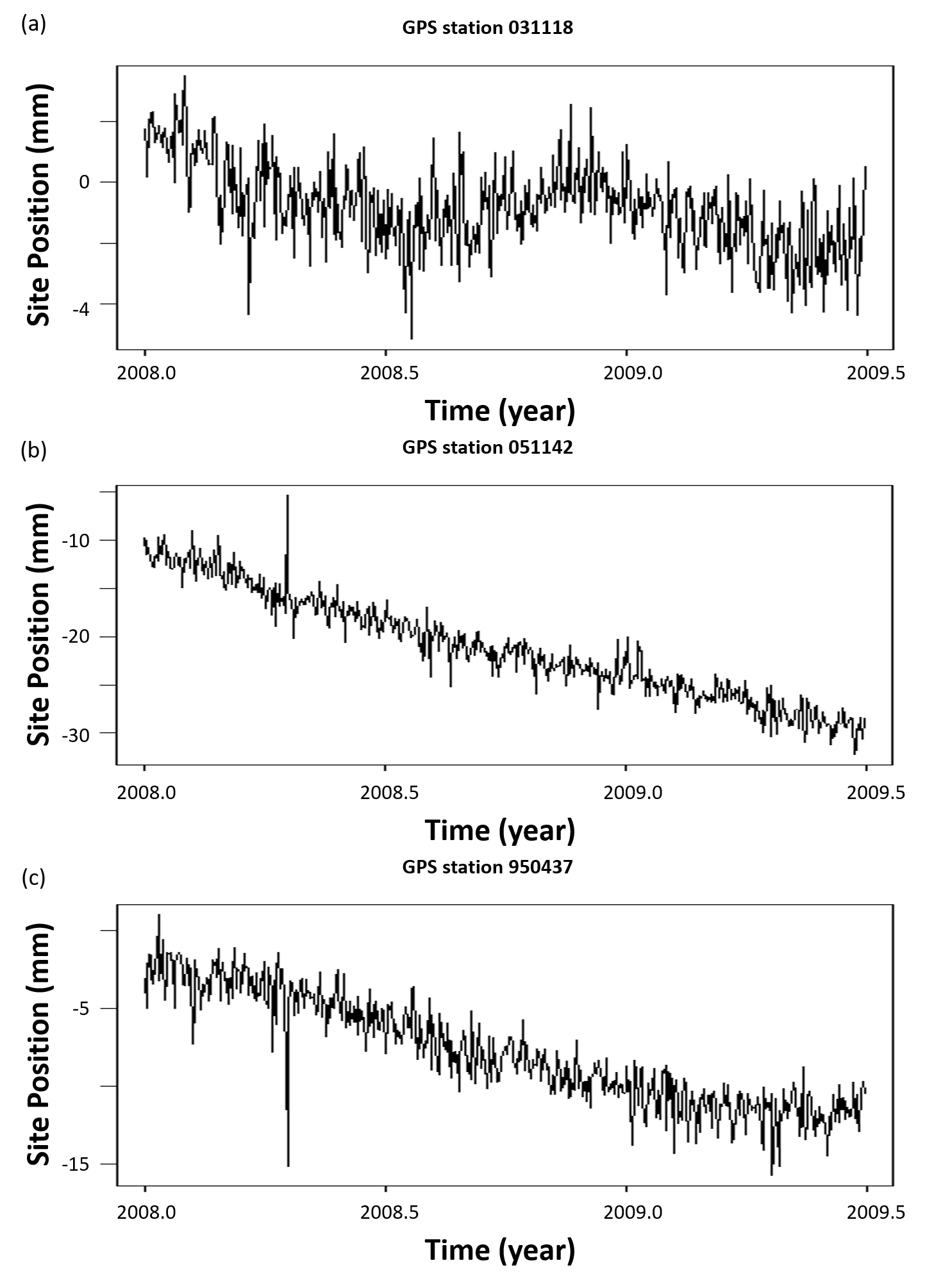}
	\begin{minipage}{0.9\textwidth}
		\textbf{Figure \theSfig.} Observed time series at different neighbouring GPS stations of station $041133$. 
	\end{minipage}
	\label{fig_appen_time_series_04}	
\end{figure}

\refstepcounter{Sfig}
\begin{figure}[htbp]	
	\centering
	\includegraphics[width=0.85\textwidth]{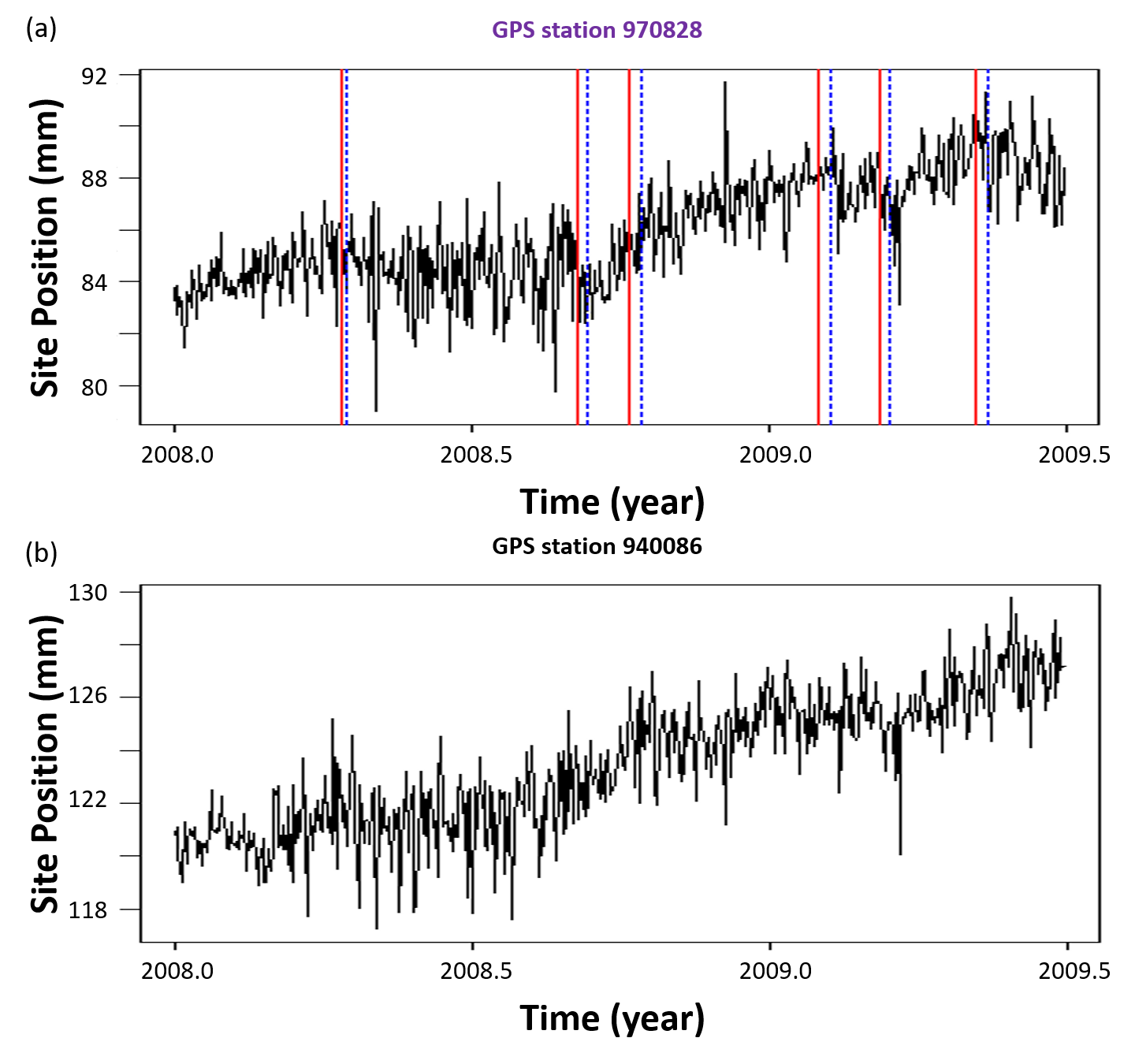}
	\begin{minipage}{0.9\textwidth}
		\textbf{Figure \theSfig.} Observed time series at the reference GPS station $970828$ and its neighbouring GPS station $940086$.  Red vertical lines: starting change-points; blue dotted vertical lines: ending change-points. 
	\end{minipage}
	\label{fig_appen_time_series_05}	
\end{figure}

\refstepcounter{Sfig}
\begin{figure}[htbp]	
	\centering
	\includegraphics[width=0.85\textwidth]{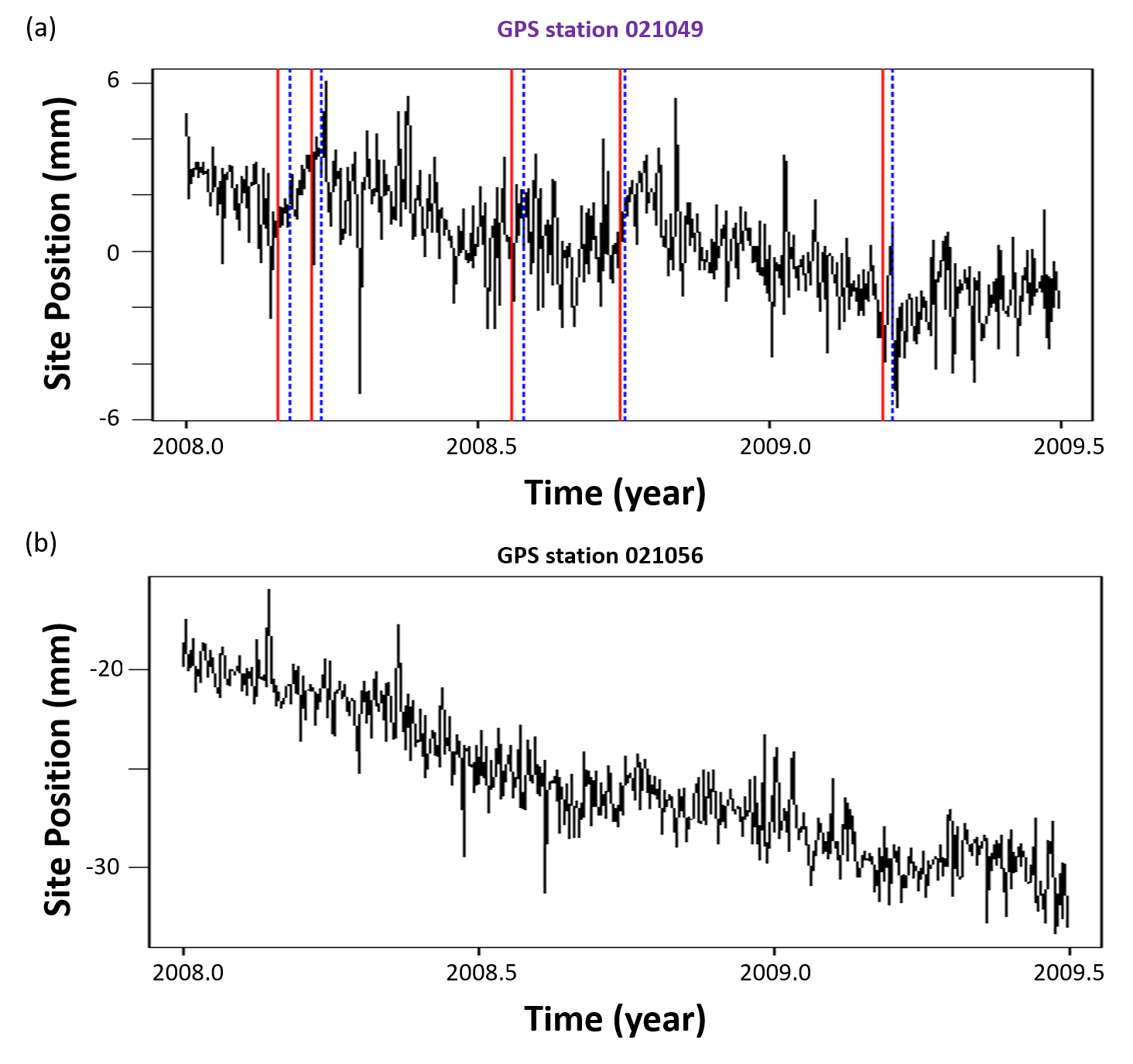}
	\begin{minipage}{0.9\textwidth} 
		\textbf{Figure \theSfig.} Observed time series at the reference GPS station $021049$ (see panel (a)) and its neighbouring GPS stations, which include $021056$ (see panel (b)), $950447$ (see Fig. \ref{fig_observed_time_series_at_three_sites} (c)), $031123$ (see Fig. \ref{fig_appen_time_series_07} (a)), $031124$ (see Fig. \ref{fig_appen_time_series_07} (b)), $031117$ (see Fig. \ref{fig_appen_time_series_07} (c)), $950436$ (see Fig. \ref{fig_appen_time_series_01} (a)), $041134$ (see Fig. \ref{fig_appen_time_series_07} (b)), $970829$ (see Fig. \ref{fig_appen_time_series_01} (c)), $021050$ (see Fig. \ref{fig_appen_time_series_03} (b)), $960681$ (see Fig. \ref{fig_appen_time_series_03} (c)).  Red vertical lines: starting change-points; blue dotted vertical lines: ending change-points.  
	\end{minipage}
	\label{fig_appen_time_series_06}	
\end{figure}

\refstepcounter{Sfig}
\begin{figure}[htbp]	
	\centering
	\includegraphics[width=0.85\textwidth]{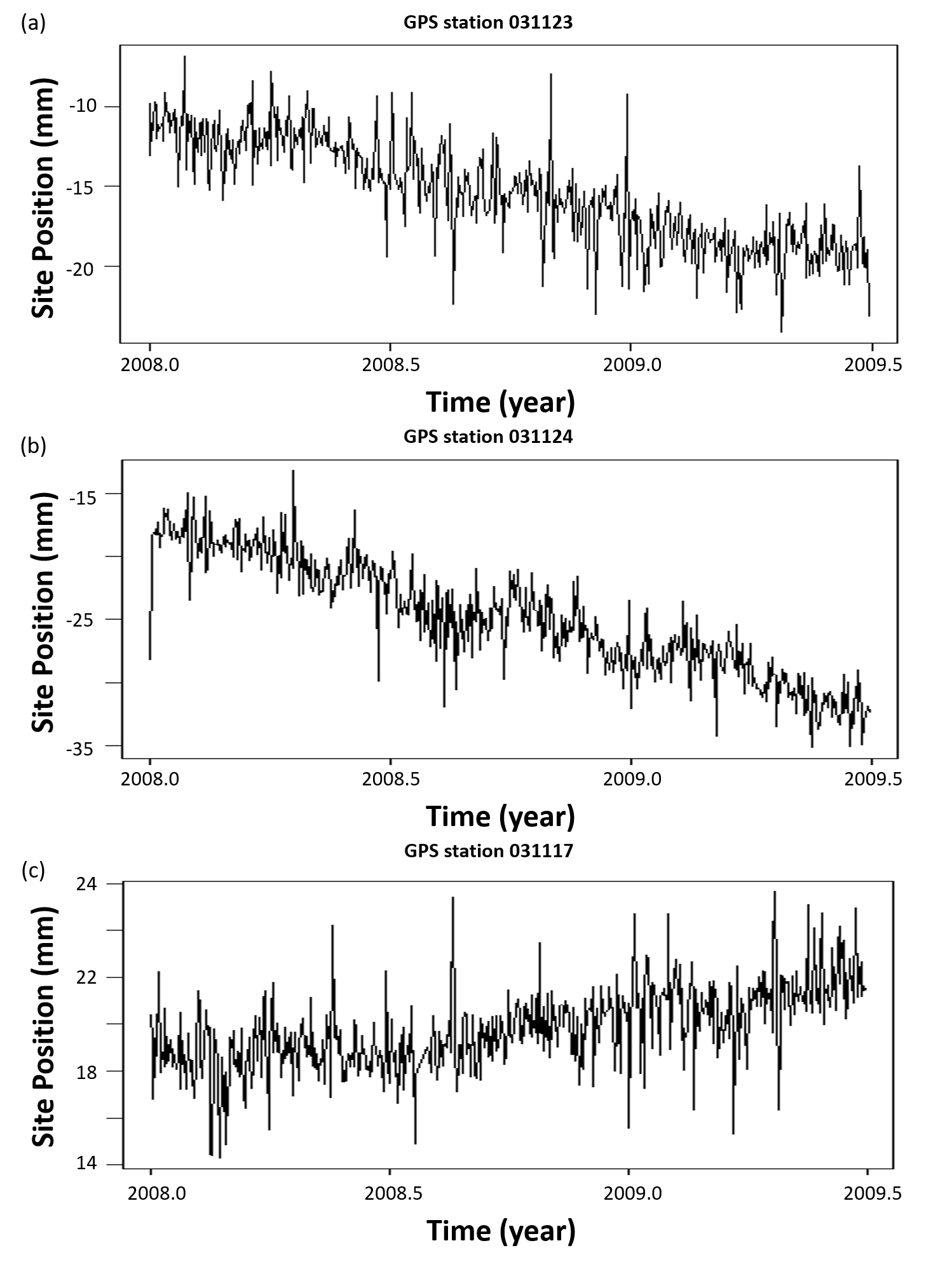}
	\begin{minipage}{0.9\textwidth}
		\textbf{Figure \theSfig.} Observed time series at different neighbouring GPS stations of station $021049$. 
	\end{minipage}
	\label{fig_appen_time_series_07}	
\end{figure}

\refstepcounter{Sfig}
\begin{figure}[htbp]		
	\centering
	\includegraphics[width=\textwidth]{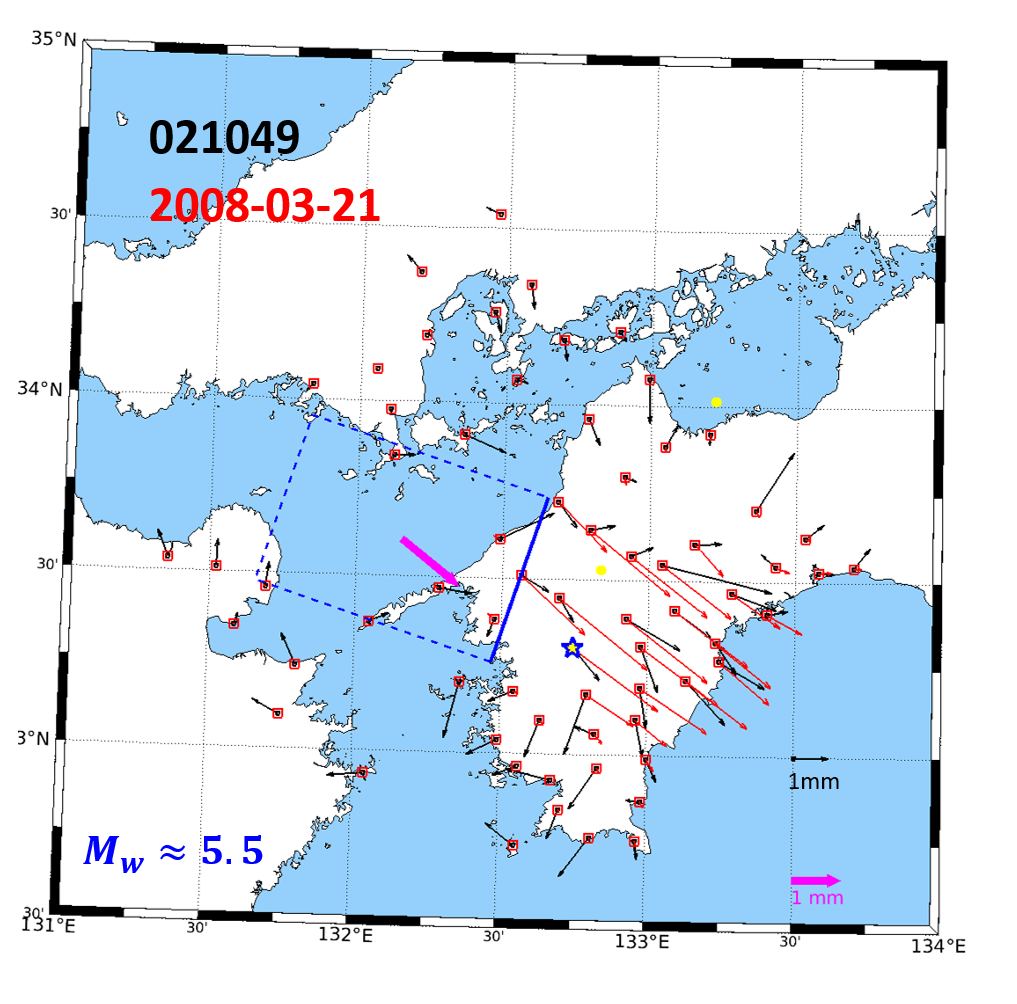}
	\begin{minipage}{0.9\textwidth}
		\textbf{Figure \theSfig.} The estimated fault model of an identified probable SSE candidate at the station $021049$. The date in red under the site name refers to the start date of this probable SSE candidate. The star in the map indicates the location of the station where this SSE candidate was identified. The black and the pink arrows in the right-bottom corner are the scale arrows for the observed displacement and the slip amount of the estimated model, respectively. The synthetic displacements by the displacement model of Okada (1985) have the same scale arrow as the observed ones. Orange dots indicate the epicentre of tremors in the episodic state $5$ days before and after the date (see the date on the left-upper corner) when this candidate was found. The blue solid line of the rectangle refers to the top edge of the estimated fault model.
	\end{minipage}	
	\label{fig_chap5_11}
\end{figure}

\refstepcounter{Sfig}
\begin{figure}[htbp]		
	\centering
	\includegraphics[width=\textwidth]{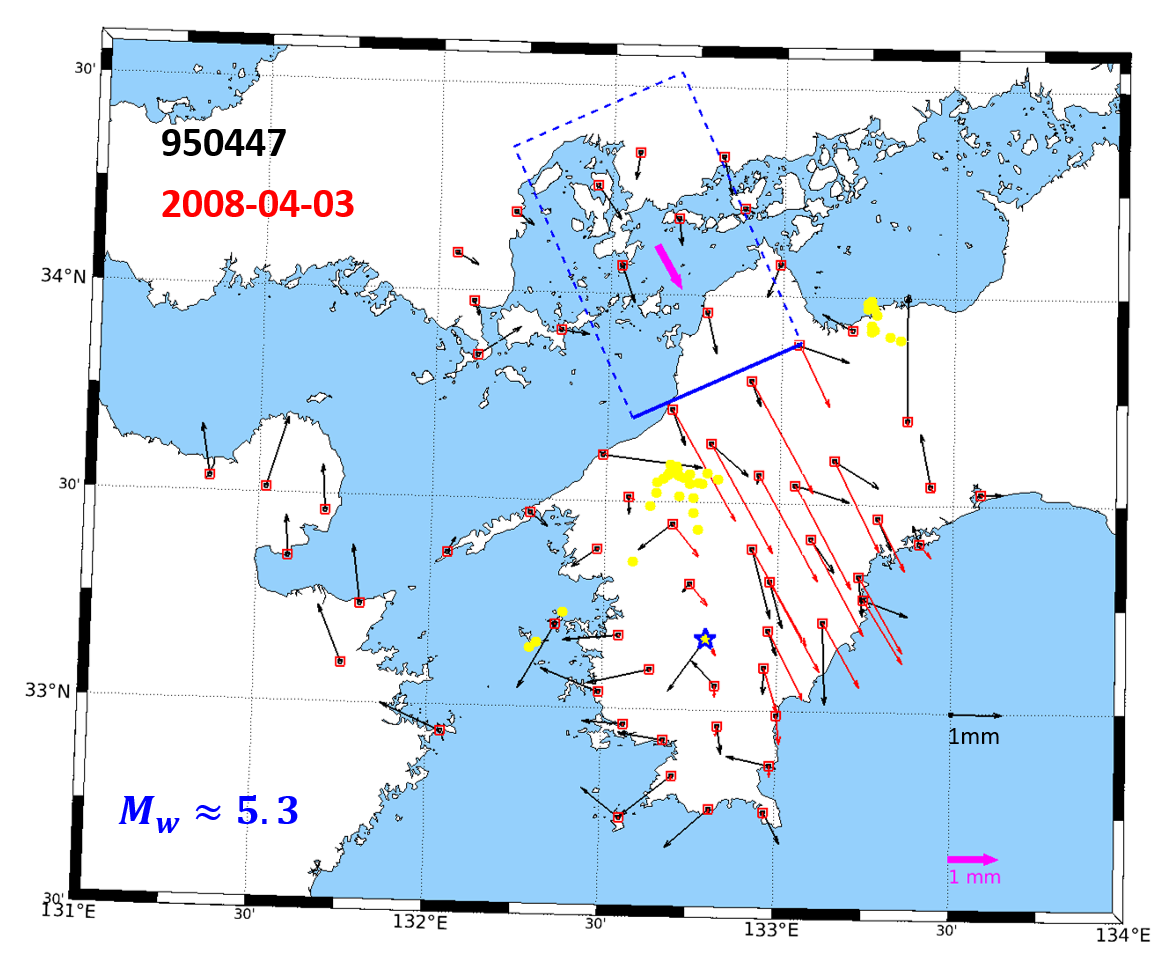}
	\begin{minipage}{0.9\textwidth}
		\textbf{Figure \theSfig.} Same as Fig. \ref{fig_chap5_11} but for a probable SSE candidate at station 950447.
	\end{minipage}
	\label{fig_chap5_13}
\end{figure}

\refstepcounter{Sfig}
\begin{figure}[htbp]		
	\centering
	\includegraphics[width=\textwidth]{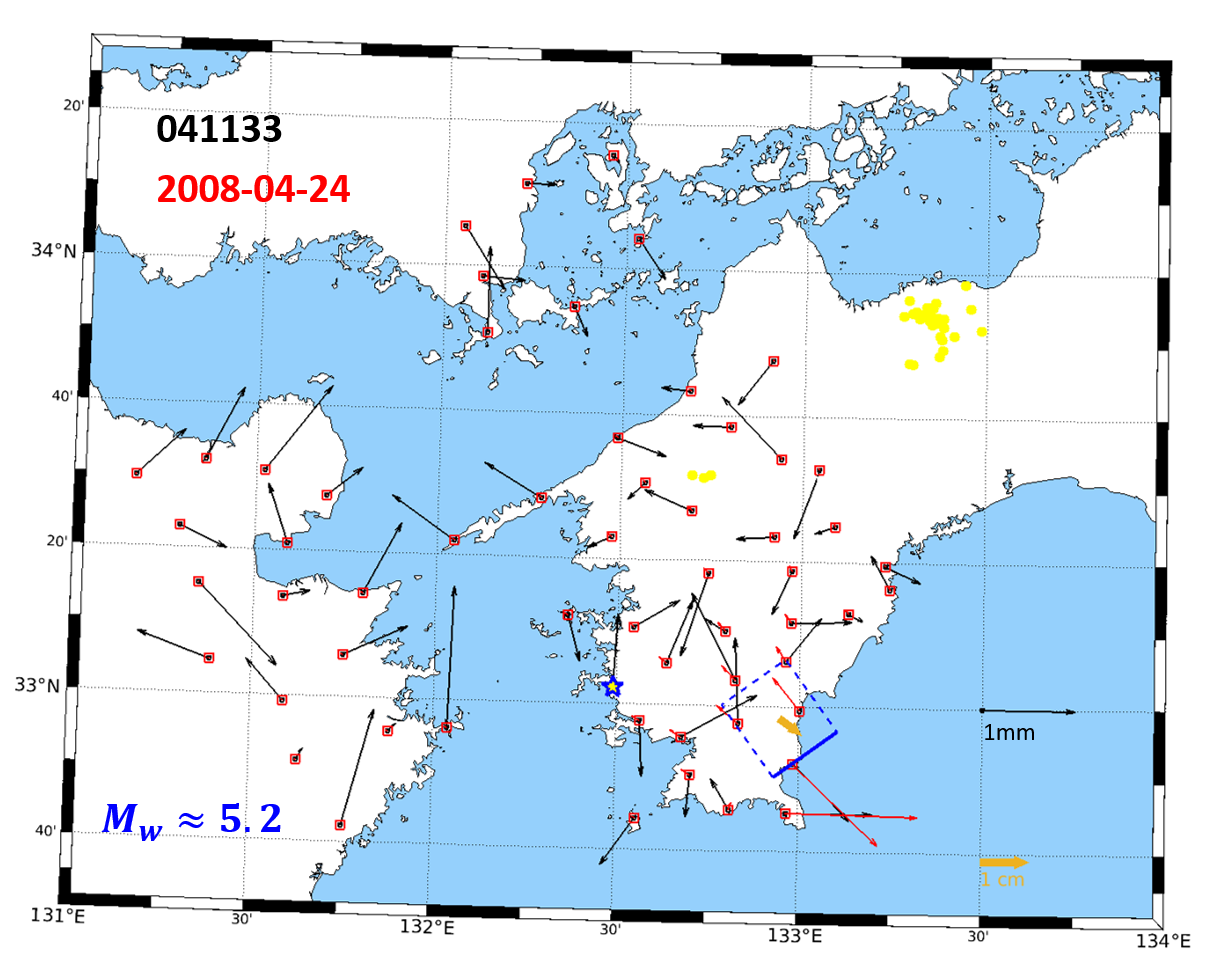}
	\begin{minipage}{0.9\textwidth}
		\textbf{Figure \theSfig.} Same as Fig. \ref{fig_chap5_11} but for a probable SSE candidate at station 041133.
	\end{minipage}	
	\label{fig_chap5_15}
\end{figure}

\refstepcounter{Sfig}
\begin{figure}[htbp]		
	\centering
	\includegraphics[width=\textwidth]{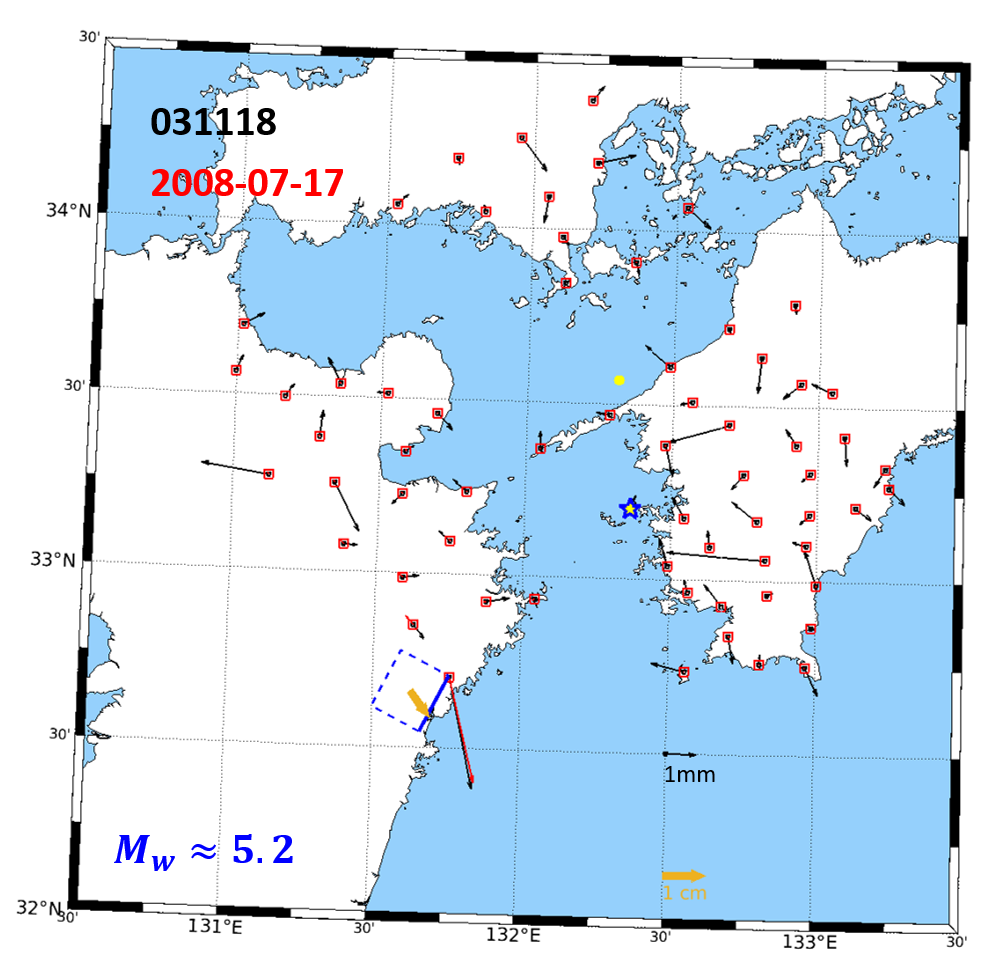}
	\begin{minipage}{0.9\textwidth}
		\textbf{Figure \theSfig.} Same as Fig. \ref{fig_chap5_11} but for a probable SSE candidate at station 031118.
	\end{minipage}	
	\label{fig_chap5_16}
\end{figure}

\refstepcounter{Sfig}
\begin{figure}[htbp]		
	\centering
	\includegraphics[width=\textwidth]{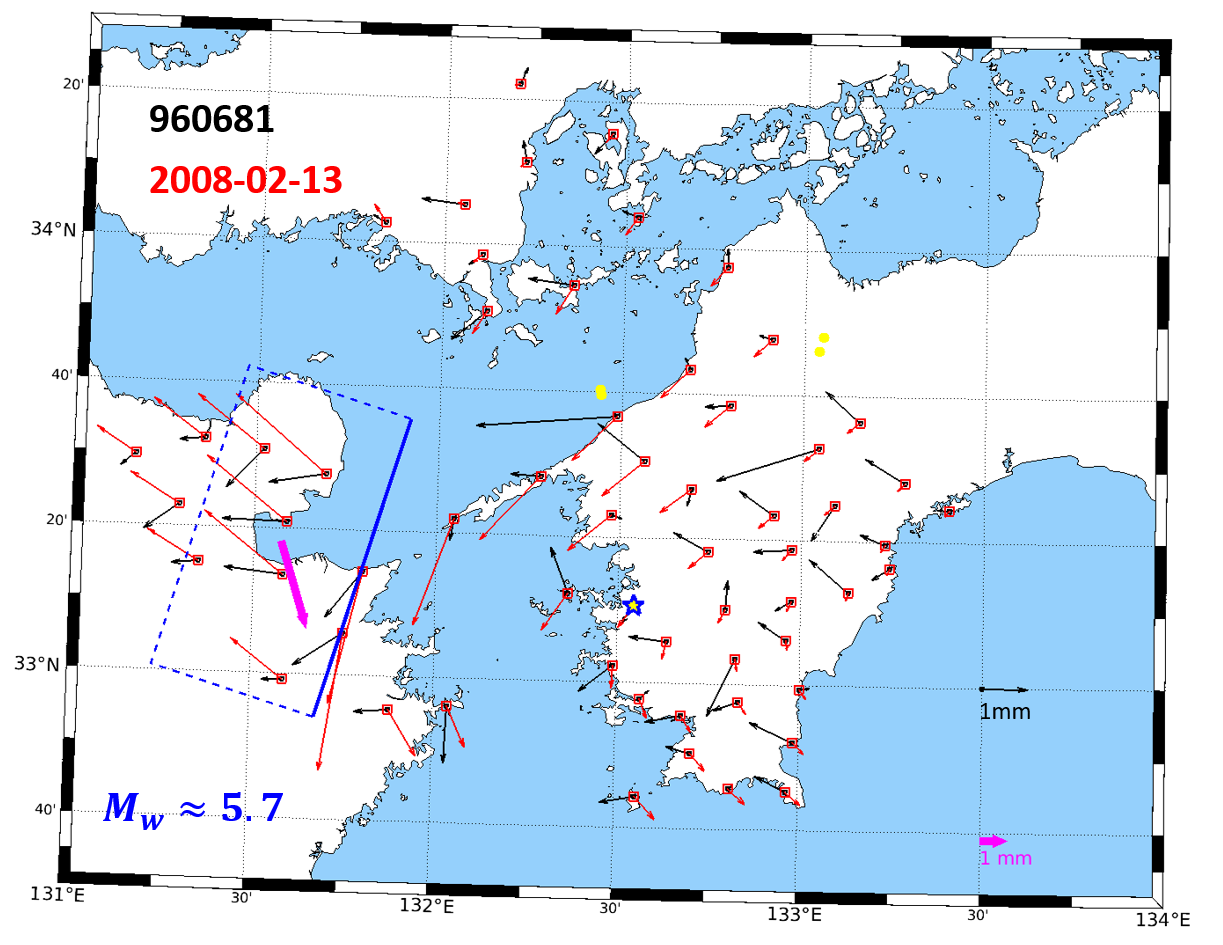}
	\begin{minipage}{0.9\textwidth}
		\textbf{Figure \theSfig.} Same as Fig. \ref{fig_chap5_11} but for a probable SSE candidate at station 960681.
	\end{minipage}	
	\label{fig_chap5_17}
\end{figure}

\refstepcounter{Sfig}
\begin{figure}[htbp]		
	\centering
	\includegraphics[width=\textwidth]{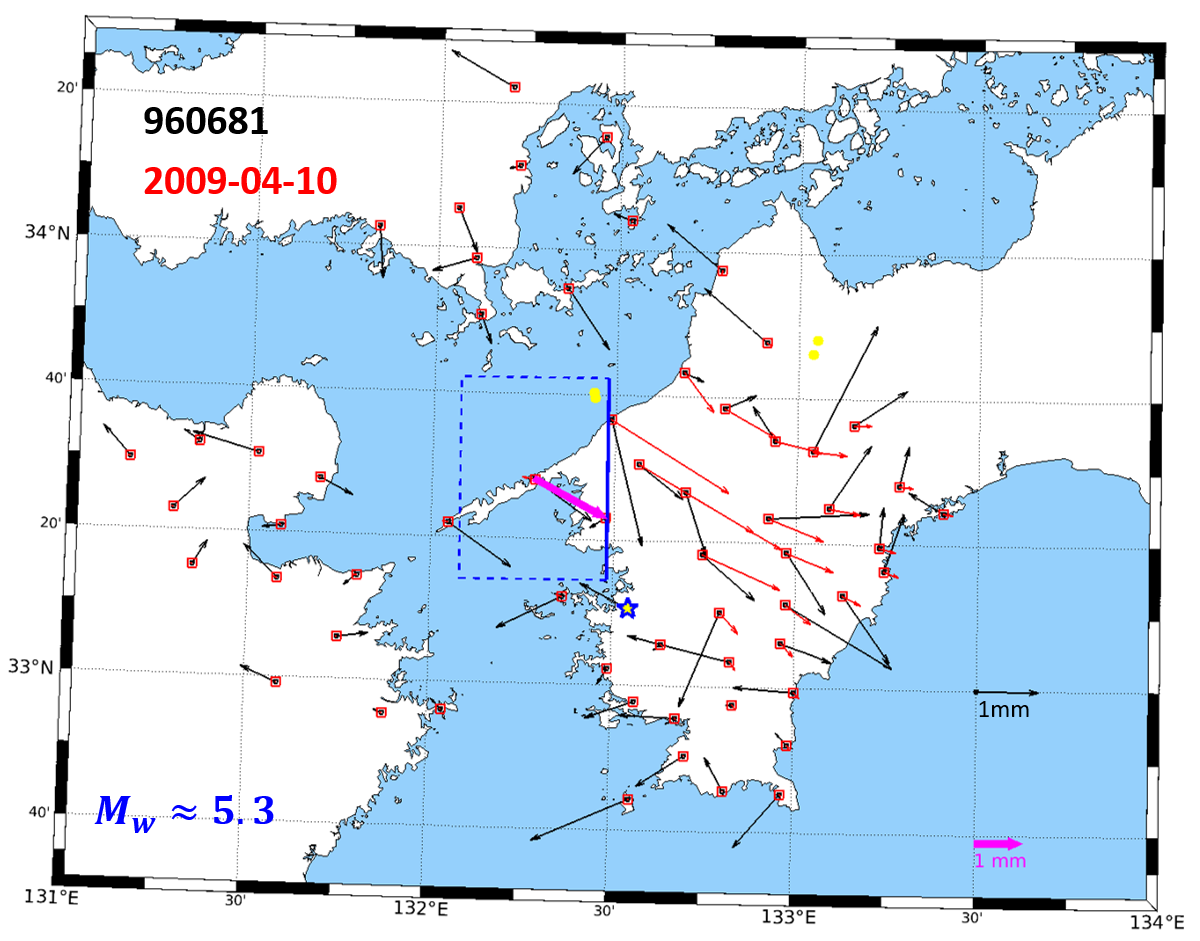}
	\begin{minipage}{0.9\textwidth}
		\textbf{Figure \theSfig.} Same as Fig. \ref{fig_chap5_11} but for a probable SSE candidate at station 960681.
	\end{minipage}	
	\label{fig_chap5_18}
\end{figure}

\refstepcounter{Sfig}
\begin{figure}[htbp]		
	\centering
	\includegraphics[width=\textwidth]{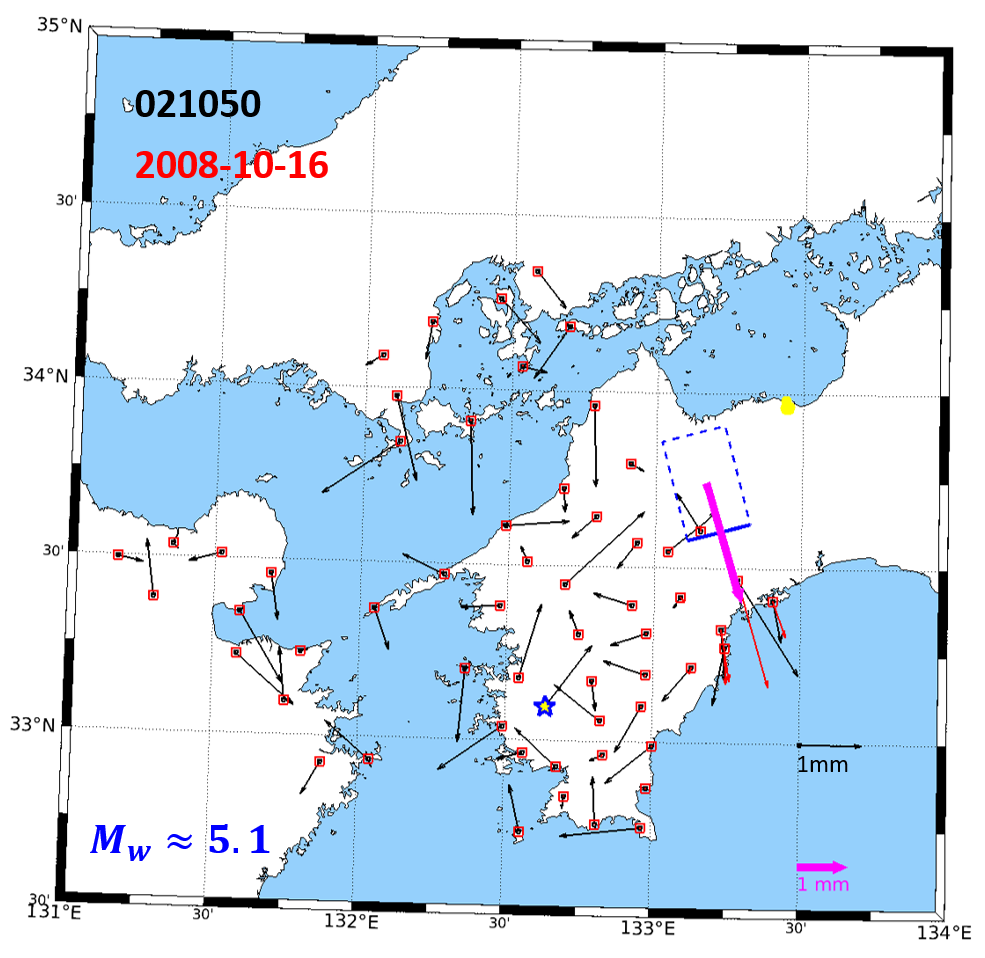}
	\begin{minipage}{0.9\textwidth}
		\textbf{Figure \theSfig.} Same as Fig. \ref{fig_chap5_11} but for a probable SSE candidate at station 021050.
	\end{minipage}	
	\label{fig_chap5_19}
\end{figure}

\refstepcounter{Sfig}
\begin{figure}[htbp]		
	\centering
	\includegraphics[width=\textwidth]{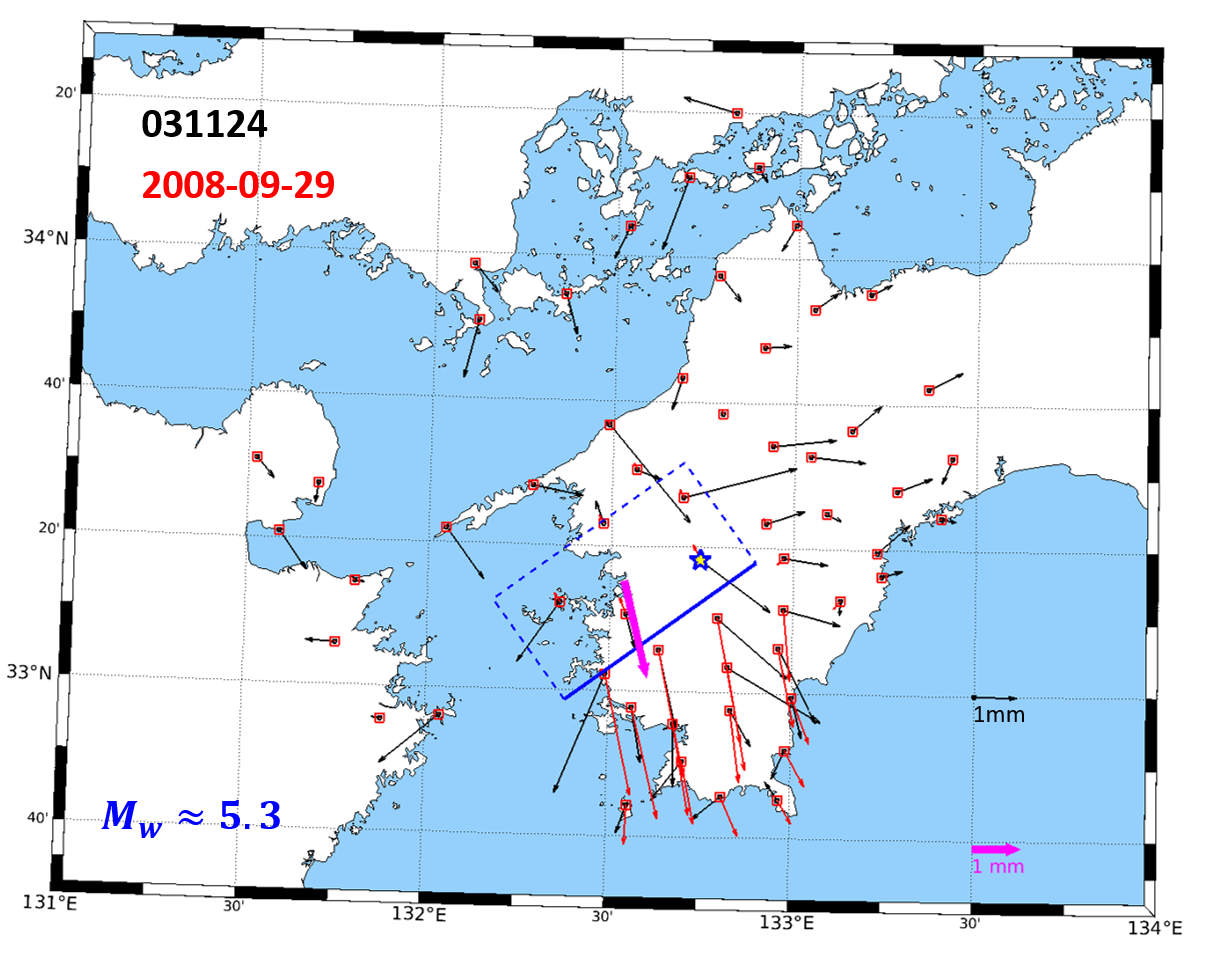}
	\begin{minipage}{0.9\textwidth}
		\textbf{Figure \theSfig.} Same as Fig. \ref{fig_chap5_11} but for a probable SSE candidate at station 031124.
	\end{minipage}	
	\label{fig_chap5_20}
\end{figure}

\refstepcounter{Sfig}
\begin{figure}[htbp]		
	\centering
	\includegraphics[width=\textwidth]{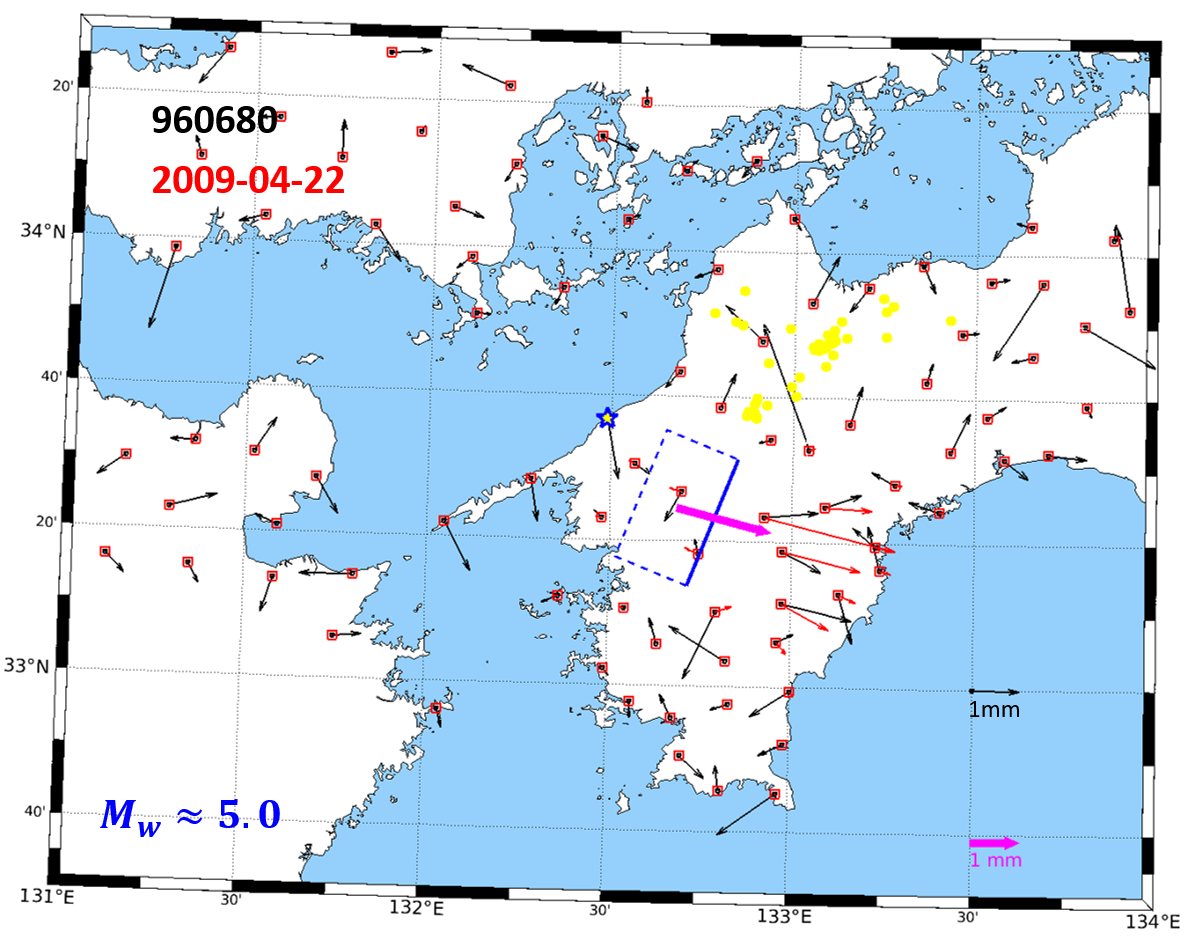}
	\begin{minipage}{0.9\textwidth}
		\textbf{Figure \theSfig.} Same as Fig. \ref{fig_chap5_11} but for a probable SSE candidate at station 960680.
	\end{minipage}	
	\label{fig_chap5_21}
\end{figure}

\refstepcounter{Sfig}
\begin{figure}[htbp]		
	\centering
	\includegraphics[width=\textwidth]{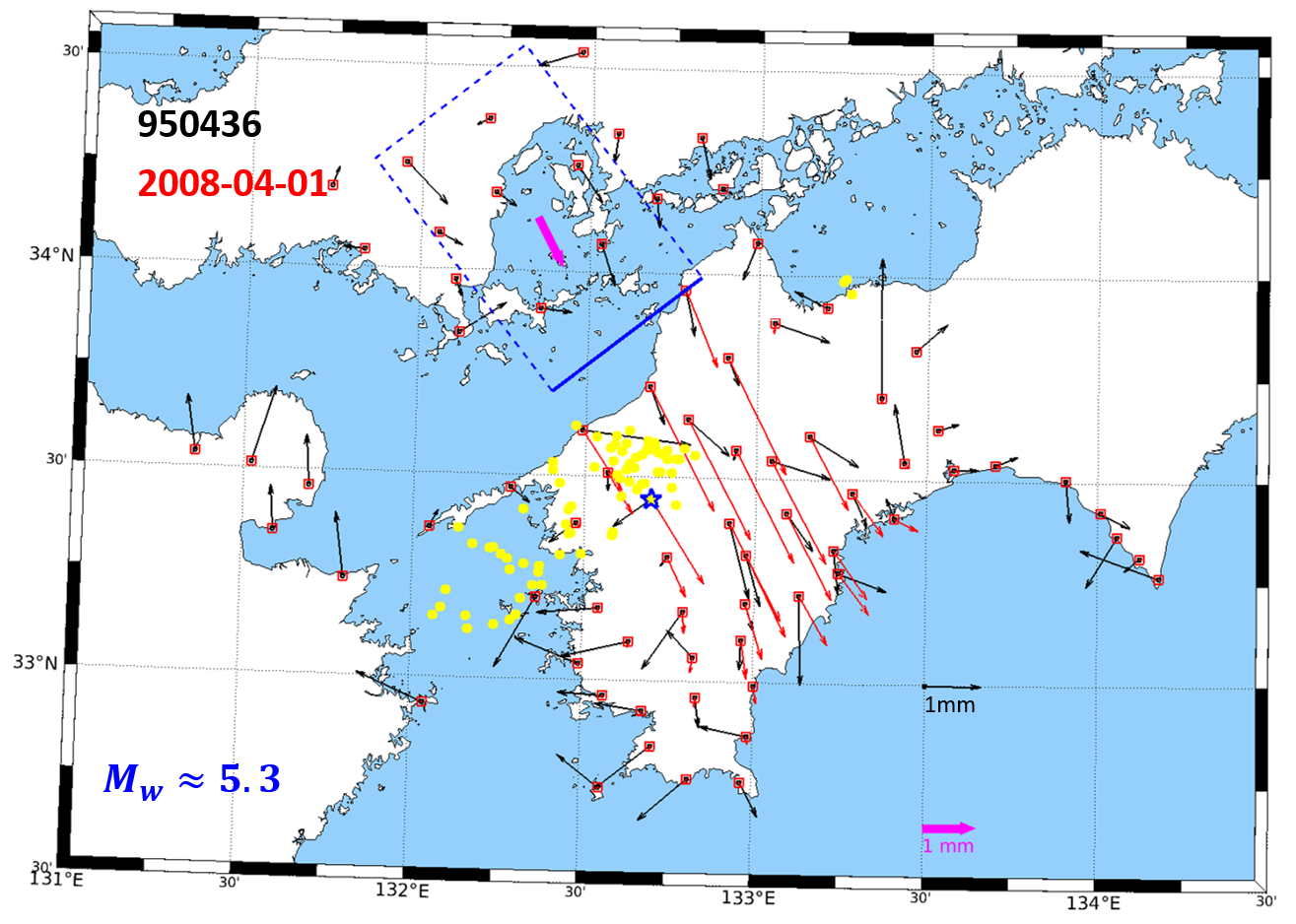}
	\begin{minipage}{0.9\textwidth}
		\textbf{Figure \theSfig.} Same as Fig. \ref{fig_chap5_11} but for a probable SSE candidate at station 950436.
	\end{minipage}	
	\label{fig_chap5_22}
\end{figure}

\refstepcounter{Sfig}
\begin{figure}[htbp]		
	\centering
	\includegraphics[width=\textwidth]{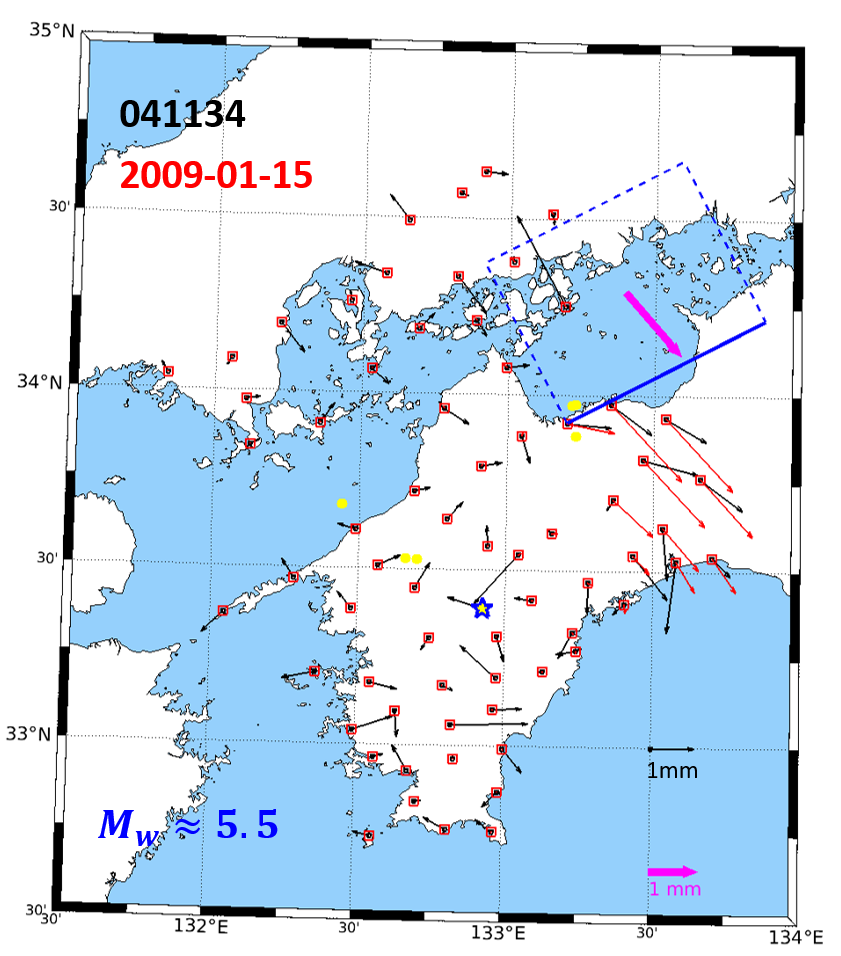}
	\begin{minipage}{0.9\textwidth}
		\textbf{Figure \theSfig.} Same as Fig. \ref{fig_chap5_11} but for a probable SSE candidate at station 041134.
	\end{minipage}	
	\label{fig_chap5_24}
\end{figure}

\refstepcounter{Sfig}
\begin{figure}[htbp]		
	\centering
	\includegraphics[width=\textwidth]{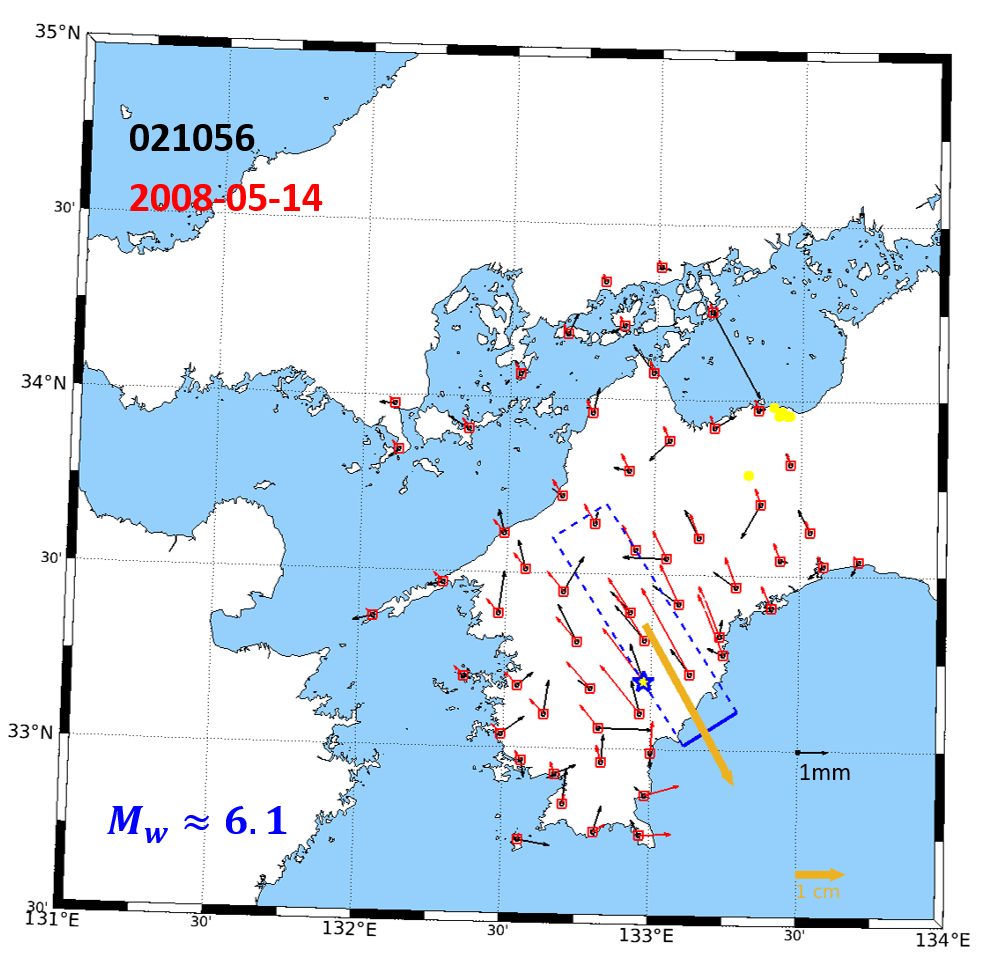}
	\begin{minipage}{0.9\textwidth}
		\textbf{Figure \theSfig.} Same as Fig. \ref{fig_chap5_11} but for a probable SSE candidate at station 021056.
	\end{minipage}	
	\label{fig_chap5_25}
\end{figure}

\refstepcounter{Sfig}
\begin{figure}[htbp]		
	\centering
	\includegraphics[width=\textwidth]{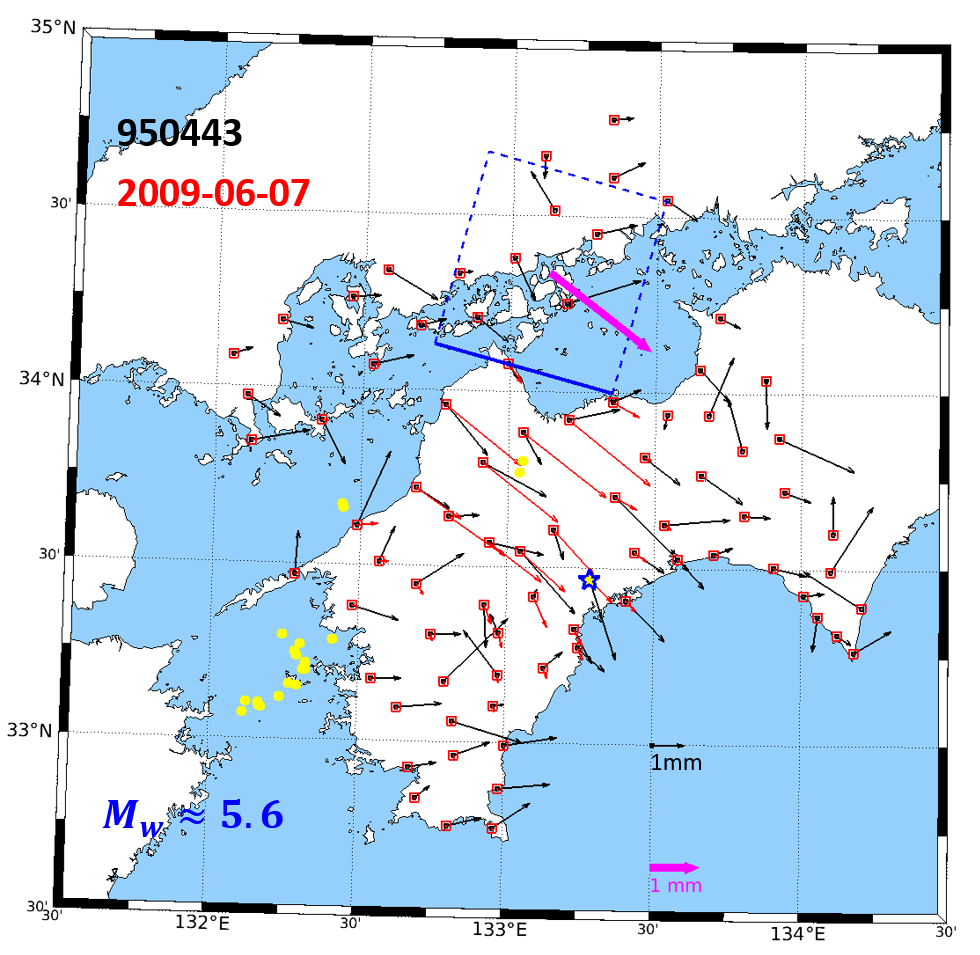}
	\begin{minipage}{0.9\textwidth}
		\textbf{Figure \theSfig.} Same as Fig. \ref{fig_chap5_11} but for a probable SSE candidate at station 950443.
	\end{minipage}	
	\label{fig_chap5_26}
\end{figure}

\refstepcounter{Sfig}
\begin{figure}[htbp]		
	\centering
	\includegraphics[width=\textwidth]{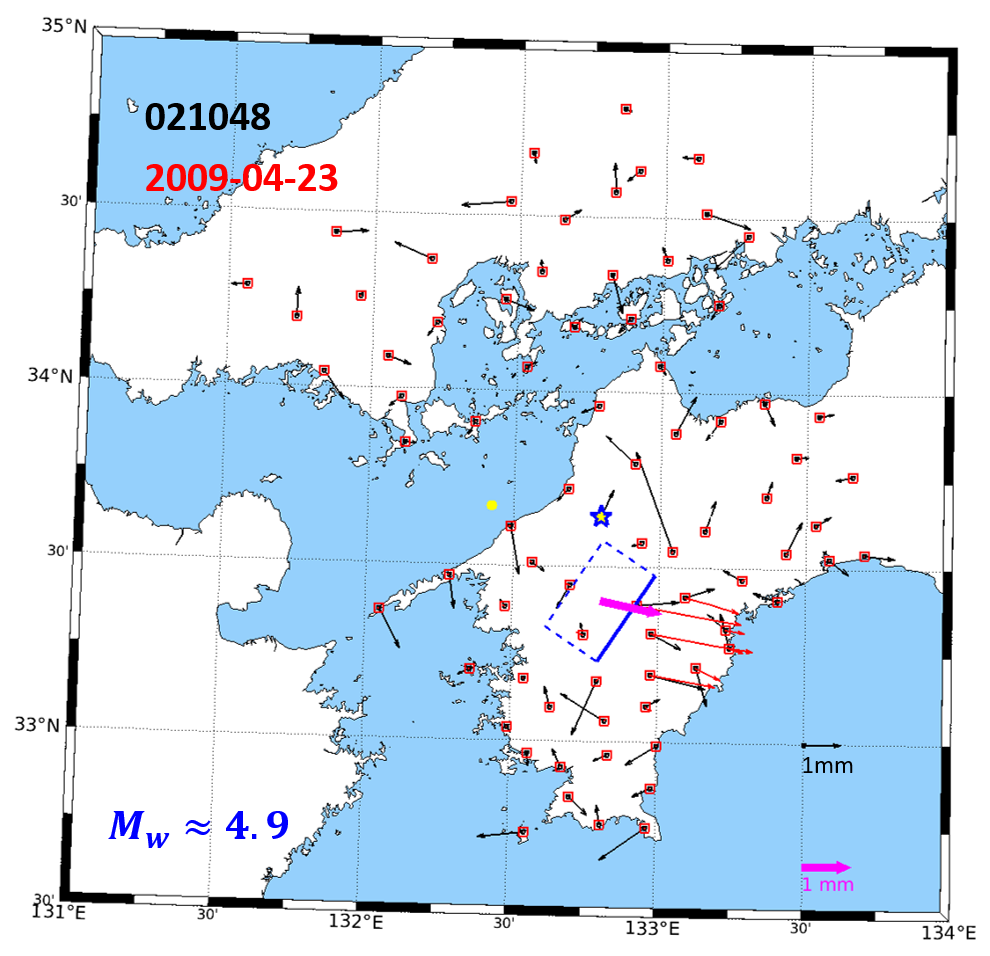}
	\begin{minipage}{0.9\textwidth} 
		\textbf{Figure \theSfig.} Same as Fig. \ref{fig_chap5_11} but for a probable SSE candidate at station 021048.
	\end{minipage}	
	\label{fig_chap5_27}
\end{figure}

\end{document}